\documentclass[usenatbib]{mn2e}
\usepackage{graphicx}

\title[LMXBs as stellar mass indicator]
{Low mass X-ray binaries as a stellar mass indicator for the host galaxy} 
 
\author[M.Gilfanov]
       {M.~Gilfanov\\
        Max-Planck-Institut f\"ur Astrophysik, 85741 Garching
            b. M\"unchen, Germany\\
	Space Research Institute, Moscow, Russia}
\date{\today}

\pubyear{2003}

\begin{document}

\maketitle

\begin{abstract}

Using results of Chandra observations of old stellar systems in 
eleven nearby galaxies of various morphological types and the census 
of LMXBs in the Milky Way, we study the population of low mass X-ray
binaries and their relation to the mass of the host galaxy.   
We show that the azimuthally averaged spatial distribution of the
number of LMXBs and, in the majority of cases, of their collective
luminosity closely follows that of the near-infrared
light. Considering galaxies  as a whole, we find that in a broad mass
range, $\log(M_*)\sim 9 - 11.5$,  the total number of LMXBs and their
combined luminosity are proportional to  the stellar mass of the host
galaxy. Within the accuracy of the light-to-mass conversion, we cannot
rule out the possibility of a weak dependence of the X/M$_*$ ratio on
morphological  type. However, the effect of such a dependence, if any,
does not exceed a factor of $\sim 1.5-2$.  

The luminosity distributions of LMXBs observed in different galaxies
are similar to each other and, with the possible exception of NGC1553,
are consistent with the average luminosity function derived from all 
data. The average XLF of LMXBs in nearby galaxies has a
complex shape and is significantly different from that of HMXBs. It
follows a power law with a differential slope of $\approx 1$ at low 
luminosities, gradually steepens at $\log(L_X)\ga 37.0-37.5$ and has a 
rather abrupt cut-off at  $\log(L_X)\sim39.0-39.5$. This value of the
cut-off luminosity is significantly, by an order of magnitude, lower
than found for high mass X-ray binaries.

\end{abstract}


\begin{table*}
\renewcommand{\arraystretch}{1.2}
\caption{The sample of nearby galaxies -- general data}
\begin{tabular}{|l|r|r|r|r||r|c|r|r|c|}
\hline
Name     & D& Morph. & Photom.& $A_e$  &  B--V  & $M_*/L_{\rm NIR}$
& $s_0$ & $m_{tot}$ & Ref.\\
        &Mpc& Type   &  Type  & arcsec &  RC3  & $M_\odot/L_\odot$
&       &           & X-ray\\
&(2)&(3)&(4)&(5)&(6)&(7)&(8)&(9)&(10)\\
\hline
NGC 4472  & 16.0 &   -4.7 &   -7.7 &  208.0 &  0.95 &  0.85 & 1.07 & 4.9 & a \\
NGC 4697  & 10.5 &   -4.7 &   -4.4 &  143.9 &  0.89 &  0.77 & 0.83 & 6.1 & b \\
NGC 5846  & 24.0 &   -4.7 &   -4.3 &  125.4 &  0.96 &  0.86 & 1.05 & 6.4 & c \\
M84       & 17.0 &   -4.2 &   -8.0 &  101.9 &  0.94 &  0.83 & 1.07 & 5.9 & d \\
NGC 1553  & 24.2 &   -2.3 &    2.1 &  131.3 &  0.87 &  0.75 & 0.81 & 5.8 & e \\
Cen A     &  3.5 &   -2.1 &   10.1 &  238.9 &  0.88 &  0.76 & 0.52 & 4.5 & f \\
NGC 1316  & 18.6 &   -1.7 &   -9.9 &  161.5 &  0.87 &  0.75 & 0.96 & 5.3 & g \\
NGC 1291  &  8.9 &    0.1 &  -10.2 &  109.2 &  0.91 &  0.80 & 1.09 & 5.5 & h \\
M81       &  3.6 &    2.4 &    2.2 &  396.4 &  0.82 &  0.70 & 0.64 & 3.8 & i \\
M31       &  0.8 &    3.0 &    5.4 & 2501.2 &  0.68 &  0.56 & 0.53 & 0.5 & k\\
M101      &  7.2 &    5.9 &    5.9 &  688.9 &  0.44 &  0.39 & 0.54 & 6.9 & j\\
\hline
\end{tabular}\\
\flushleft
(2) -- distance; 
(3), (4) -- Morphological and photometric type 
(HyperLeda Database, \citet{leda}, http://www-obs.univ-lion1.fr/hypercat); 
(5) -- B-band circular effective aperture (diameter) 
\citep[RC3 catalog,][]{rc3}; 
(6) -- B-V color index corrected for galactic and internal extinction 
\citep[RC3 catalog,][]{rc3};
(7) -- K-band mass-to-light ratio calculated as described in
subsection \ref{sec:nir_mass};
(8), (9) -- slope and asymptotic total K-band magnitude obtained from the
near-infrared growth curve fit as described in subsection
\ref{sec:nir_gcurve};
References for X-ray data: 
(a) -- \citet{ngc4472}, 
(b) -- \citet{ngc4697}, 
(c) -- \citet{ngc5846}, 
(d) -- \citet{m84}, 
(e) -- \citet{ngc1553}, 
(f) -- \citet{cena}, 
(g) -- \citet{ngc1316}, 
(h) -- \citet{ngc1291}, 
(i) -- \citet{m81}, 
(j) -- \citet{m31},
(k) -- \citet{m101}. 
\label{tab:sample}
\end{table*}

\section{Introduction}
\label{sec:intro}

X-ray binaries are known to be an important contributor to the
emission of the host galaxy in the X-ray energy domain \citep[see
e.g. a recent review by][]{fabbiano2003}. In the
absence of an actively accreting supermassive black hole and/or a
large amount of hot gas, they 
account for the major fraction of the host galaxy's X-ray luminosity,
as illustrated by the example of the Milky Way.  
Depending on the mass of the donor star, X-ray binaries are subdivided
into low, $M_{\rm opt}\la 1$ M$_\odot$, and high, $M_{\rm opt}\ga 8$  
M$_\odot$, mass X-ray binaries (LMXB and HMXB respectively) having
very different evolutionary time scales. The life time of high
mass X-ray binaries is limited by the nuclear time scale of the
massive donor star, $\la 10^6-10^7$ years, i.e. is comparable to the
duration of a star formation event. The X-ray active phase of a
low mass X-ray binary is delayed with respect to formation of the
compact object by the nuclear evolution time scale of the donor star
and/or binary orbit decay time scale,  $\sim 10^9-10^{10}$ years
\citep{xrb_evol_rev}. The duration of the subsequent X-ray
active phase can be of the  same order or shorter \citep{pods2002,
pfahl2003}.   
These numbers define different relations of high and low mass X-ray
binaries to the star formation process. In a naive  picture, short
living HMXBs provide prompt emission during and shortly after the star
formation event. The LMXBs ignition/life time exceeds, by several orders
of magnitude, the characteristic time 
scale of a star formation event and  might be comparable to the life
time of the host galaxy. Consequently, one might expect that their
population is defined by the cumulative effect of the star formation
episodes experienced by the host galaxy throughout its lifetime, 
i.e. is proportional to its total stellar mass. \citet{grimm1} studied
the population of X-ray binaries in the Milky Way and provided a
preliminary calibration of the LMXB--M$_*$ and HMXB--SFR relations. 

In a more realistic approach, one should consider  details of
the formation and evolution of X-ray binaries and take into account a 
continuous distribution of the masses of the donor stars, which
depends on the star  formation history of the host galaxy. The X-ray
emission from X-ray binaries following a star formation event should
be a continuous function of time, 
dominated by HMXBs at the early times and by LMXBs at the later times
\citep[e.g.][]{ghosh01} with a possible contribution of intermediate
mass X-ray binaries (IMXB) in between \citep{pods2002,
pfahl2003}. The role of IMXBs in other
galaxies is not well studied, neither theoretically nor
observationally. In the case of the Milky Way galaxy, the  
intermediate mass range,  $M_{\rm opt}\sim 2-5 $ M$_\odot$, is
sparsely populated. In addition, a number of effects, such as  
metallicity variations, shape of the IMF, etc can complicate the
picture. 

Various correlations between the X-ray and optical/far-infrared
properties of galaxies have been noted and studied in the past
\citep[e.g.][]{griffiths90, david92}. Although the X-ray data lacked
sufficient spatial resolution and adequate energy coverage,
\citet{david92} suggested that the existence of such correlations could be
understood in a two component model consisting of an old and young
population having different relations to the current star formation
activity and stellar content of the galaxy.

The Chandra observatory, thanks to its sub-arcsecond angular
resolution, presented for the first time an opportunity to study
the population of X-ray binaries in nearby (within $\la 15-20$ Mpc
from the Sun) galaxies in a nearly 
confusion-free regime and to investigate their relation to the
fundamental parameters of the host galaxy, such as stellar 
mass and star formation rate. Optical identifications of the
compact sources in the nearby galaxies are (potentially) available
only for the most nearby galaxies and there is a limited possibility to 
distinguish between LMXBs and HMXBs in other galaxies, based on their
X-ray emission in the Chandra bandpass. However, in the naive
picture, outlined above, one might expect, that at high star formation
rates, e.g. in young distant star-forming galaxies or in starbursts in
merging and interacting galaxies, the population of compact sources is  
dominated by HMXBs, whereas in old stellar systems, e.g elliptical
galaxies or bulges of spiral galaxies, the primary contribution comes
from LMXBs. Quantitatively, one might use the well studied population
of X-ray binaries in the Milky Way to approximately calibrate the
relative  abundance of HMXB and LMXB as a function of the  SFR/M$_*$
ratio \citep{grimm1, grimm2}.

\begin{table*}
\renewcommand{\arraystretch}{1.4}
\caption{The sample of nearby galaxies -- X-ray and near-infrared data}
\begin{tabular}{|l|r|r|r|c|r|r|r|r|c|c|c|}
\hline
%%%
%Name   & $a_X$ & $L_{\rm NIR}$ & $M_*$ & $L_{X,min}$  & $N_X$ & $L_X$& 
%$\Delta N_X$ & $\Delta L_X$ & $N_X/M_*$ & $L_X/M_*$ & R \\
%& arcsec& $10^{11}$ L$_\odot$ & $10^{11}$ M$_\odot$ &
%$10^{38}$erg/s & & $10^{39}$erg/s & & $10^{39}$erg/s & &  & \\\
%%%
Name   & $a_X$ & $L_{\rm NIR}$ & $M_*$ & $L_{X,min}$  & $N_X$ & $L_X$& 
$\Delta N_X$ & $\Delta L_X$ & $N_X/M_*$ & $L_X/M_*$ & mode \\
& (2) & (3) & (4) & (5) & (6) & (7) & (8) & (9) & (10) & (11) & (12) \\
\hline
NGC 4472  &  50--200 & 18.70 & 15.82 & 1.0 & $28.7 \pm 5.6$ & $ 87.2_{-10.2}^{+11.9}$ & 196.0 &  57.0 & $14.2 \pm  2.8$ & $ 9.1_{- 1.1}^{+ 1.2}$ & 0.97 \\
NGC 4697  &  30--300 &  4.87 &  3.77 & 1.0 & $12.1 \pm 4.0$ & $ 24.2_{- 4.2}^{+ 5.2}$ &  82.7 &  24.0 & $25.2 \pm  8.3$ & $12.8_{- 2.2}^{+ 2.7}$ & 0.93 \\
NGC 5846  & 100--300 & 10.90 &  9.36 & 3.0 & $ 4.9 \pm 3.0$ & $ 25.7_{- 3.2}^{+ 4.7}$ & 153.2 &  65.7 & $16.9 \pm 10.4$ & $ 9.8_{- 1.2}^{+ 1.8}$ & 0.91 \\
M84      &  30--200 & 13.10 & 10.92 & 1.0 & $19.1 \pm 4.8$ & $ 54.9_{- 7.6}^{+ 9.2}$ & 130.6 &  38.0 & $13.7 \pm  3.4$ & $ 8.5_{- 1.2}^{+ 1.4}$ & 0.95 \\
NGC 1553  &  30--300 & 43.90 & 32.94 & 2.0 & $18.8 \pm 5.0$ & $125.0_{-12.3}^{+15.0}$ & 313.0 & 117.1 & $10.1 \pm  2.7$ & $ 7.4_{- 0.7}^{+ 0.9}$ & 0.97 \\
NGC 1553  &  30--100 & 20.50 & 15.38 & 2.0 & $ 7.7 \pm 2.9$ & $ 23.8_{- 3.2}^{+ 4.1}$ & 127.8 &  47.8 & $ 8.8 \pm  3.3$ & $ 4.7_{- 0.6}^{+ 0.8}$ & 0.94 \\
%NGC 1553  & 100--300 & 23.40 & 17.56 & 2.0 & $11.1 \pm 4.1$ & $104.0_{-12.4}^{+15.7}$ & 185.2 &  69.3 & $11.2 \pm  4.1$ & $ 9.9_{- 1.2}^{+ 1.5}$ & 0.95 \\
Cen A  & 100--300 &  1.84 &  1.40 & 0.1 & $36.6 \pm 6.3$ & $ 16.3_{- 4.0}^{+ 5.2}$ &   0.0 &   0.0 & $26.1 \pm  4.5$ & $11.6_{- 2.8}^{+ 3.7}$ & 0.89 \\
NGC 1316  &  30--150 & 18.60 & 13.96 & 1.0 & $23.1 \pm 5.0$ & $ 56.0_{- 7.2}^{+ 8.6}$ & 157.4 &  45.8 & $12.9 \pm  2.8$ & $ 7.3_{- 0.9}^{+ 1.1}$ & 0.96 \\
NGC 1291  &  30--200 &  4.70 &  3.74 & 0.3 & $27.7 \pm 5.5$ & $ 28.2_{- 5.5}^{+ 6.8}$ &  39.0 &   6.9 & $17.8 \pm  3.5$ & $ 9.4_{- 1.8}^{+ 2.3}$ & 0.93 \\
M81      &  30--200 &  3.07 &  2.14 & 0.1 & $22.1 \pm 4.9$ & $ 13.2_{- 3.9}^{+ 6.1}$ &   0.0 &   0.0 & $10.3 \pm  2.3$ & $ 6.2_{- 1.8}^{+ 2.9}$ & 0.81 \\
M31      &  30--500 &  1.86 &  1.05 & 0.1 & $19.8 \pm 4.5$ & $  7.4_{- 2.3}^{+ 3.7}$ &   0.0 &   0.0 & $18.9 \pm  4.3$ & $ 7.1_{- 2.2}^{+ 3.5}$ & 0.79 \\
M101     &  30--120 &  0.30 &  0.12 & 0.1 & $ 3.3 \pm 2.2$ & $  0.9_{- 0.3}^{+ 1.5}$ &   0.0 &   0.0 & $27.7 \pm 18.3$ & $ 7.2_{- 2.7}^{+12.5}$ & 0.36 \\
Milky Way  &(a)&  4.02 &  2.27 & 0.1 & $15.0 \pm 3.9$ & $  7.9_{- 2.5}^{+ 5.3}$ &   0.0 &   0.0 & $ 6.6 \pm  1.7$ & $ 3.5_{- 1.1}^{+ 2.3}$ & 0.70 \\
\hline
\end{tabular}\\
\flushleft
(2) -- X-ray aperture (diameter) range used for the source selection, arcsec;
(3) -- K-band luminosity in the aperture range $a_X$ specified in the
previous column, $10^{10}$  L$_\odot$; 
(4) -- stellar mass in units of $10^{10}$  M$_\odot$ in the aperture
range  $a_X$, computed from K-band luminosity and infrared mass-to-light
ratio (subsection \ref{sec:nir_mass}, Table \ref{tab:sample});
(5) -- lower X-ray luminosity limit used for the source selection, 
$10^{38}$erg/s;
(6), (7) -- number of sources detected in the aperture range  $a_X$
with luminosity exceeding $L_{X,min}$ and their total luminosity in
units of $10^{38}$erg/s;
(8) -- number of sources between $10^{37}$ erg/s
and lower luminosity limit $L_{X,min}$ estimated using the average
luminosity function as described in section \ref{sec:x_nir_ratios};
(9) -- the same as (8) but for the total luminosity, $10^{38}$erg/s;
(10) -- ratio of the total number of sources with $L_X>10^{37}$ erg/s
to the stellar mass, sources per $10^{10}$ M$_\odot$;
(11) -- $L_X$--stellar mass ratio, $10^{38}$ erg/s per $10^{10}$ M$_\odot$;
(12) -- ratio of most probable value of the total luminosity (mode of
the probability distribution) to the expectation mean (section
\ref{sec:stat})\\
(a) -- using the LMXB sources located at $X>0$ and at the projected
galactocentric distance of 1--10 kpc (section \ref{sec:mw_gcurve}); the NIR
luminosity was estimated correspondingly, using the NIR growth curve
of M31.
\label{tab:sample2}
\label{tab:xray}
\end{table*}

Based on the Chandra observations of the nearby star forming galaxies
and studies of the high mass X-ray binaries  in the
Milky Way and SMC, \citet{grimm2}  explored quantitatively the relation 
between the population of HMXB sources and the current value of the
star formation rate in the host galaxy. They found that, in the broad
range of the star formation regimes and rates, the X-ray luminosity
distribution of HMXBs in $\log(L_X)\sim 35.5-40.5$ luminosity range can
be approximately described by a ``universal'' luminosity function,
whose  shape is a $L_X^{-1.6}$ power law with a cut-off at
$\log(L_X)\sim 40.5$ and with a normalization proportional to the star  
formation rate.  They showed that the $L_X$--SFR relation between
total luminosity of HMXBs and the star formation rate derived for the 
nearby galaxies holds for distant star forming galaxies at redshifts
up to $\sim 1.3$ in the Hubble Deep Field North.

In the present paper we study the population of LMXBs and its relation
to the stellar mass of the host galaxy, using results of Chandra
observations of nearby galaxies of various morphological types,
ranging from ellipticals to spirals. 
The structure of the paper is as follows. In section
\ref{sec:sample} we describe our sample and details of the X-ray and  
near-infrared data. In particular, in the subsection \ref{sec:cxb} we
point out the importance of contamination by the  CXB sources and in
the subsection \ref{sec:stat} we discuss the influence of statistical
effects on the total luminosity of X-ray binaries in a galaxy. 
In section \ref{sec:gcurves} we
present  X-ray and near-infrared growth curves and demonstrate that
the spatial  distribution of X-ray binaries closely follows that of
the near-infrared light. In section \ref{sec:xlf} we show that
the luminosity distributions in different galaxies from our sample
have similar shape and derive the average luminosity function of
LMXBs. In section \ref{sec:x_nir_ratios} we show that the total number
of LMXBs and their collective luminosity is proportional to the stellar
mass of the host galaxy. In section \ref{sec:discussion} we discuss
various aspects and astrophysical applications of the obtained results
as well as peculiar case of NGC1553 and in section \ref{sec:summary}
we summarize our findings.

\section{The sample}
\label{sec:sample}

Our sample includes 11 nearby galaxies of various morphological types
observed with Chandra. To minimize the contribution of high mass X-ray
binaries, we focus our study on old stellar systems -- E, S0 galaxies
and the bulges of spiral galaxies, typically having negligible star
formation activity.\footnote{
We compare bulge and disk population of X-ray sources in
spiral galaxies in section \ref{sec:bulge_disk}} 
The list of galaxies along with their general optical/NIR 
characteristics is given in Table \ref{tab:sample}. This sample of
external nearby galaxies was complemented by the population of low
mass X-ray binaries in the Milky Way, based on the results of
\citet{grimm1}.

\begin{figure*}
\hbox{
\resizebox{0.5\hsize}{!}{\includegraphics{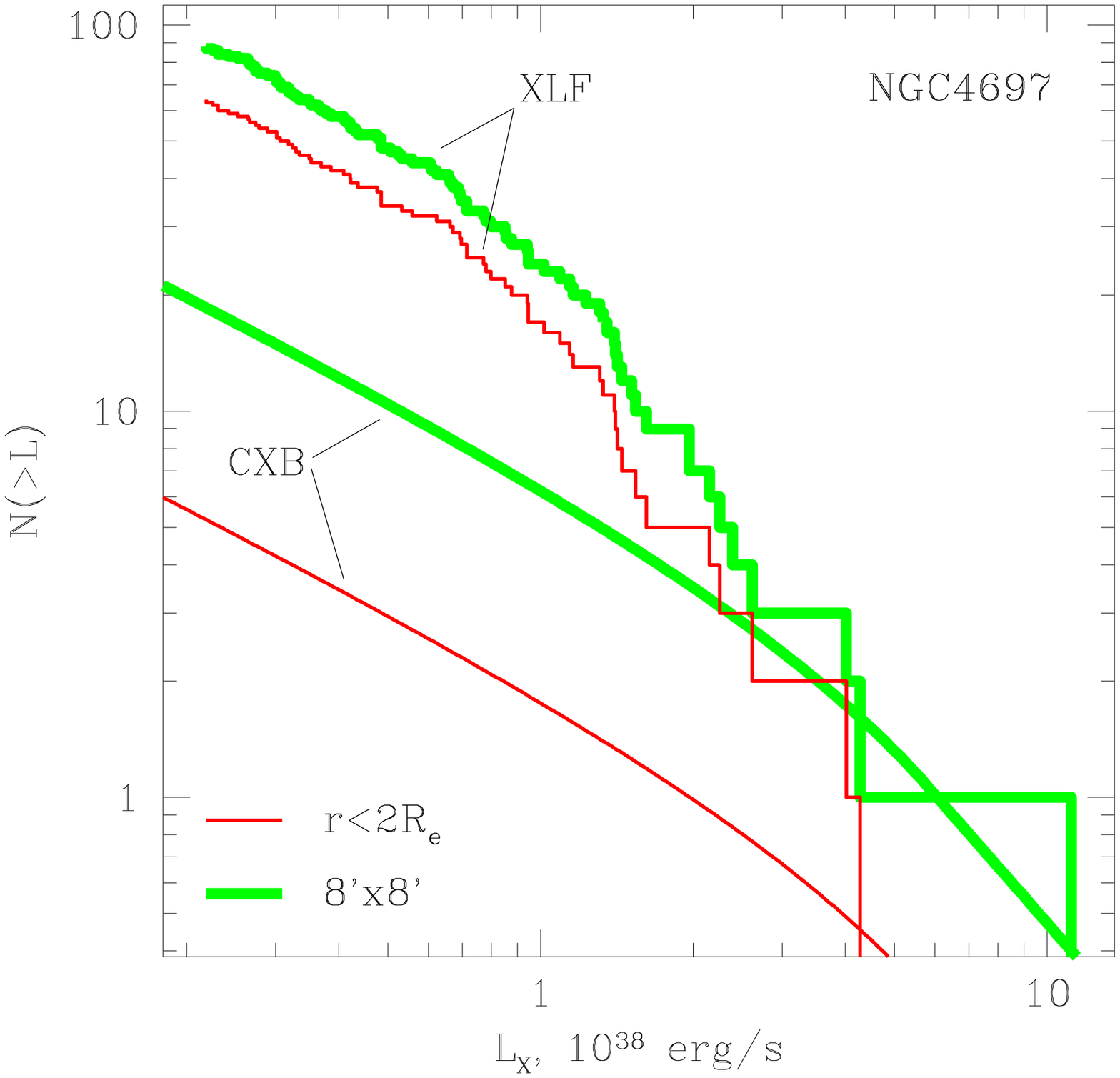}}
\resizebox{0.5\hsize}{!}{\includegraphics{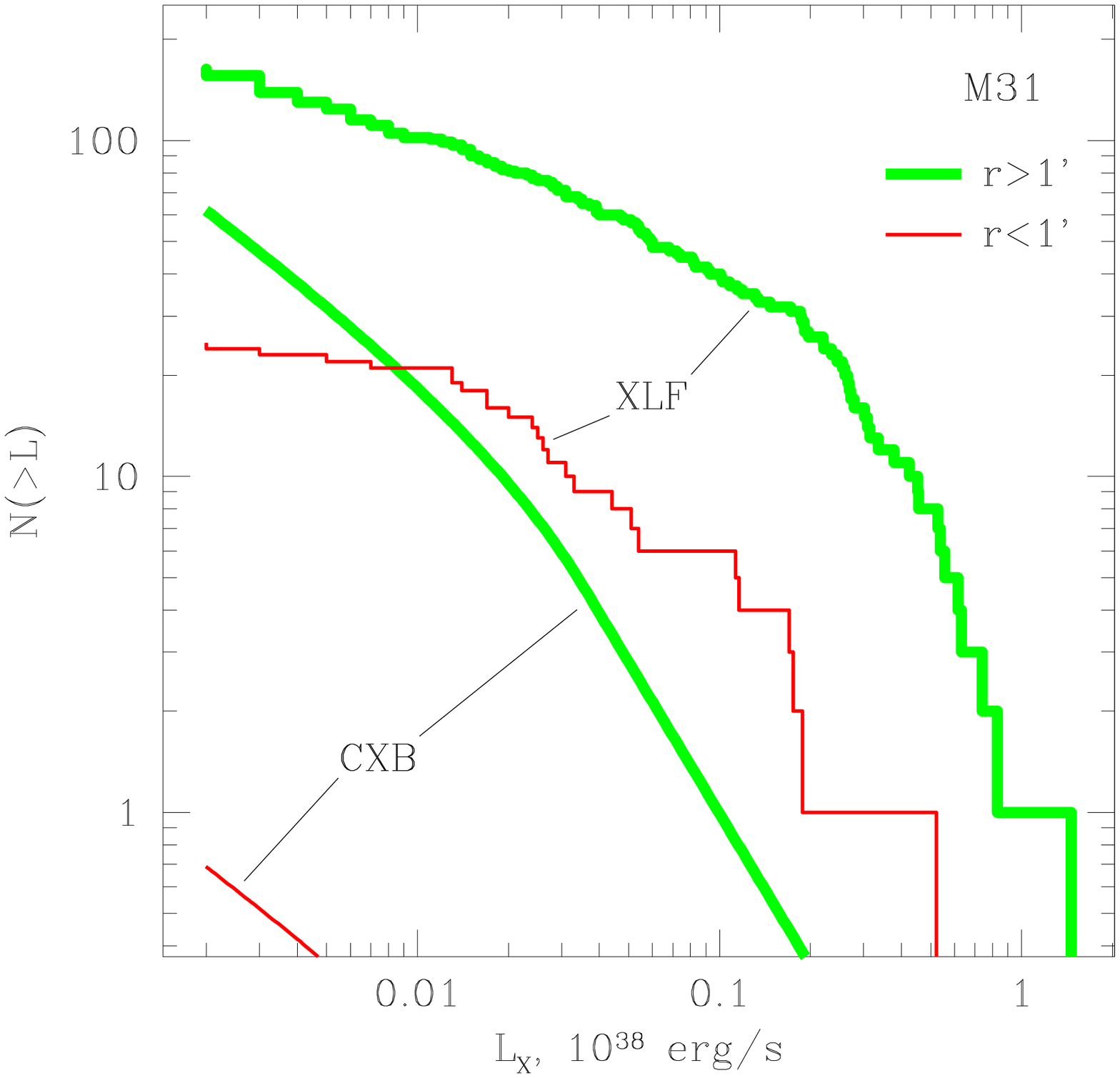}}
}
\caption{Illustration of importance of the  contribution of CXB
sources to observed XLFs of galaxies, depending on the shape
of the luminosity function and the distance. 
{\em Left:} XLF of elliptical galaxy NGC4697 (distance 10.5 
Mpc) obtained in a $\sim 40$ ksec Chandra observation
\citep{ngc4697}. The histograms show luminosity functions of the
central 2 effective radii (thin histogram) and for entire ACIS-S3 chip
($\sim 8\arcmin\times 8\arcmin$, thick grey  histogram), the solid
lines show expected contribution of the background sources. Importance
of the CXB contribution on the bright end of the luminosity function
is apparent at large aperture.  
{\em Right:} Chandra ACIS-I observations of M31 \citep{m31} (distance
0.78 Mpc). The thin and thick lines and histograms correspond to
central $1\arcmin$ and it's exterior respectively. 
}
\label{fig:lf_cxb}
\end{figure*}

\subsection{X-ray data}
\label{sec:xray_data}

To study the spatial and luminosity distribution of X-ray binaries in
11 external galaxies, we used published lists of X-ray sources 
detected by Chandra. The references to the original X-ray publications
are given in Table \ref{tab:sample}.  
The Chandra source lists were filtered as follows.
If the central X-ray source  associated with the galactic nucleus was
present, it was excluded from the list. Identified 
foreground and background sources as well as sources having
statistically significant spatial extent were also excluded. 
With two exceptions we adopted the distances from the original
X-ray publications. For NGC4697 we corrected the distance to 10.5 Mpc, 
the value determined recently by \citet{dist4697} based on planetary
nebulae study. This distance is consistent with surface brightness
fluctuation analysis and is significantly smaller than usually assumed 
value of 24 Mpc. For NGC5846 we used the distance of 24 Mpc as
determined by \citet{dist_5846} from I-band surface brightness
fluctuation analysis.

The source detection in the original Chandra publications was carried
out in different energy bands, and the count rates were converted to
the energy flux under different assumptions regarding the source
spectrum. This does not affect the growth curve analysis presented in
section \ref{sec:gcurves} but might introduce additional systematic
dispersion in studying the luminosity functions and X-ray-to-mass
ratios. We estimate that in our sample of galaxies these uncertainties
do not exceed $\sim 20-30\%$  \citep[cf.][]{ngc4472}. 
The difference in the source detection energy range
could also introduce different spectra-dependent selection biases.
However, considering the shape of the sensitivity curve of the Chandra
ACIS detectors, and the energy ranges used for the source detection
(with lower and upper boundaries in the range 0.3--0.5 and 7-10 keV) we 
do not expect significant systematic differences in the selection
effects for the source samples obtained by different authors.  

Important in constructing X-ray growth curves and luminosity functions
is the completeness limit of the source lists. The point source
sensitivity is affected by a number of factors.  Most important
in the context of studying the population of compact sources in 
galaxies are the degradation of the point spread function at large
off-axis angles and the diffuse X-ray emission, the latter being
especially significant for gas-rich elliptical galaxies.
These effects might result in non-uniform sensitivity across the
Chandra field of view and lead to apparent flattening of the
luminosity function at the low luminosity end and distortion of the
radial profile/growth curve \citep[Fig.\ref{fig:incompl}, see
also][]{m84,ngc1316}.  
Whenever completeness analysis was performed in the original
publication, the derived value of the completeness limit was used. In
all other cases we assumed the completeness limit of $\approx 2-3$
times of the detection sensitivity.

\subsection{Contribution of CXB sources}
\label{sec:cxb}

In the central regions of the galaxies, the surface density of the
compact sources is sufficiently high that the contribution of 
CXB sources (background AGNs) can be safely neglected. In outer
parts, however, the surface density of X-ray binaries  becomes
comparable to or smaller than the average density of CXB sources
(Fig.\ref{fig:profiles})  and
the contribution of the latter needs to be taken into account. This
plays an important role in studying the growth curves and X-ray
luminosity functions within large aperture 
\citep[c.f.][]{ngc1553, m84, ngc1316}.

To estimate the contribution of the CXB sources, we used results of
the Chandra 1 Msec survey of the Chandra Deep Field South
\citep{rosati02}. As the energy range for source detection 
included $\sim 0.5-2$ keV energy range where the Chandra sensitivity
peaks, we used the soft band counts, rescaling the flux to the energy
band of the source list under consideration. 
Based on the results of the Chandra CDF-S survey the log(N)--log(S) 
distribution of the CXB sources in the 0.5--2 keV energy range  was
assumed to have the form: 
\begin{eqnarray}
\frac{dN}{dS}=\left\{ \begin{array}{ll}
K S^{-\alpha_1}	& \mbox{$S<S_b$}\\
K S_b^{\alpha_2-\alpha_1}S^{-\alpha_2}	& \mbox{$S\ge S_b$}
\end{array}
\right.
\label{eq:cxb}
\end{eqnarray}
where the break flux is $S_b=1.4\cdot 10^{-14}$ erg/s/cm$^2$, the
differential slope before the break is $\alpha_1=1.63\pm0.13$ and
equals to Euclidean value after the break, $\alpha_2=2.5$. The
normalization $K=291\pm20$ corresponds to $N(S>2\cdot
10^{-15})=380\pm80$ sources/deg$^2$. Note, that both $S$ and $S_b$ in
eq.(\ref{eq:cxb}) refer to the 0.5--2 keV flux and are expressed in
units of $2\cdot 10^{-15}$ erg/s/cm$^2$. 
In order to use the above log(N)--log(S) relation to estimate the number
of CXB sources $N_{CXB}(S>S_{E_1-E_2})$, the threshold flux
$S_{E_1-E_2}$ should be transformed to the 0.5--2 keV energy band and
corrected for difference in the assumed spectral shape. 

The importance of the CXB source contribution, especially when
working with large aperture, significantly exceeding the effective
radius of the 
galaxy, is illustrated in Fig.\ref{fig:lf_cxb}. Due to the presence of
a break in the log(N)--log(S) for background sources (corresponding to
a luminosity of  $1.7\cdot 10^{38}(d/10~{\rm Mpc})^2$ erg/s) at which
the slope changes by $\Delta\alpha\sim 0.9-1$, the contribution of the
CXB sources depends critically on the galaxy distance.
For relatively distant galaxies, $D\ga 10-15$ Mpc, it is more prominent
at the high luminosity end of the XLF.   
Although existence of the ultra-luminous X-ray sources with
luminosities exceeding $\sim 10^{38}-10^{39}$ erg/s is beyond any
doubts, Fig.\ref{fig:lf_cxb} (left panel) demonstrates  that
some fraction of them might be in fact background
AGNs.\footnote{Note that the brightest ULX sources observed in the
star-forming galaxies with $\log(L_X)\ga 39.5-40$ are less subject to
CXB contamination than ``dimmer'' ULXs in ellipticals, having
typically $\log(L_X)\sim 38.5-39.5$.} For nearby galaxies on the other hand,
the contribution of CXB sources is more significant
in the lower luminosity end of the XLF (Fig.\ref{fig:lf_cxb}, right
panel). This can, in particular, distort the slope of the
XLF at low $L_X$ in the outer parts of the galaxies.

An important factor that precludes the precise subtraction of the CXB 
contribution is clustering of the  background sources on various
angular scales, including sub-arcmin 
scales \cite[e.g][]{vikhl95}  leading to significant field-to-field  
variations in their log(N)-log(S) distribution
\citep[e.g.][]{rosati02,elais}.  
Another factor that complicates accounting for CXB contribution to the
total luminosity is the effects of small numbers statistics
\citep{grimm2,stat}, discussed in more detail in subsection
\ref{sec:stat}.

\begin{figure*}
\hbox{
\resizebox{0.5\hsize}{!}{\includegraphics{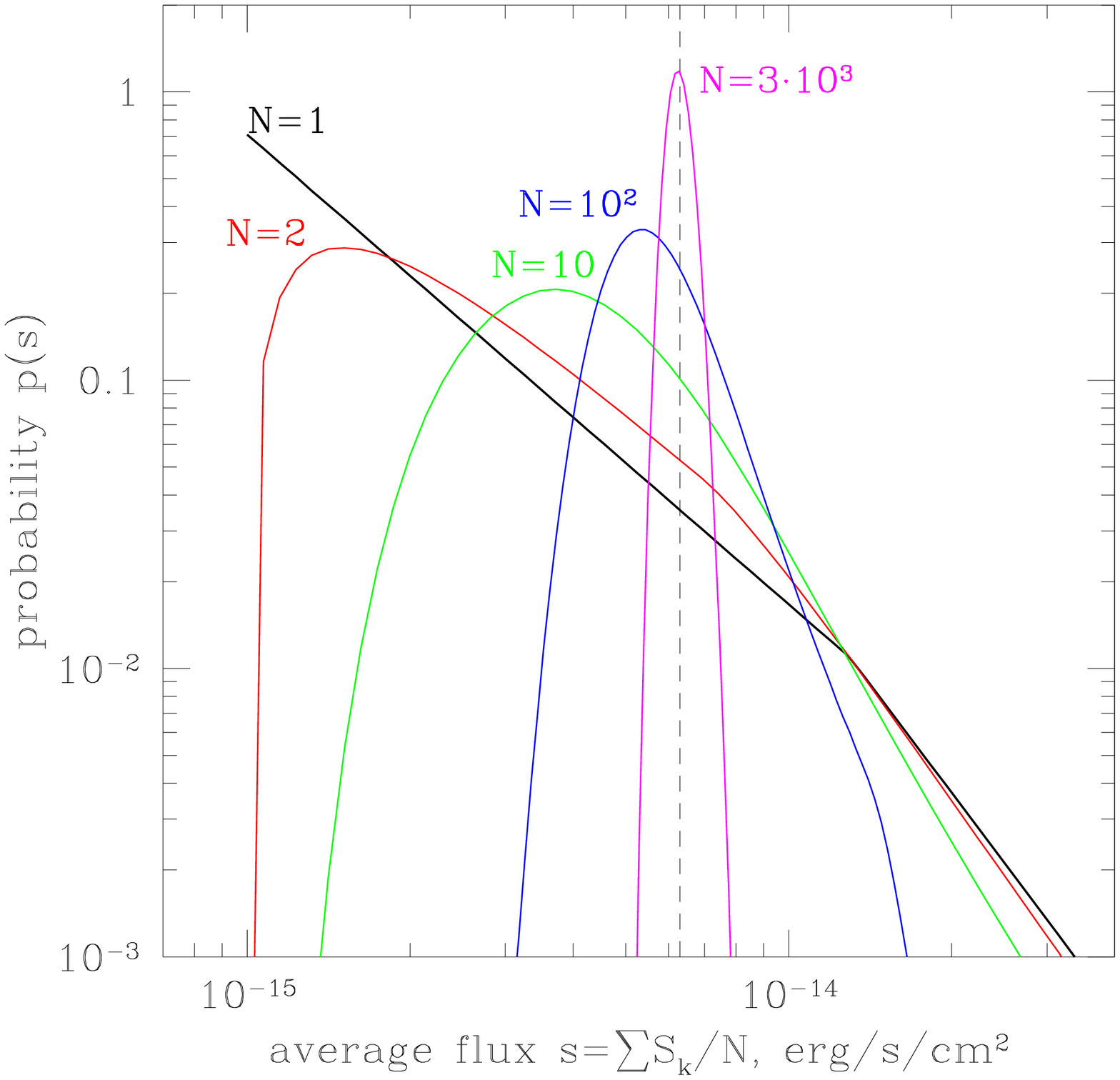}}
\resizebox{0.5\hsize}{!}{\includegraphics{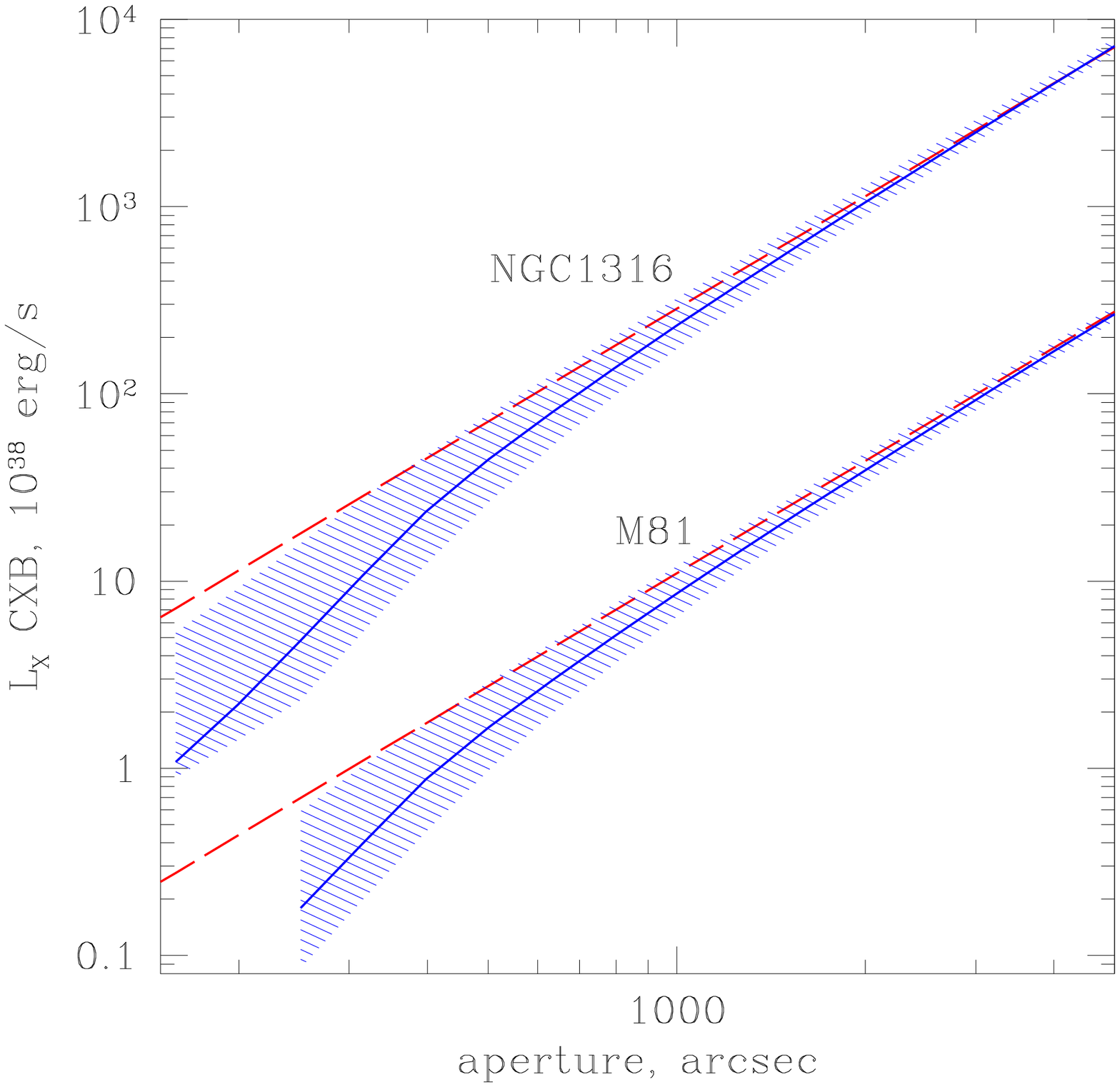}}
}
\caption{Illustration of the effects of small number statistics, using
the integrated flux of the CXB sources as an example.  
{\em Left:} Probability distributions of  average CXB source 
flux for various numbers of detected sources $N$. The assumed detection
threshold  was $10^{-15}$ erg/s/cm$^2$ and the probability
distribution of the flux of one source is the CXB log(N)--log(S) as
given by eq.(\ref{eq:cxb}). The vertical dashed line is the expectation
mean (cf.eq.(\ref{eq:ltot_mean})).
{\em Right:} Combined apparent luminosity of the CXB sources detected
above the  sensitivity limit $L_{\rm min}$ as a function of aperture
for NGC1316 
(D=18.6 Mpc, $L_{\rm min}=10^{38}$ erg/s) and M81 (D=3.6 Mpc, $L_{\rm
min}=10^{37}$ erg/s). The luminosity was computed assuming
that CXB sources belong to the galaxy under consideration (i.e. are
located at the same distance). The dashed straight lines show 
expectation mean, eq.(\ref{eq:ltot_mean}) (cf. vertical dashed line in
the left panel). The solid curved lines show the luminosity
corresponding to the mode of the distribution, i.e. 
the most probable value of the integrated luminosity of the CXB
sources. The shaded areas around the solid curves show 67\% intrinsic
uncertainty. 
}
\label{fig:stat_cxb}
\end{figure*}

\subsection{Stellar mass}
\label{sec:nir_mass}

It is well-known, that the mass-to-light ratio in the near-infrared
band is significantly less subject to the stellar population dependent
variations than at the optical wavelength. Moreover, near-infrared
light is much less affected by extinction: $A_B/A_K\sim
10$, where $A_K$ and $A_B$ are the extinction in the K- and B-band
respectively \citep[e.g.][]{extinction}. These make the near-infrared
light a more robust estimator of the stellar mass than emission at
shorter wavelengths. 

Although to a lesser degree than optical, the
near-infrared mass-to-light ratio bears some sensitivity (probably up
to a factor of $\sim 2$, \citet{m2l}) on the properties of the stellar 
population i.e. on the star formation history of the galaxy. To
compensate for this effect, we use the fact that variations of the
mass-to-light ratio correlate tightly with the optical color of the
galaxy \citep[e.g][]{jarle2001, m2l, sdss}.  
Specifically, the results of galaxy evolution modeling by
\citet{m2l} relate the K-band mass-to-light ratio $M_*/L_K$ to the
$B-V$ optical color:   
\begin{eqnarray}
\log(M_*/L_K)=-0.692+0.652(B-V)
\label{eq:m2l}
\end{eqnarray}
Although \citet{m2l} modeled disk systems in spiral
galaxies, the slope of the $M_*/L$--color correlation is
remarkably robust with respect to the uncertainties in the stellar
population and details of the galaxy evolution, provided there is no
systematic variation of the initial mass function with galactic
type. Although the details of the IMF do not change the slope of the
$M_*/L$--color relation,  they significantly affect the zero point in 
eq.(\ref{eq:m2l}). A second factor affecting the $M_*/L$--color relation
is strong recent bursts of star formation. This, however, is
irrelevant in the context of old stellar systems in the nearby
galaxies studied in the present paper. 
To conclude, the simple color based correction described by
eq.(\ref{eq:m2l}) should, to the first approximation, account for the
main trend in the stellar population dependent variations of the
$M_*/L$ ratio.

The extinction corrected B--V colors for the galaxies  from our sample,
adopted from the RC3 catalog \citep{rc3} and corresponding K-band
mass-to-light ratios computed using eq.(\ref{eq:m2l}) are listed in
Table \ref{tab:sample}.

\subsection{NIR multi-aperture photometry}
\label{sec:nir_data}

The near-infrared multi-aperture photometry data were
taken from the ``Catalog of Infrared Observations, Edition 5''
\citep*{gezari99}. For all galaxies except two we used K-band  
measurements. In the case of M31 and M81, the data were insufficient
for constructing meaningful growth curves in 
the K-band. Therefore we used H-band data and
transformed it to K-band, using average NIR colors for normal galaxies
$\overline{H-K}=0.2$ \citep{fioc99}. For several galaxies (NGC1316,
NGC4472, NGC4697, M84/NGC4374, Cen A/NGC5128 and NGC5846) we
complemented the \citet{gezari99} by the data from \citet{pahre99}. 
Absolute K-band magnitude of the Sun was assumed to be equal to
$M_{K,\odot}=3.39$. Multi-aperture photometry was
used to construct the near-infrared growth curves, as described in
section \ref{sec:nir_gcurve}.

\subsection{Statistics of small numbers -- collective luminosity of a
population of discrete sources}
\label{sec:stat}

In many astrophysically relevant situations, a problem arises to count
or measure the collective luminosity of an ensemble of discrete
sources with a power law (or similar) luminosity function. Rigorous
mathematical treatment of this problem, including the formulae
convenient for practical applications,
is given in \citet{stat}. Below we present a brief qualitative
consideration in the context of  this paper.

Let us consider a population of compact sources (X--ray binaries in a
galaxy or CXB sources projecting inside a galaxy)  whose luminosity or
flux distribution is a power law with differential slope $\alpha$ and
cut-off at $L_{\rm cut}$: 
\begin{eqnarray}
\frac{dN}{dL}=\left\{ \begin{array}{ll}
K L^{-\alpha}	& \mbox{$L<L_{\rm cut}$}\\
0	& \mbox{$L\ge L_{\rm cut}$}
\end{array}
\right.
\label{eq:p1}
\end{eqnarray}
The normalization of this distribution $K$ might depend on
such quantities as the mass of the galaxy (in the case of X-ray
binaries) or the sky area (in the case of the CXB sources).

The quantity of interest is the total luminosity of all sources in the
galaxy with luminosity exceeding a given detection limit $L_{\rm min}$:
\begin{eqnarray}
L_{\rm tot}=\sum_{L_k>L_{\rm min}} L_k
\end{eqnarray}

A seemingly obvious expression for the total luminosity can be
obtained by integrating the luminosity distribution (\ref{eq:p1}):
\begin{eqnarray}
\left  <L_{\rm tot} \right > =
\int_{L_{\rm min}}^{L_{\rm cut}} \frac{dN}{dL}\,L\,dL
\propto K
\label{eq:ltot_mean}
\end{eqnarray}
This equation defines the expectation mean for $L_{\rm tot}$ and 
implies, that e.g. the total luminosity of the CXB sources is directly
proportional to the sky area.

However, in the case of a ``small'' number of sources
(the threshold value depending on the slope of the luminosity function
and values of $L_{\rm min}$ and $L_{\rm cut}$), 
the probability distribution $p(L_{\rm tot})$ might be strongly
asymmetric, as illustrated by the left panel in
Fig.\ref{fig:stat_cxb}. 
Because of the skewness of $p(L_{\rm tot})$, the most probable value of
$L_{\rm tot}$ -- the mode of the distribution, is not equal to
the expectation mean defined by eq.(\ref{eq:ltot_mean}). Importantly,
it is the mode of the distribution $p(L_{\rm tot})$, that would be most
likely measured in an  arbitrarily  chosen galaxy.\footnote{ 
Obviously in the case of e.g. a flat ($dN/dL=$const) or Gaussian  flux
distribution, the most probable value of $L_{\rm tot}$ always equals 
the expectation mean defined by eq.(\ref{eq:ltot_mean}).
}

The difference between these two quantities is further illustrated in
the right panel in Fig.\ref{fig:stat_cxb}  which shows the expected
total 
luminosity of the CXB sources detected above the sensitivity threshold
in the Chandra observations of NGC1316 and M81, as a function of
aperture.  As evident from Fig.\ref{fig:stat_cxb}, the value of the
total flux/luminosity of the CXB sources which will be most likely
detected in these observations inside, for example, the effective
diameter of the 
galaxy, $\sim 100\arcsec-200\arcsec$, deviates significantly from the
expectation value given by eq.(\ref{eq:ltot_mean}). It is correctly
predicted  by eq.(\ref{eq:ltot_mean}) only for sufficiently large
aperture (sky area), i.e. when sufficient number of sources are
detected above the sensitivity threshold.   

The mode of the probability distribution (the solid line in
the right panel of Fig.\ref{fig:stat_cxb}) is the value of the
total luminosity which will most likely be measured in an arbitrarily 
chosen galaxy. If a number of  observations of many 
(different) galaxies are performed, the obtained values of $L_{\rm
tot}$ will obey the probability distribution depicted in the left panel
of Fig.\ref{fig:stat_cxb}. Consequently, strong and asymmetric dispersion
among the measured values of $L_{tot}$ should be expected, due to
strongly asymmetric shape of the probability distribution. On rare
occasions, for some of the galaxies, large values of the total
luminosity would be observed, corresponding to the tail of the
$p(L_{\rm tot})$. The average of the measured values of 
$L_{tot}$ will be equal to the expectation mean given by
eq.(\ref{eq:ltot_mean}).

\section{X-ray and NIR growth curves}
\label{sec:gcurves}

\subsection{X-ray growth curves}
\label{sec:xray_gcurve}

In order to study the spatial distribution of X-ray binaries, we
utilize  circular aperture growth curves. The growth curves were
constructed 
from the filtered source lists with the appropriate completeness limits
applied, as detailed in subsection \ref{sec:xray_data}. They
describe the number of sources, $N_X(<a)$, located inside
circular aperture  (diameter) $a$ and their total luminosity,
$L_X(<a)$.  

Two considerations were taken into account in the growth curves
analysis. First, the detection sensitivity might exhibit significant 
non-uniformity across the galaxy image and typically worsens towards
the center of the galaxy. If not properly corrected for, this effect
can lead to distortions in the apparent surface density 
of X-ray sources, as discussed in subsection \ref{sec:xray_data}. 
Precise correction of the growth curves requires knowledge of the
$\log(N)-\log(S)$ distribution. Another possible way to minimize this
effect is to increase the lower luminosity limit for the source
selection, $L_{\rm min}$. This, however, 
results in a significant decrease in the number of 
sources available for study. Instead, we chose to exclude the very
central parts of the galaxies, $\sim$ few tens of arcsec, from
the analysis. An additional advantage of this approach is that it
alleviates, to some extent, the problem of confusion which might become
relevant in the very centers of even relatively nearby
galaxies (indeed, an angular resolution of $1\arcsec$ at the distance
of 17 Mpc is equivalent to $\sim 30\arcmin$ at the distance of our
Galactic Center). 

Secondly, in the outer parts of the galaxies, the contribution of the
CXB sources becomes important. Its accurate subtraction
is complicated by field-to-field variations and, in
the case of the X-ray luminosity growth curves, by the effects of
small number statistics.  We therefore excluded from the analysis
outer parts of the galaxies, where the surface density of the CXB
sources becomes comparable with the surface density of X-ray
binaries.

\subsection{NIR growth curves}
\label{sec:nir_gcurve}

The near-infrared growth curves were constructed from the K-- and
H--band multi-aperture photometry data following the approach of
\cite{fioc99}. Using the fact that the near-infrared magnitudes
correlate almost linearly (but with a slope different from 1)  
with optical B-band magnitudes, the NIR growth curve can be
approximated by
\begin{eqnarray}
m(a)=m_T+s_0\,B(X,T_p)\\
X=\log(a/A_e)\nonumber
\end{eqnarray}
where $m(a)$ is the near-infrared magnitude inside the aperture
(diameter)  $a$, $m_T$ is the total NIR magnitude of the galaxy,
$B(X,T_p)$ is the 
B-band growth curve of the galaxy as a function of the dimensionless
aperture $X$ and the B-band photometric type $T_p$, $s_0$ is a
parameter characterizing the slope of the NIR -- B-band correlation,
and $A_e$ is the B-band 
effective (half-light) diameter of the galaxy. The shape function
$B(X,T_p)$ was defined by \citet{leda} as photometric type-dependent 
extrapolation between the de Vaucouleurs and the exponential
profiles (see also Appendix A in \citealt{fioc99}). 

The B-band photometric types of the galaxies were adopted from the
HiperLeda catalog \citep{leda}. The effective diameters $A_e$ are from  
RC3 catalog \citep{rc3} and are listed in  Table \ref{tab:sample}. 
The two unknown parameters of the NIR growth curves -- 
total magnitude $m_T$ and slope $s_0$ were determined from the
weighted least-square fits to the NIR multi-aperture photometry.
The weights were defined as the inverse square of the deviation of the  
given measurement from the best fit growth curve and were determined
via an iterative procedure as described in \citet{fioc99}.

The values of  $s_0$ and $m_T$ obtained from the fits to NIR data are
listed in the Table \ref{tab:sample}.  For several galaxies 
independent determinations of the total NIR magnitude were available, 
and we compared our best fit values of $m_T$ with those published
elsewhere and generally found reasonable consistency, within 
$\sim 0.1-0.4$ magnitude. This is well within the typical disagreement
between different measurements and does not significantly compromise
the following comparison of the X-ray and NIR growth curves since the
NIR photometric measurements in most cases overlap
considerably in the aperture with the X-ray data. 

In the case of the Cen A galaxy, no $A_e$ is given in the RC3 catalog.
An attempt to use the value from  HiperLeda catalog 
($A_e\approx 752 \arcsec$) resulted in a bad fit to the near-infrared
photometric data and the value of the total K-band magnitude lies
significantly outside the range of other measurements $m_T\sim 4.4-5.0$ 
\citep{pahre99}.  We therefore adjusted the value of $A_e$ to achieve
a good fit and a consistent value of the total magnitude. Listed in
the Table \ref{tab:sample} is the resulting value of $A_e\approx 239
\arcsec$.

\begin{figure*}
\centerline{
\vbox{
\hbox{
\resizebox{0.30\hsize}{!}{\includegraphics{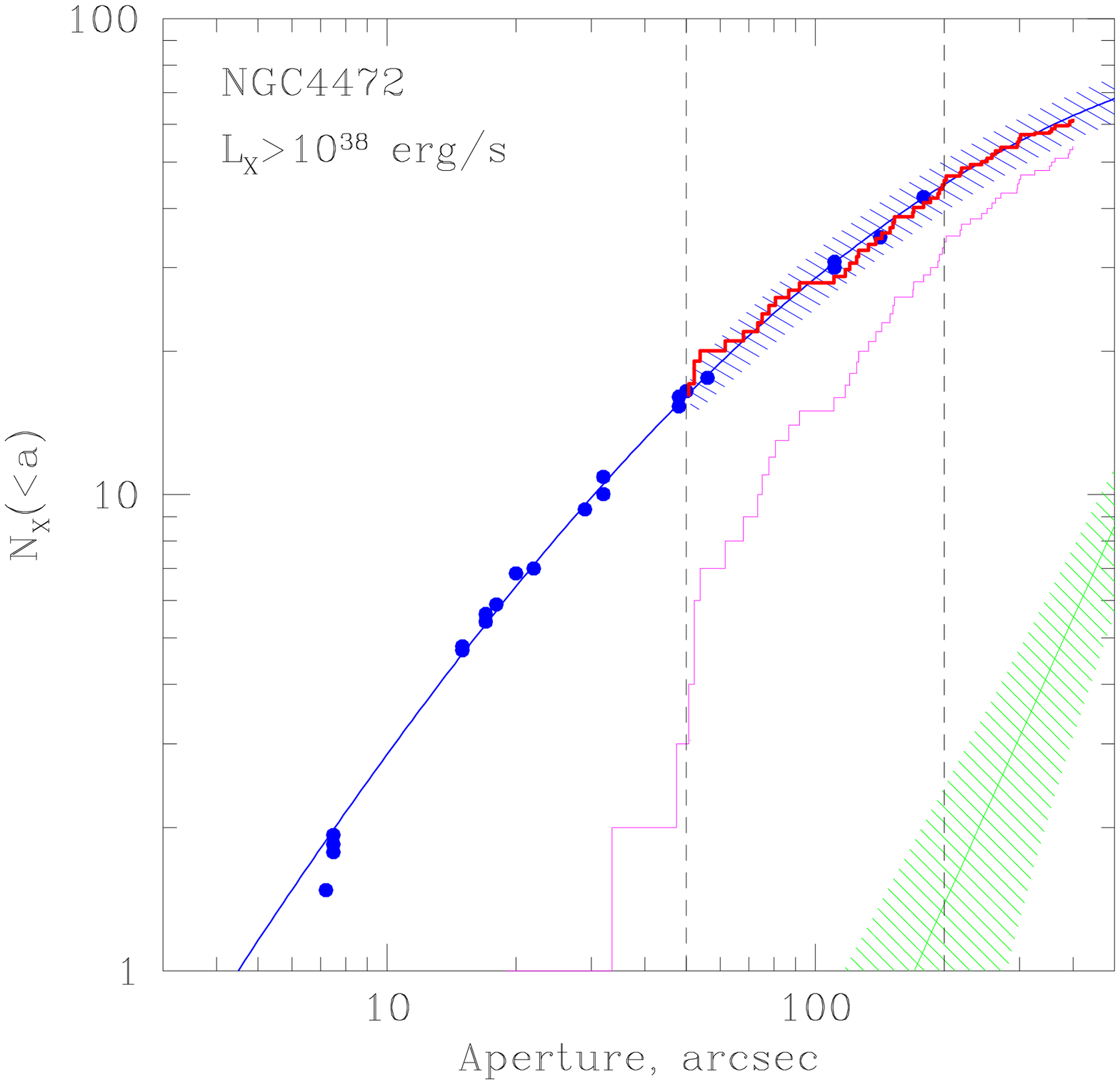}}
\resizebox{0.30\hsize}{!}{\includegraphics{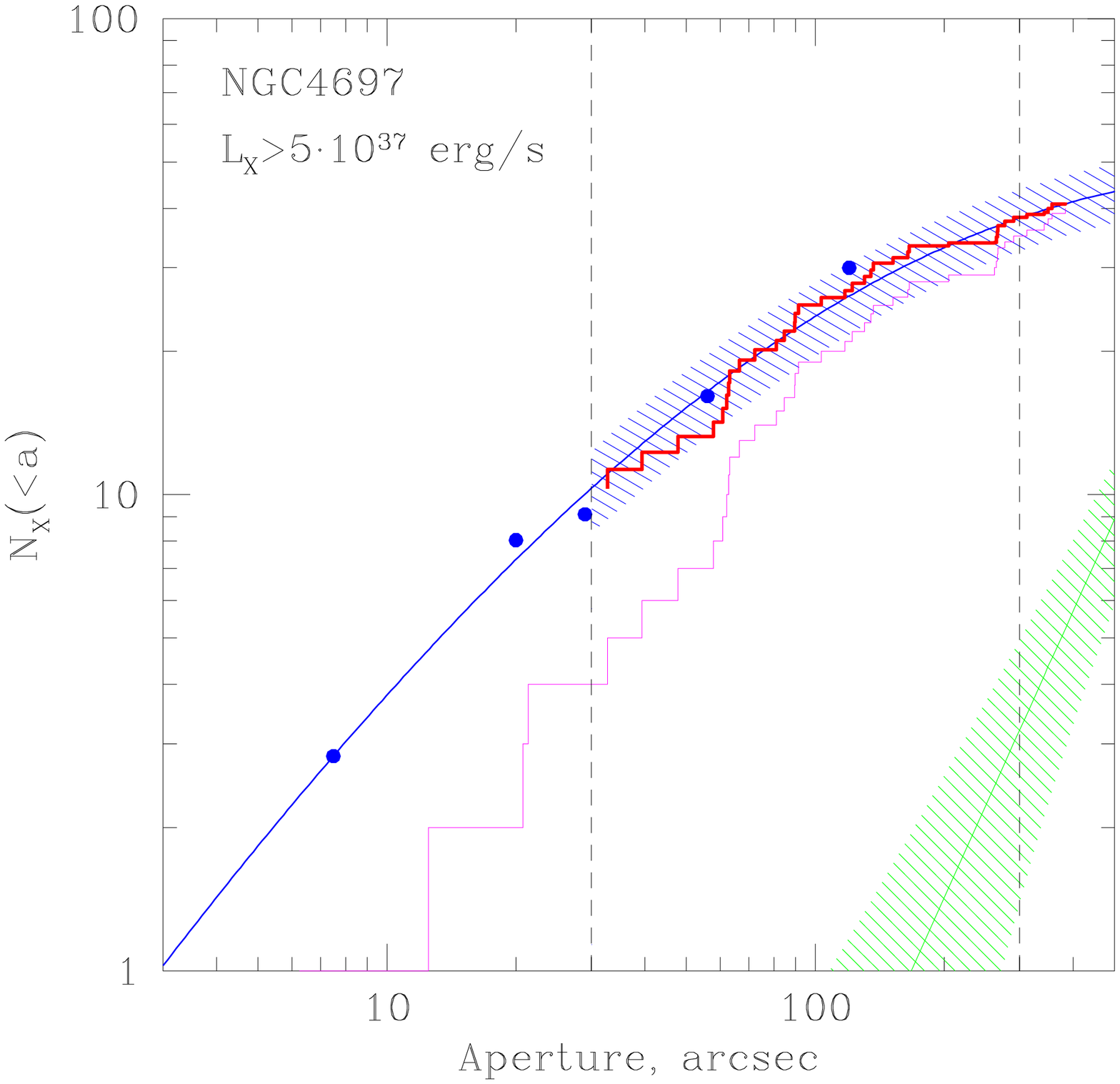}}
\resizebox{0.30\hsize}{!}{\includegraphics{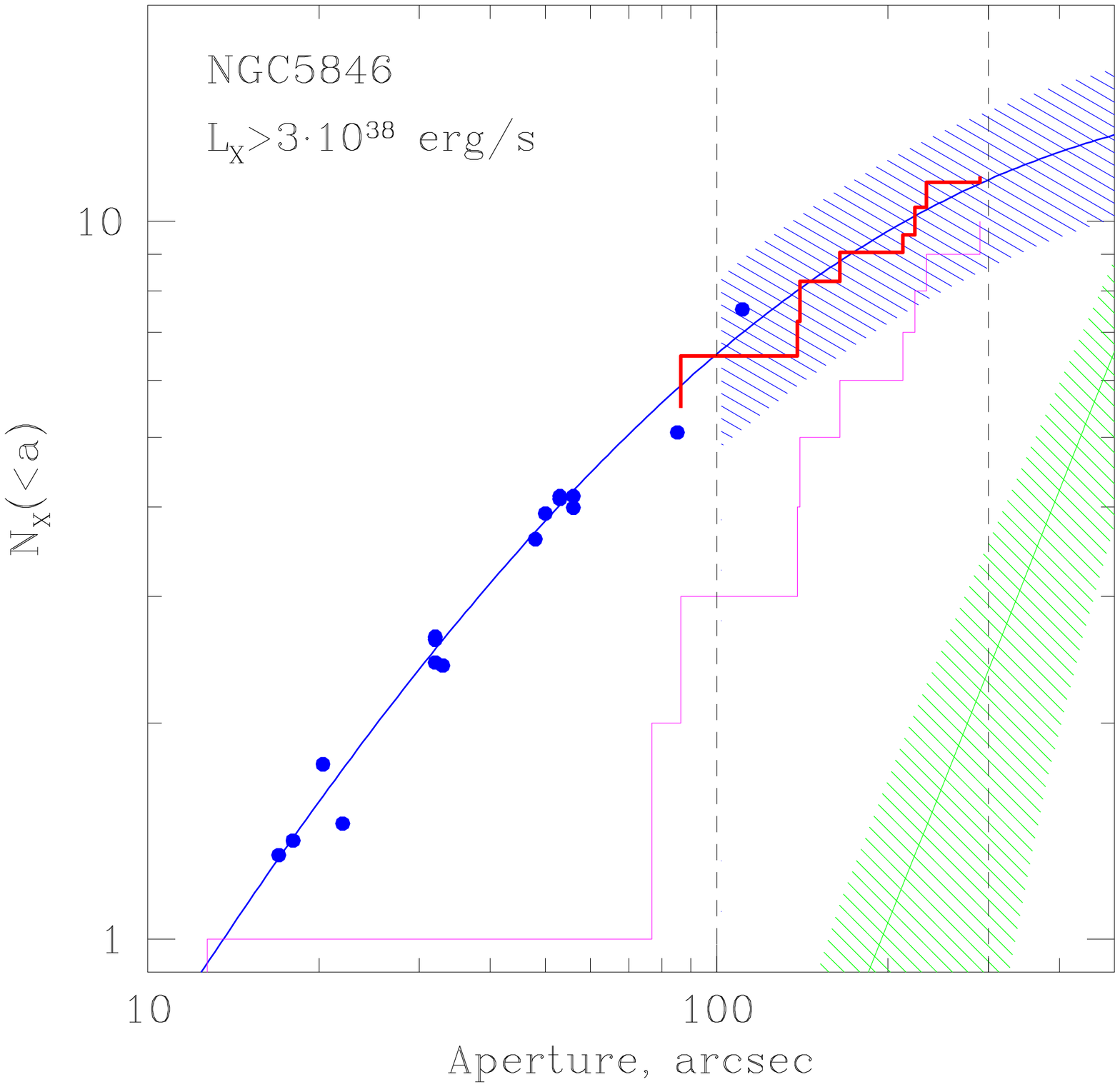}}
}
\hbox{
\resizebox{0.30\hsize}{!}{\includegraphics{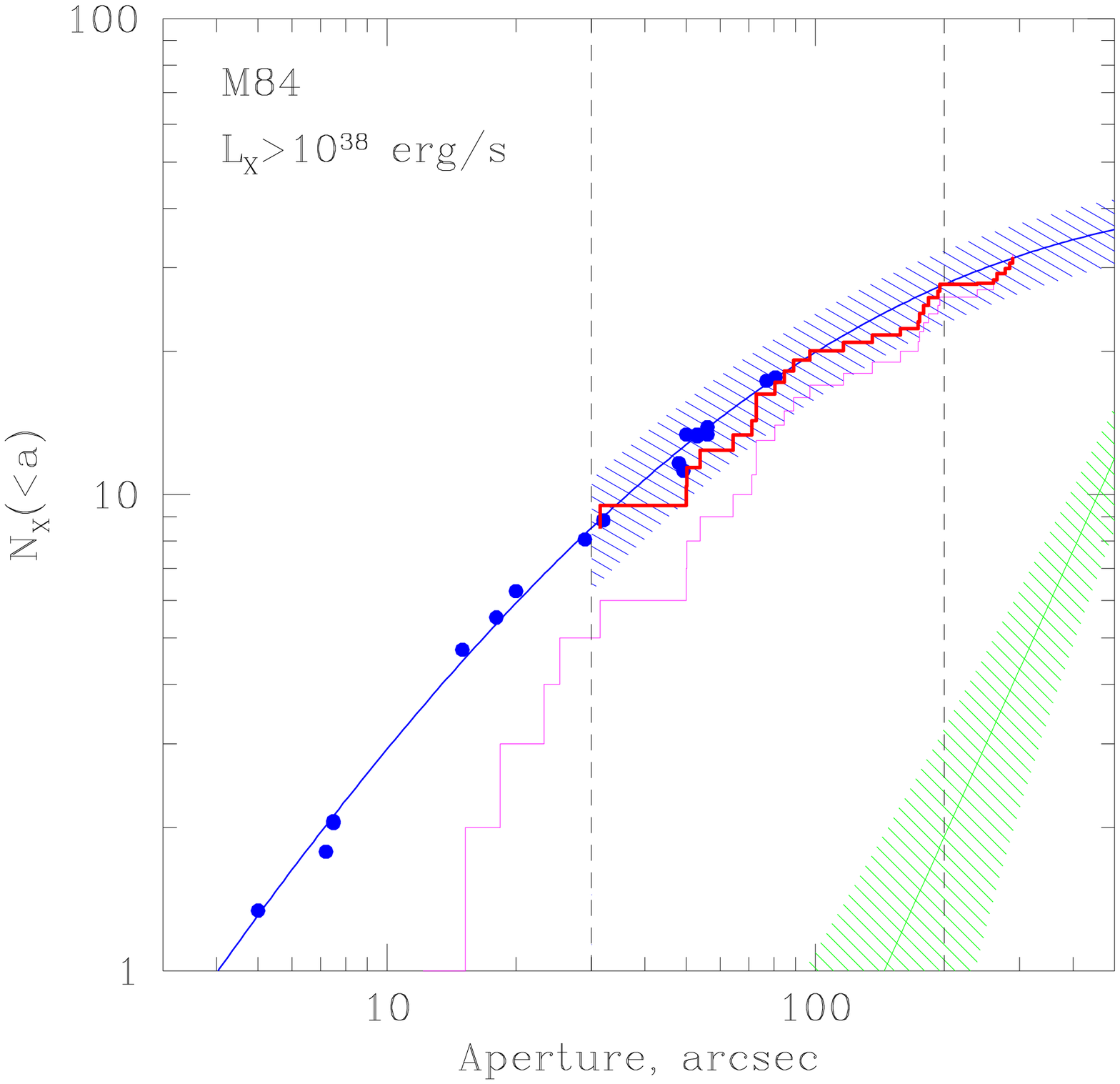}}
\resizebox{0.30\hsize}{!}{\includegraphics{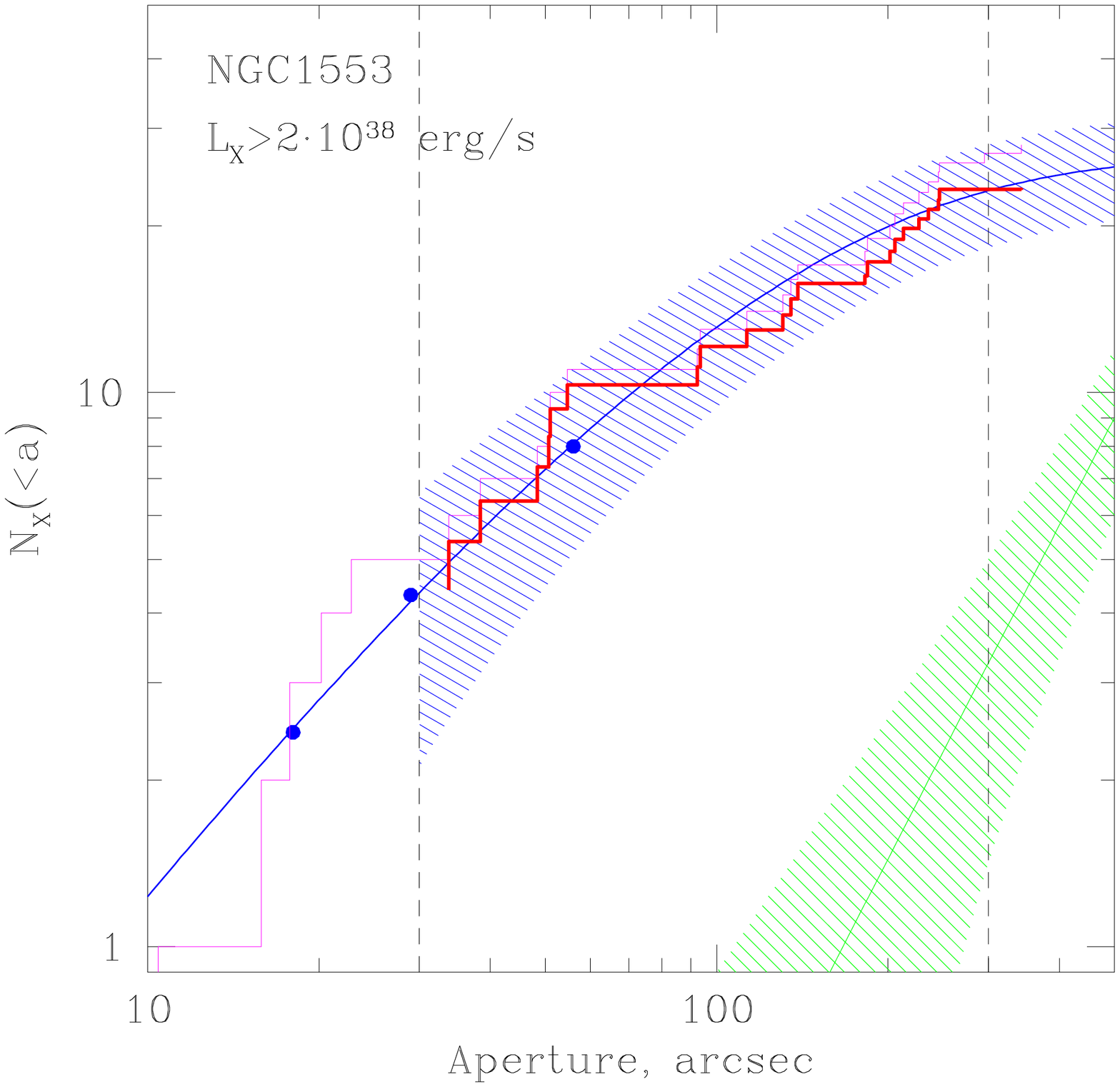}}
\resizebox{0.30\hsize}{!}{\includegraphics{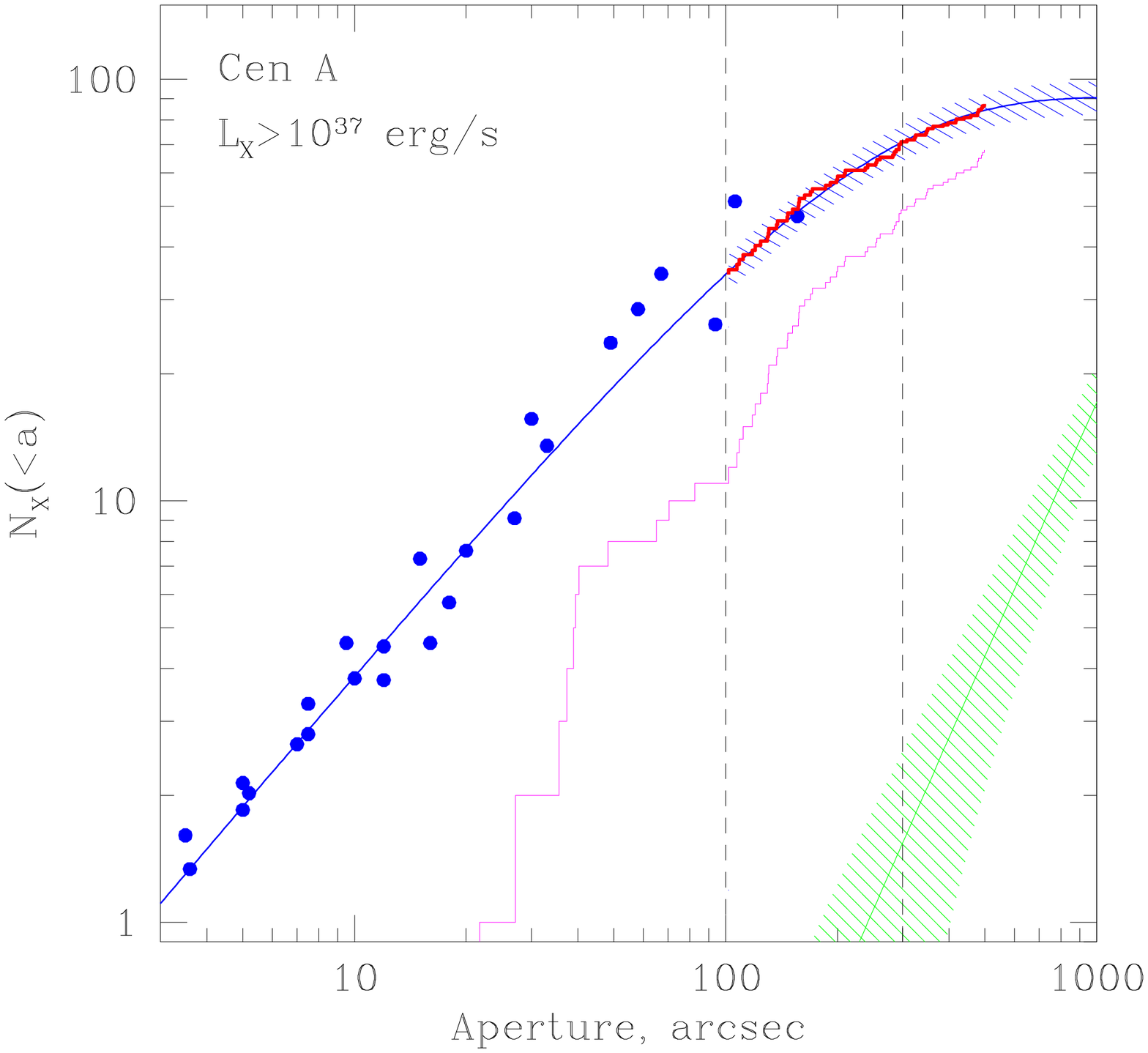}}
}
\hbox{
\resizebox{0.30\hsize}{!}{\includegraphics{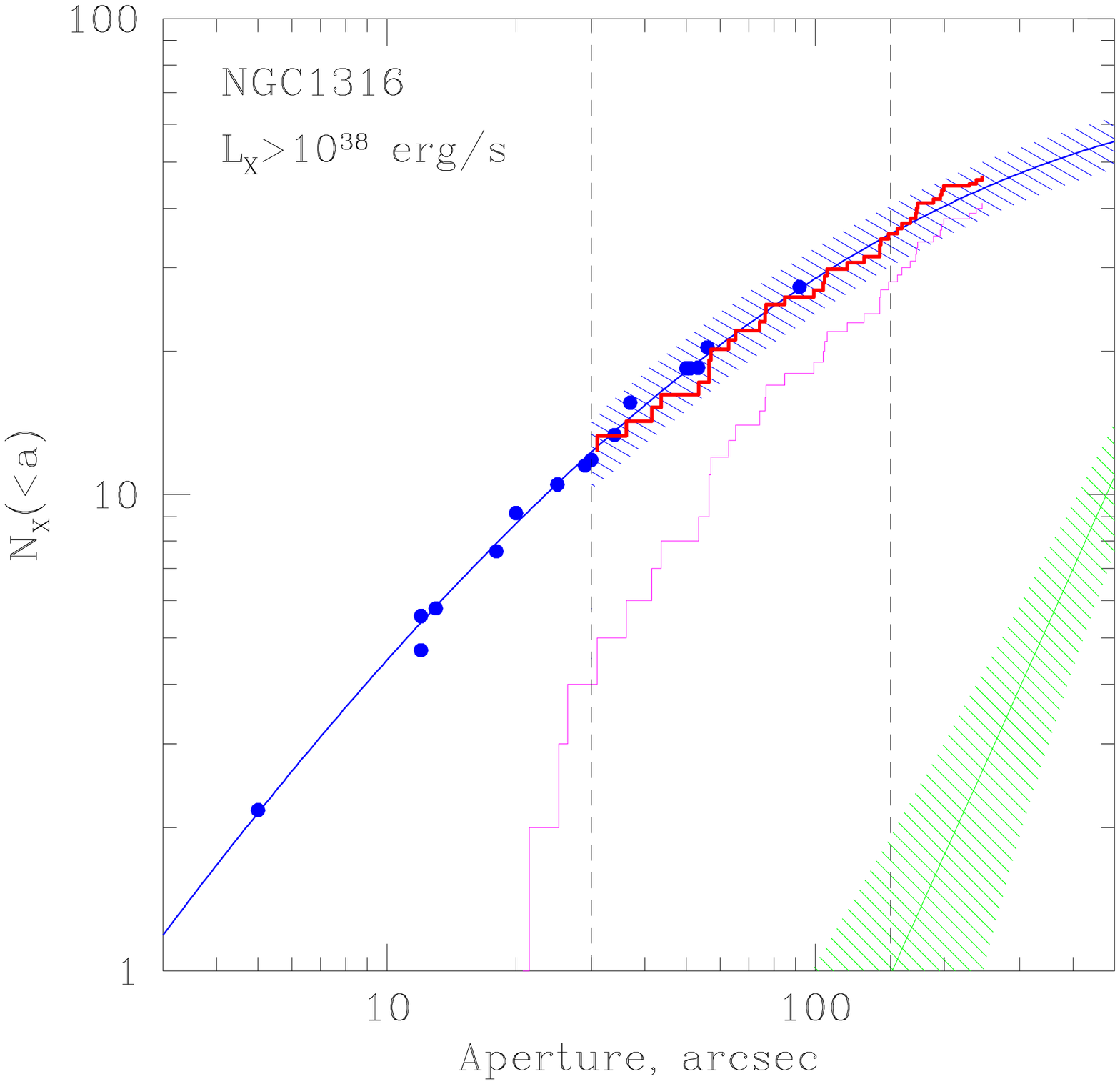}}
\resizebox{0.30\hsize}{!}{\includegraphics{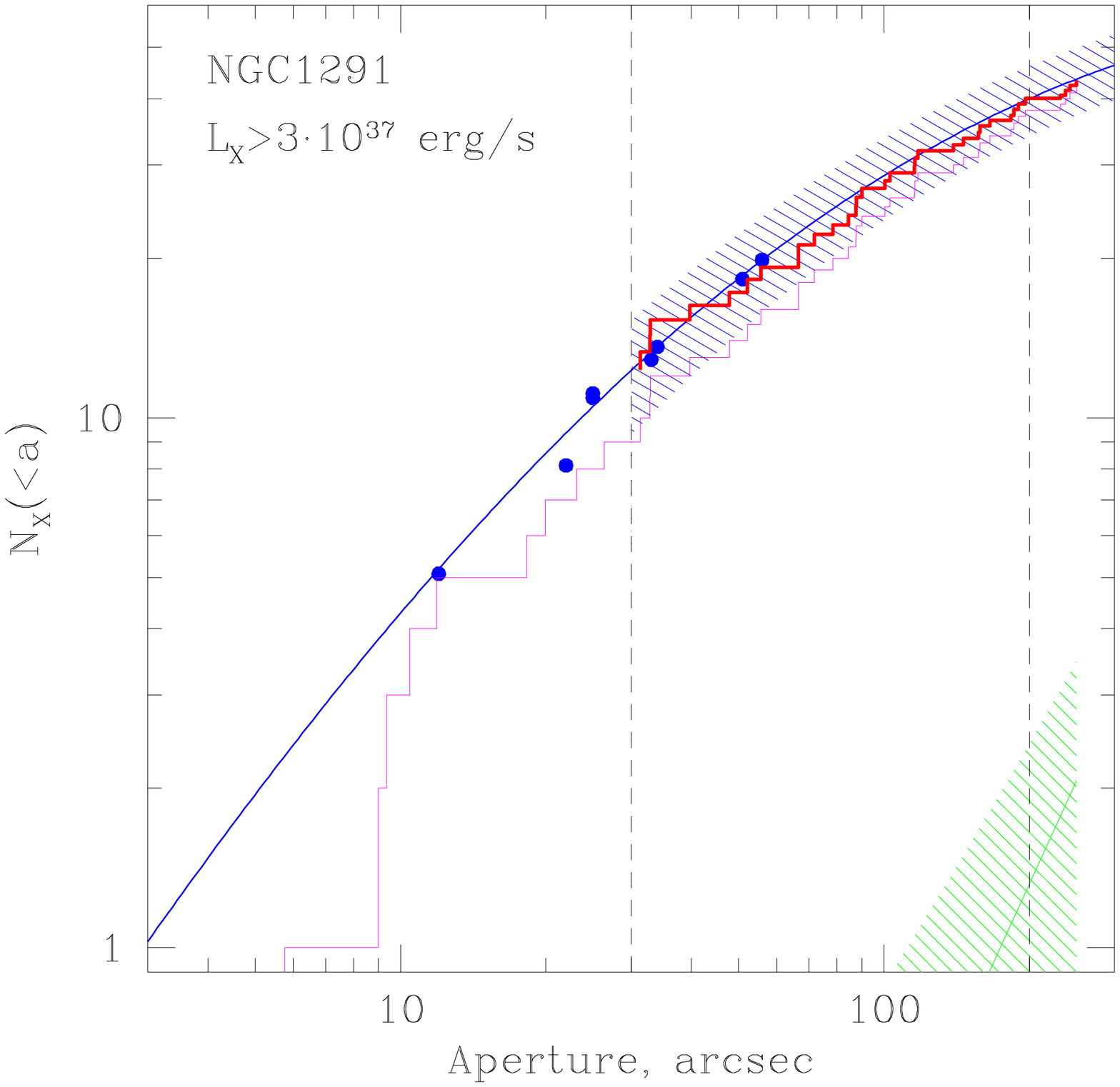}}
\resizebox{0.30\hsize}{!}{\includegraphics{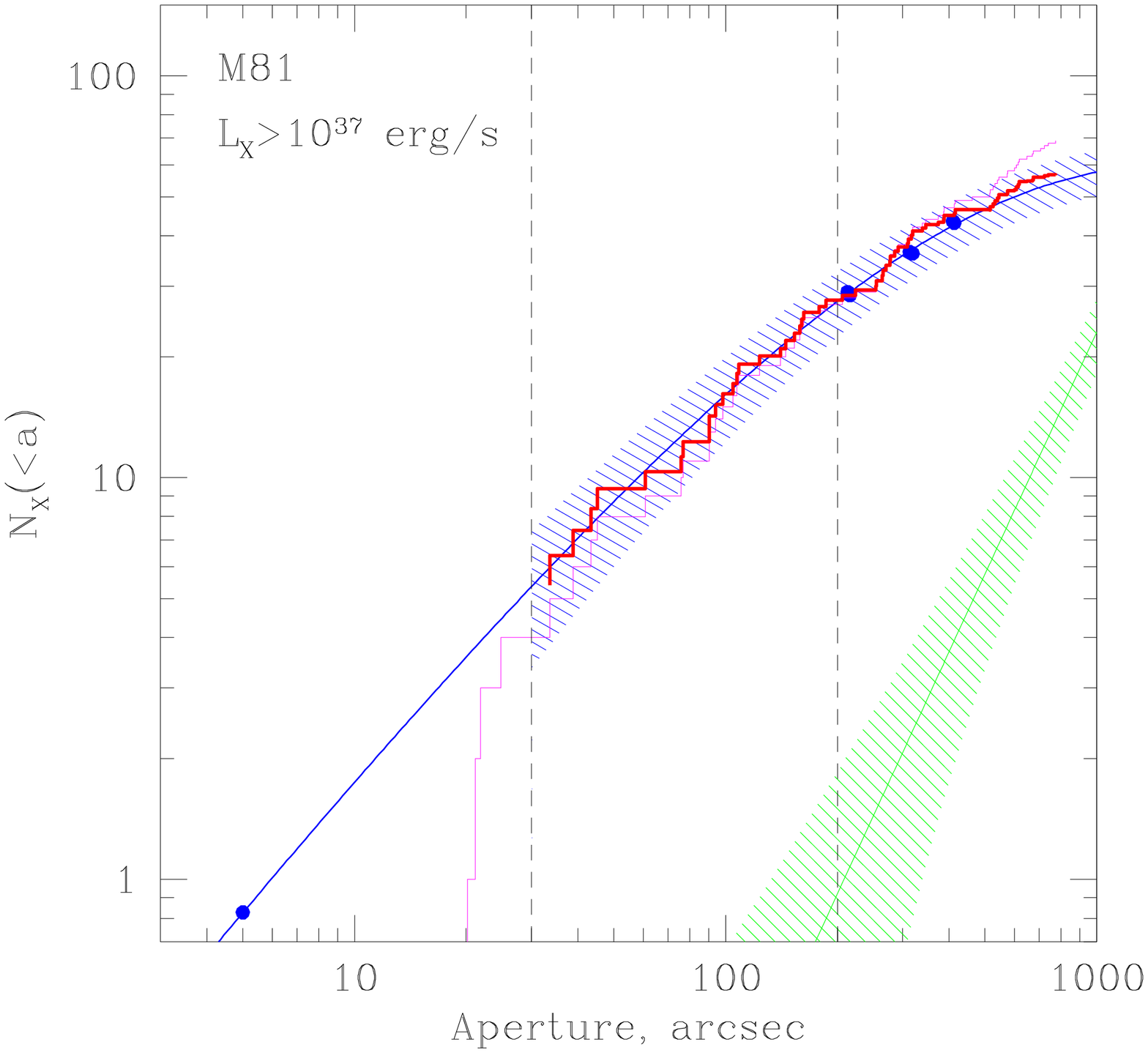}}
}
}
}
\centerline{
\vbox{
\hbox{
\resizebox{0.30\hsize}{!}{\includegraphics{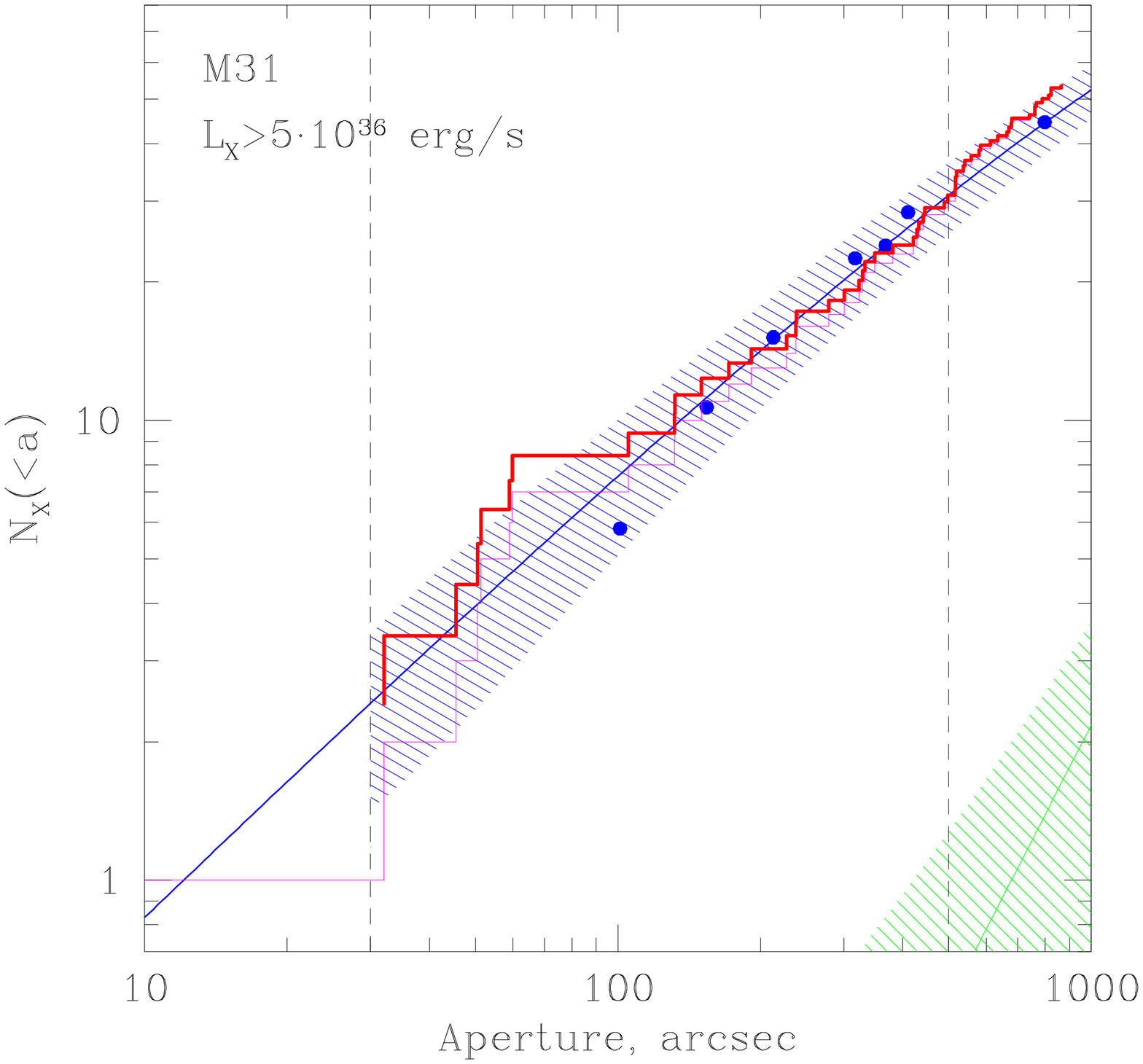}}
\resizebox{0.30\hsize}{!}{\includegraphics{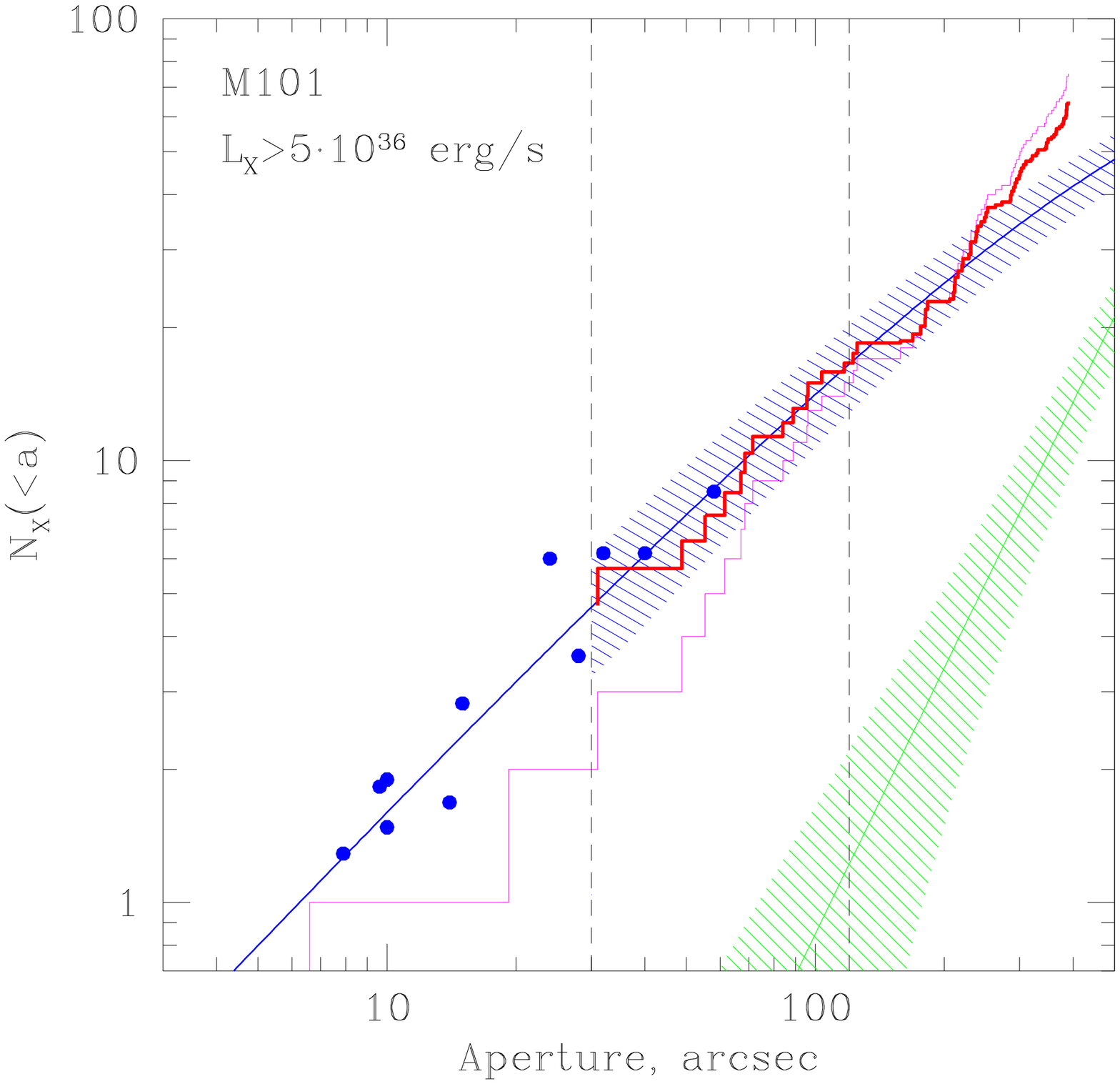}}
}
}
}
\caption{X-ray, $N_X(<a)$, (thick histogram) and near-infrared K-band
(thin solid line) growth curves. The thin histograms show apparent growth
curves for the number of sources with luminosity exceeding
$L_{X,min}$, whose value is indicated in the upper-left corner of each
plot.  
The thin solid line and shaded region in the lower-right corner of
each panel show the CXB growth curve and its 67\% statistical
uncertainty.  
The solid circles are multi-aperture NIR photometry measurements, the
thin solid curve and shaded area are the best fit NIR growth curve and its
67\% statistical uncertainty.The NIR data and the growth curve were
multiplied by   $N_X/L_{NIR}$ ratios determined in the annulus with
boundaries $a_{X,1}$ and $a_{X,2}$ that are shown by the vertical
dashed lines.  
The  thick histogram shows  the final corrected X-ray growth curve
(see section \ref{sec:nir_x_gcurve} for details).
}
\label{fig:gcurves_nx}
\end{figure*}

\begin{figure*}
\centerline{
\vbox{
\hbox{
\resizebox{0.30\hsize}{!}{\includegraphics{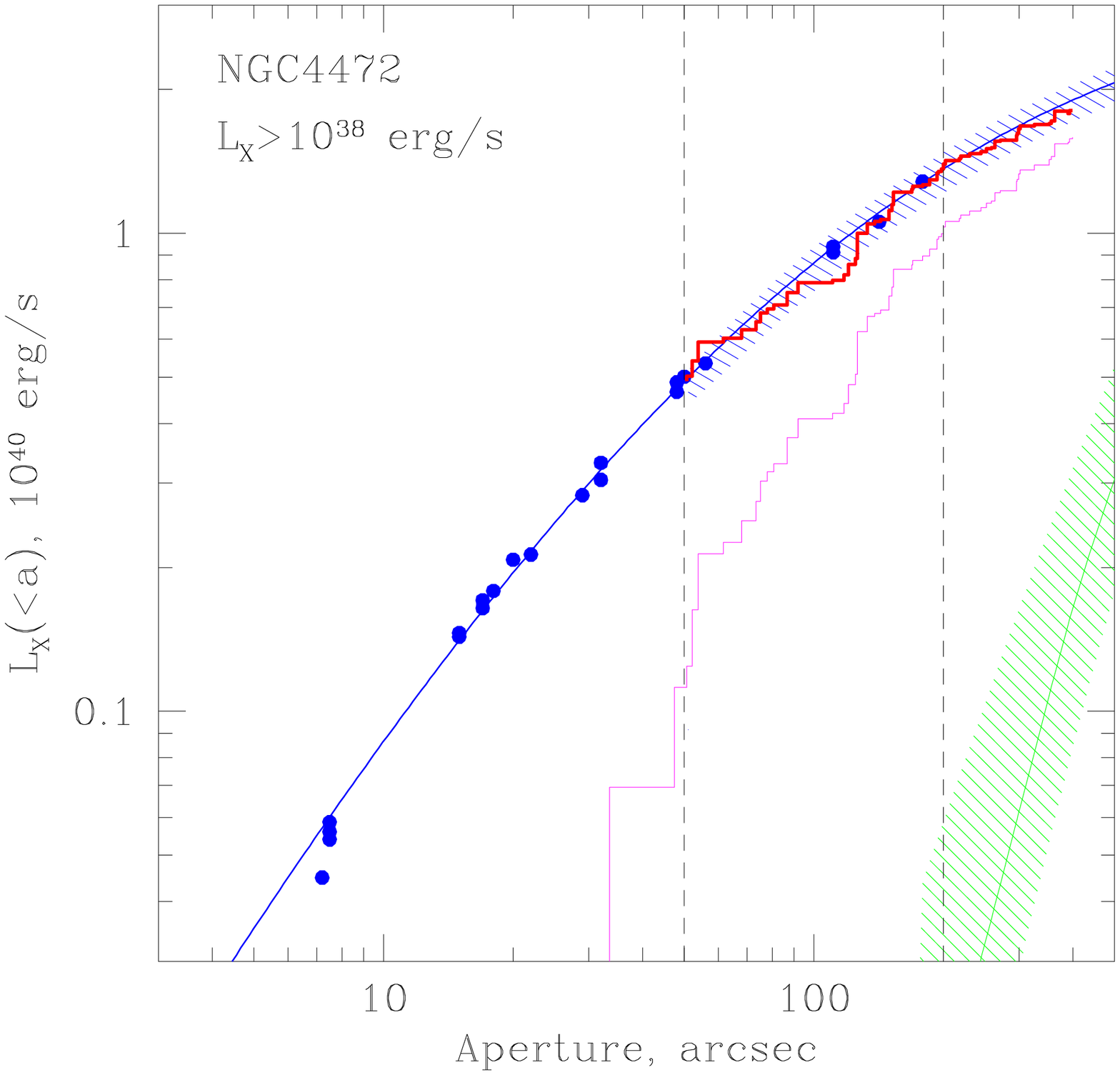}}
\resizebox{0.30\hsize}{!}{\includegraphics{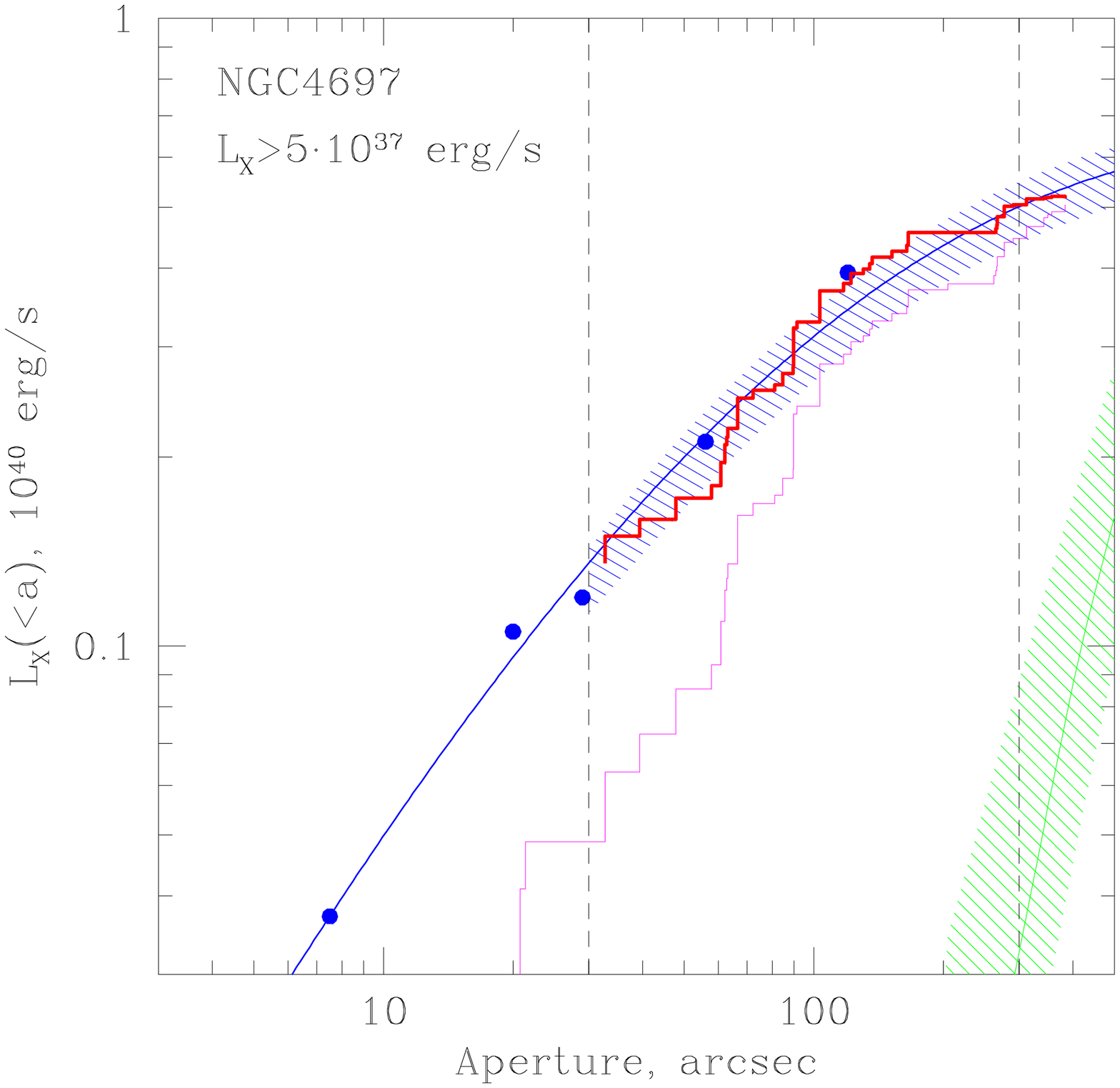}}
\resizebox{0.30\hsize}{!}{\includegraphics{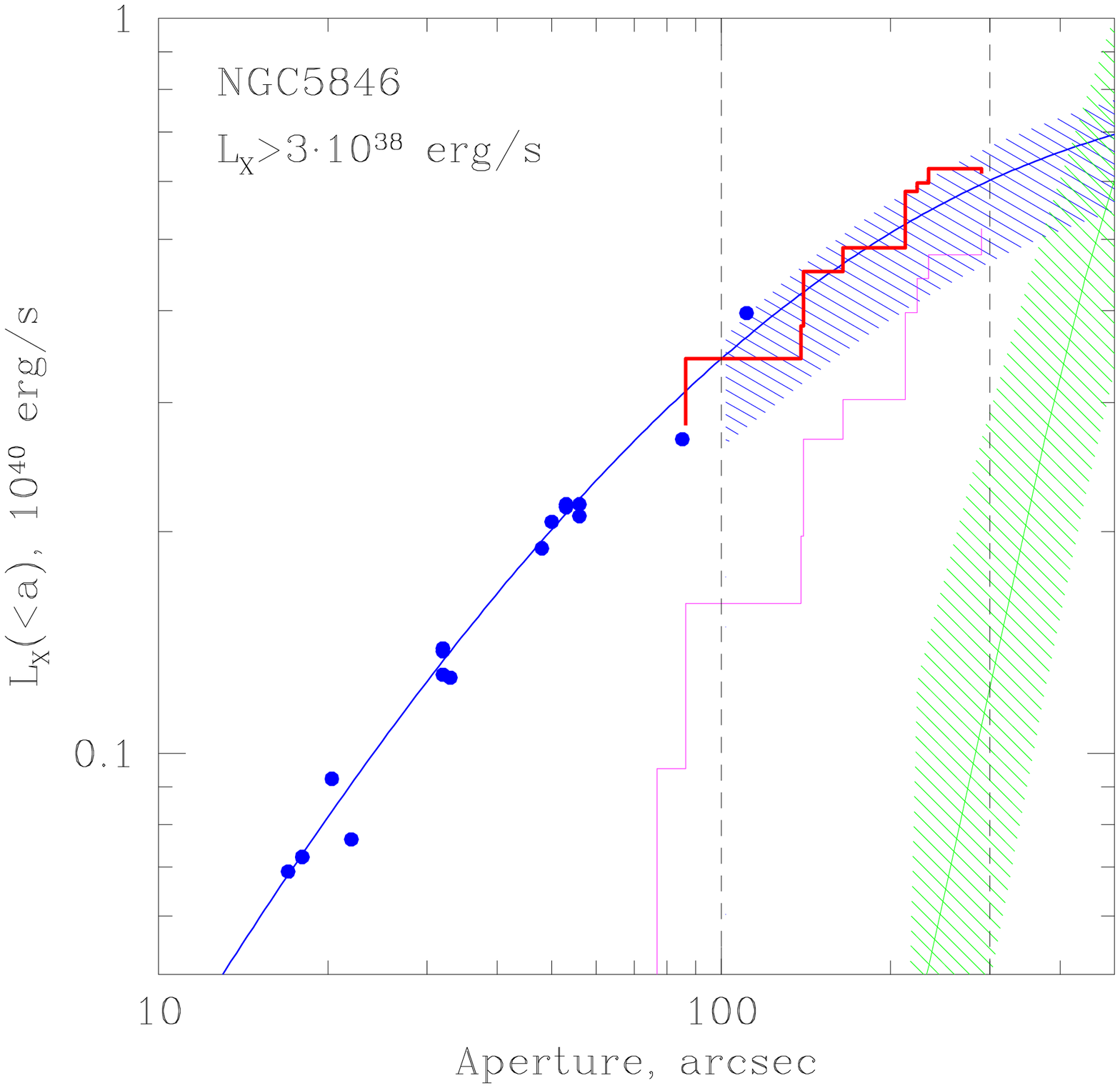}}
}
\hbox{
\resizebox{0.30\hsize}{!}{\includegraphics{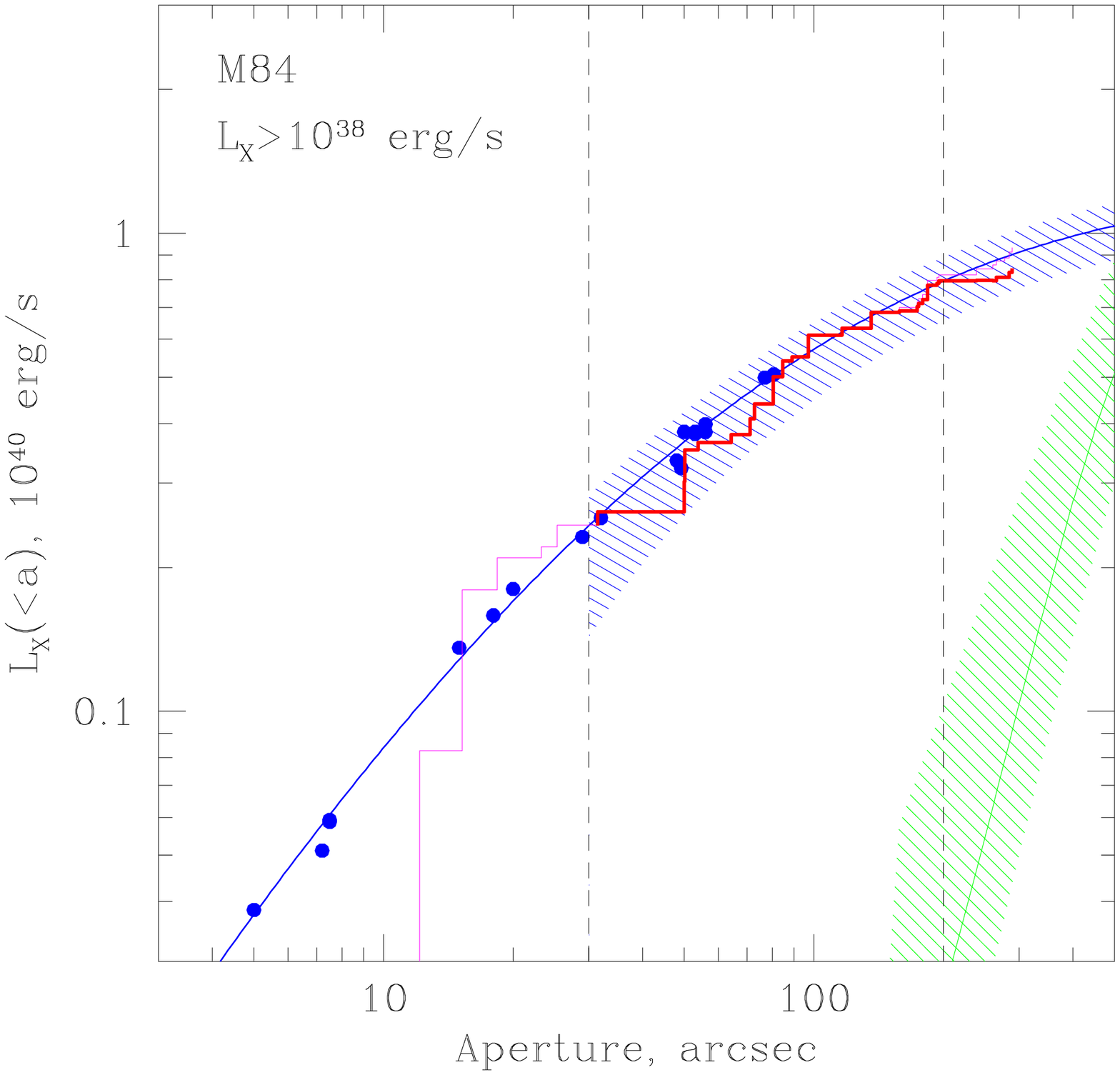}}
\resizebox{0.30\hsize}{!}{\includegraphics{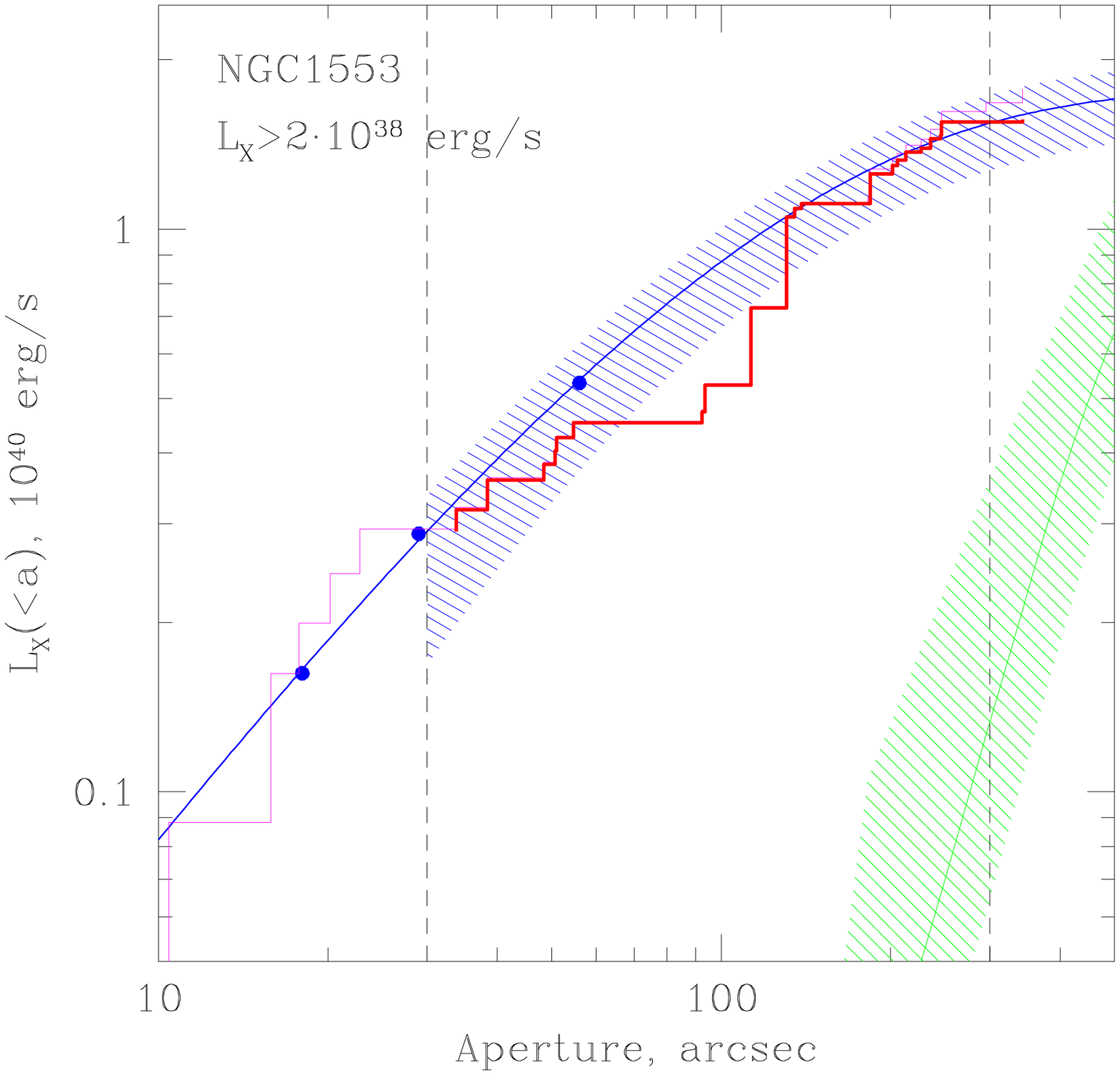}}
\resizebox{0.30\hsize}{!}{\includegraphics{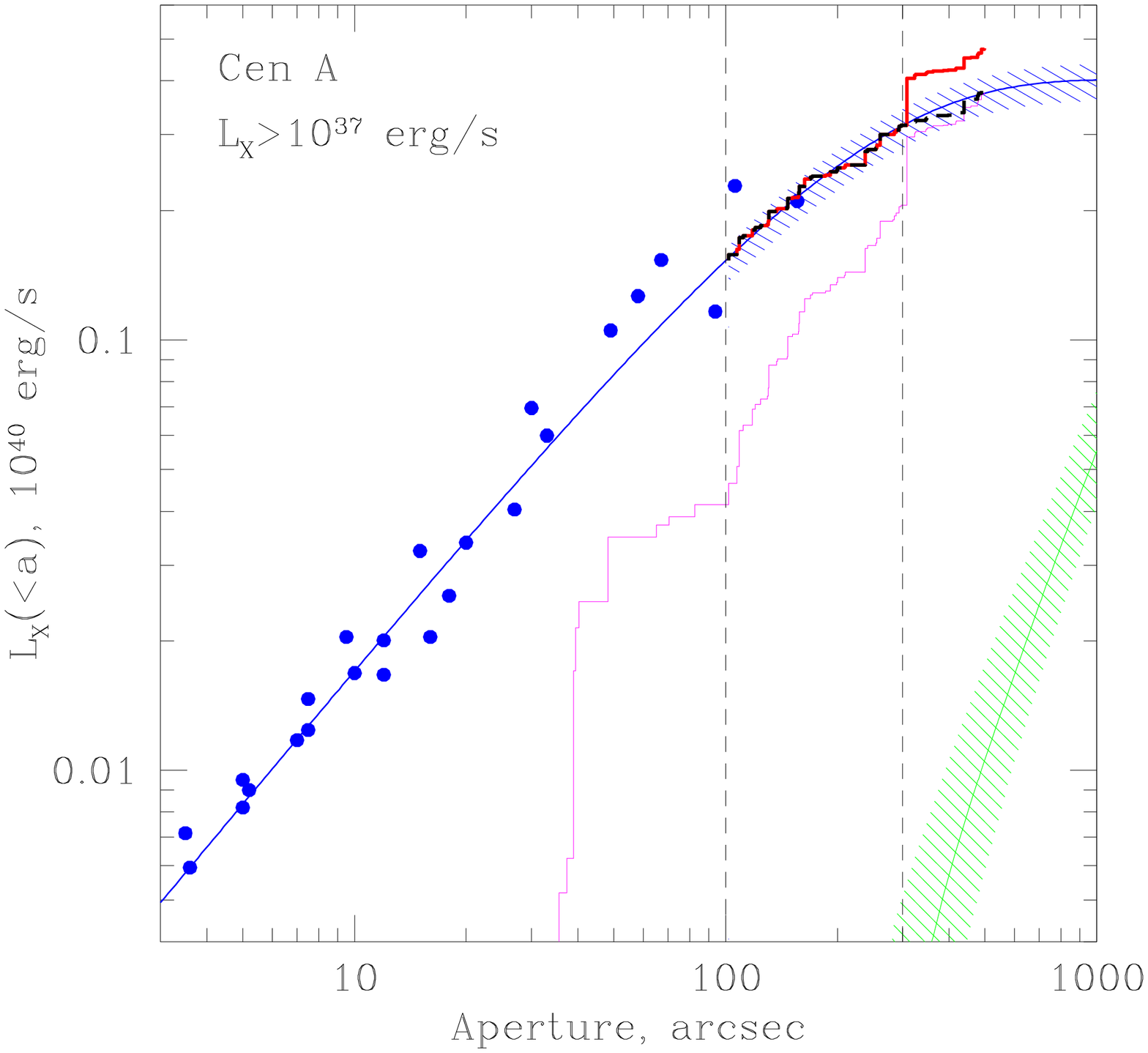}}
}
\hbox{
\resizebox{0.30\hsize}{!}{\includegraphics{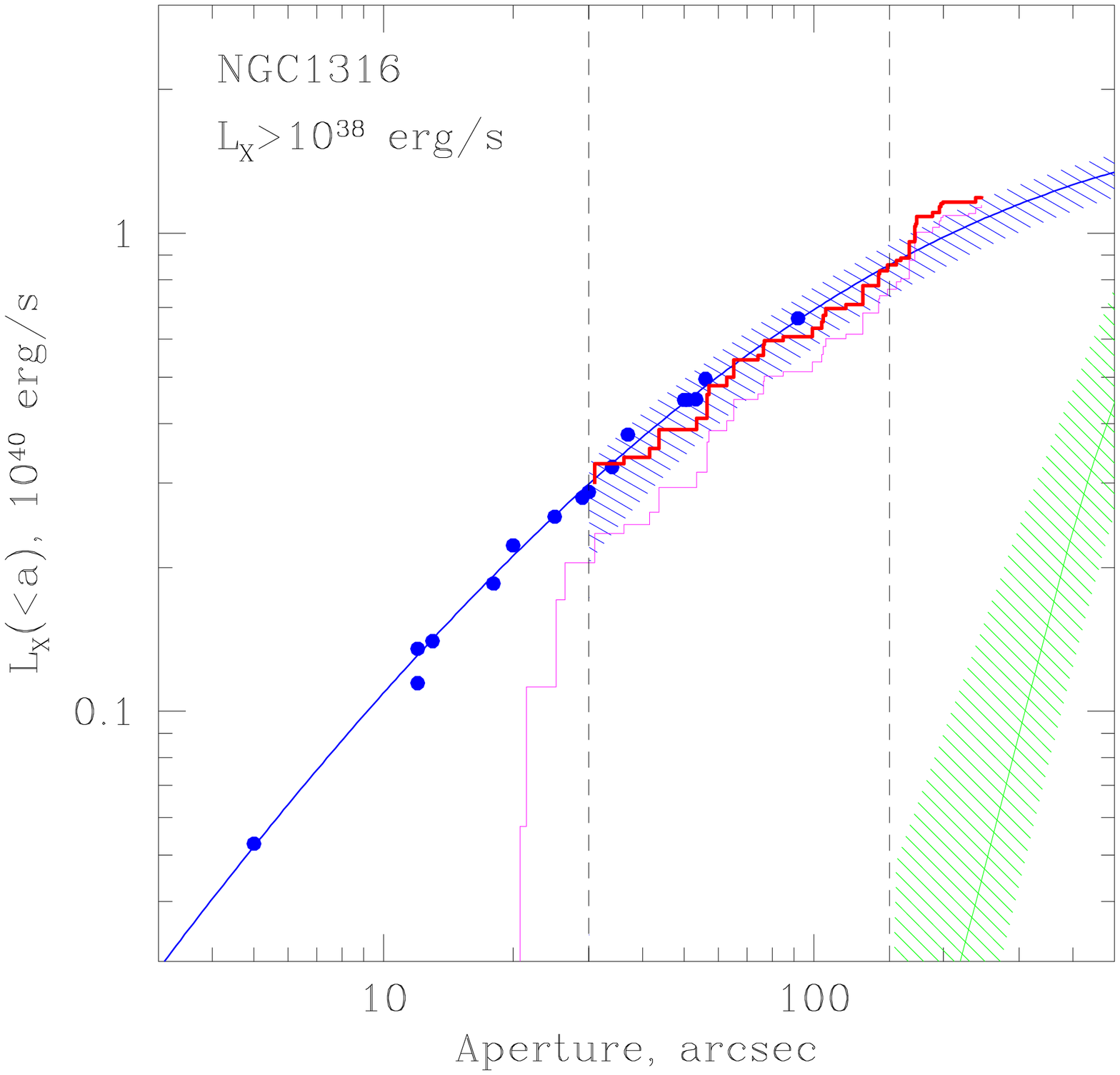}}
\resizebox{0.30\hsize}{!}{\includegraphics{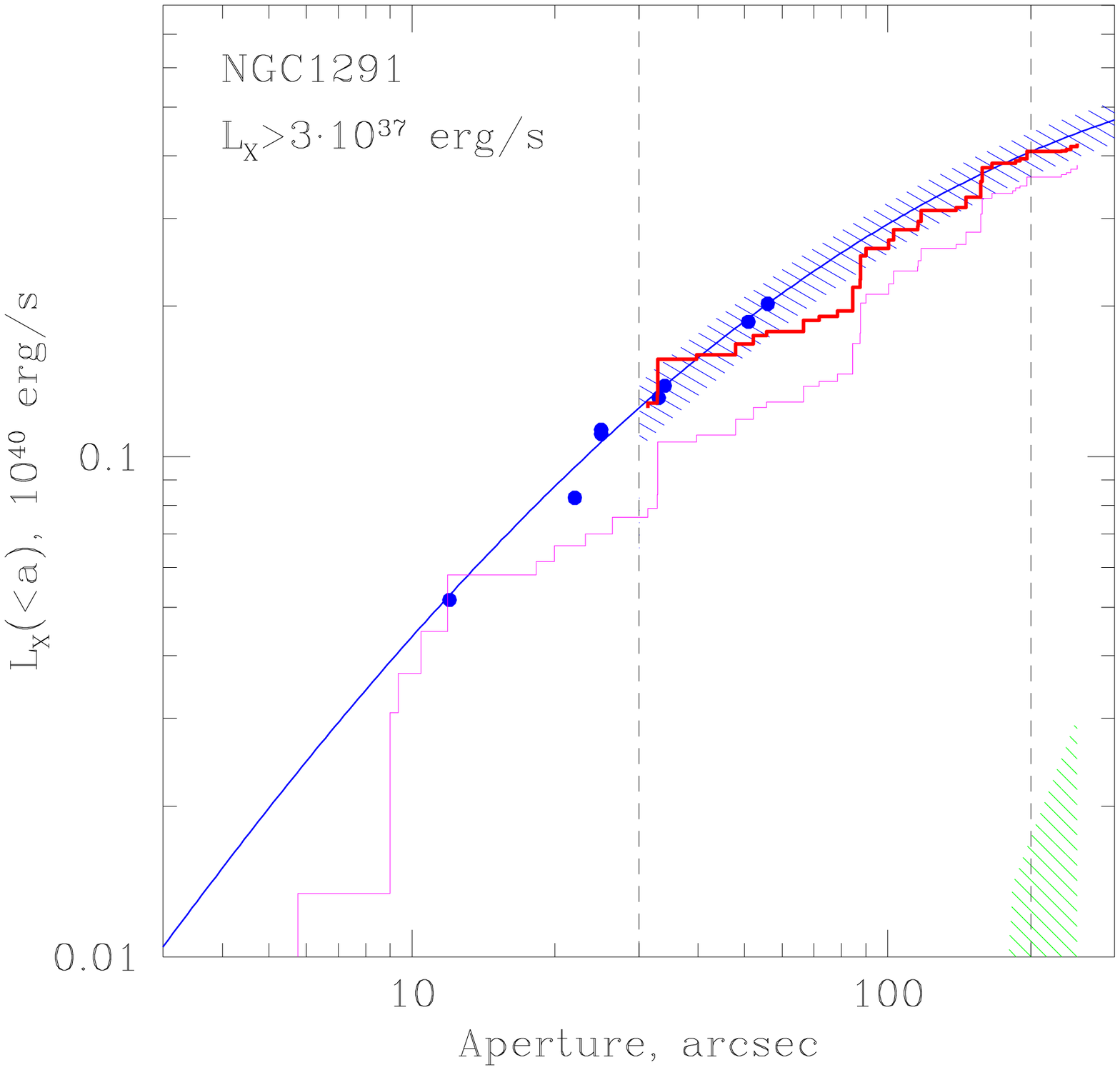}}
\resizebox{0.30\hsize}{!}{\includegraphics{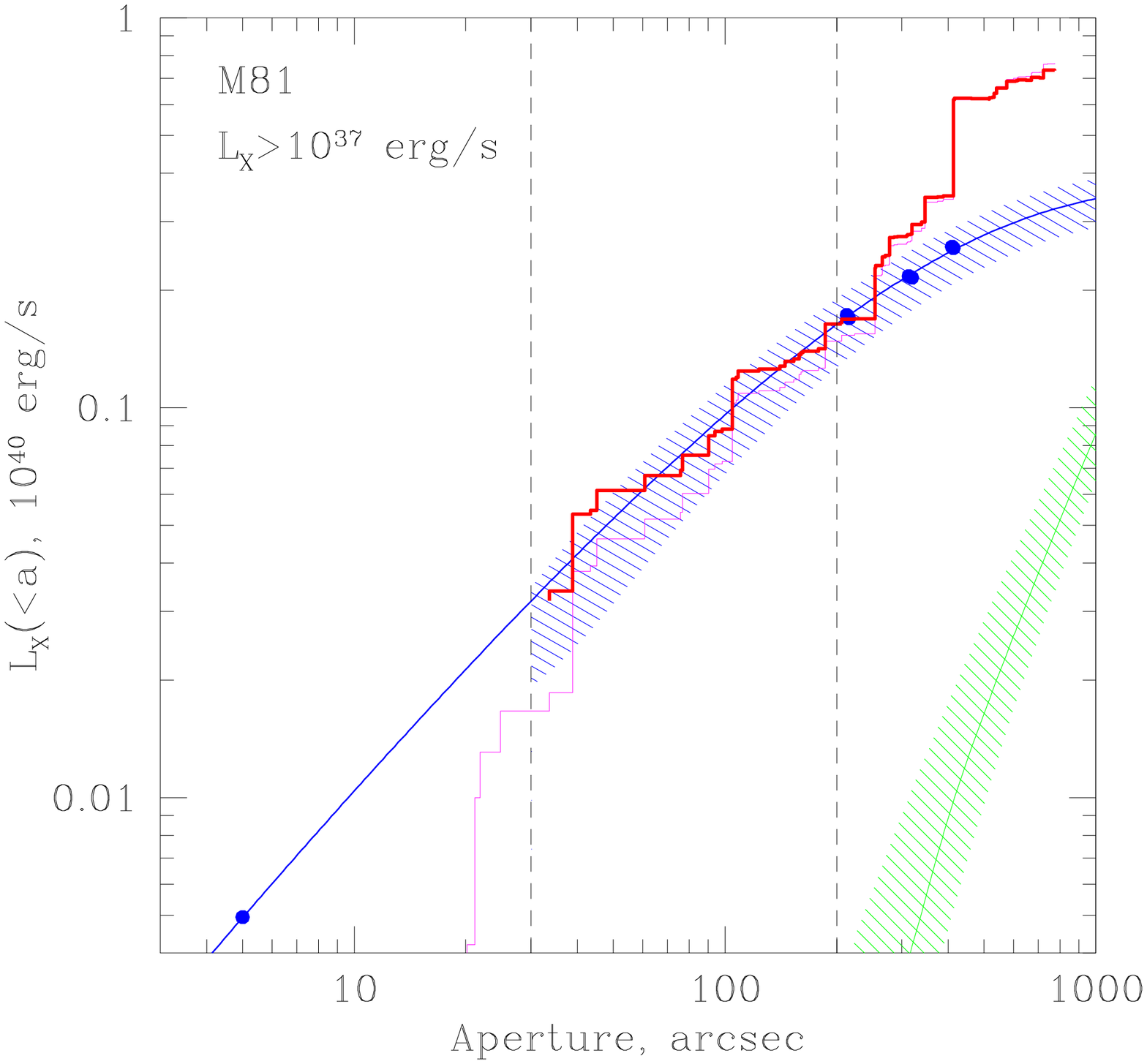}}
}
}
}
\centerline{
\vbox{
\hbox{
\resizebox{0.30\hsize}{!}{\includegraphics{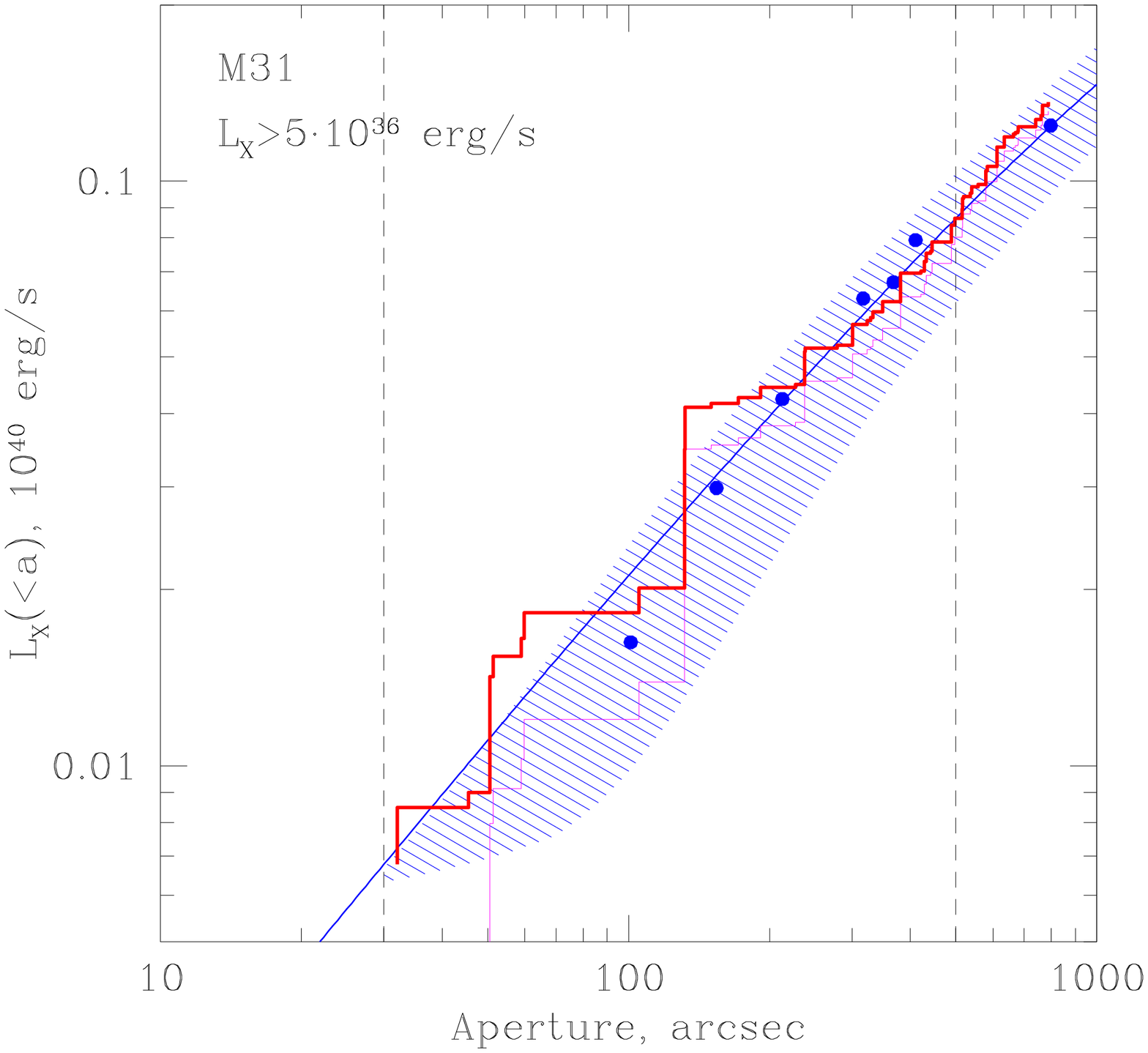}}
\resizebox{0.30\hsize}{!}{\includegraphics{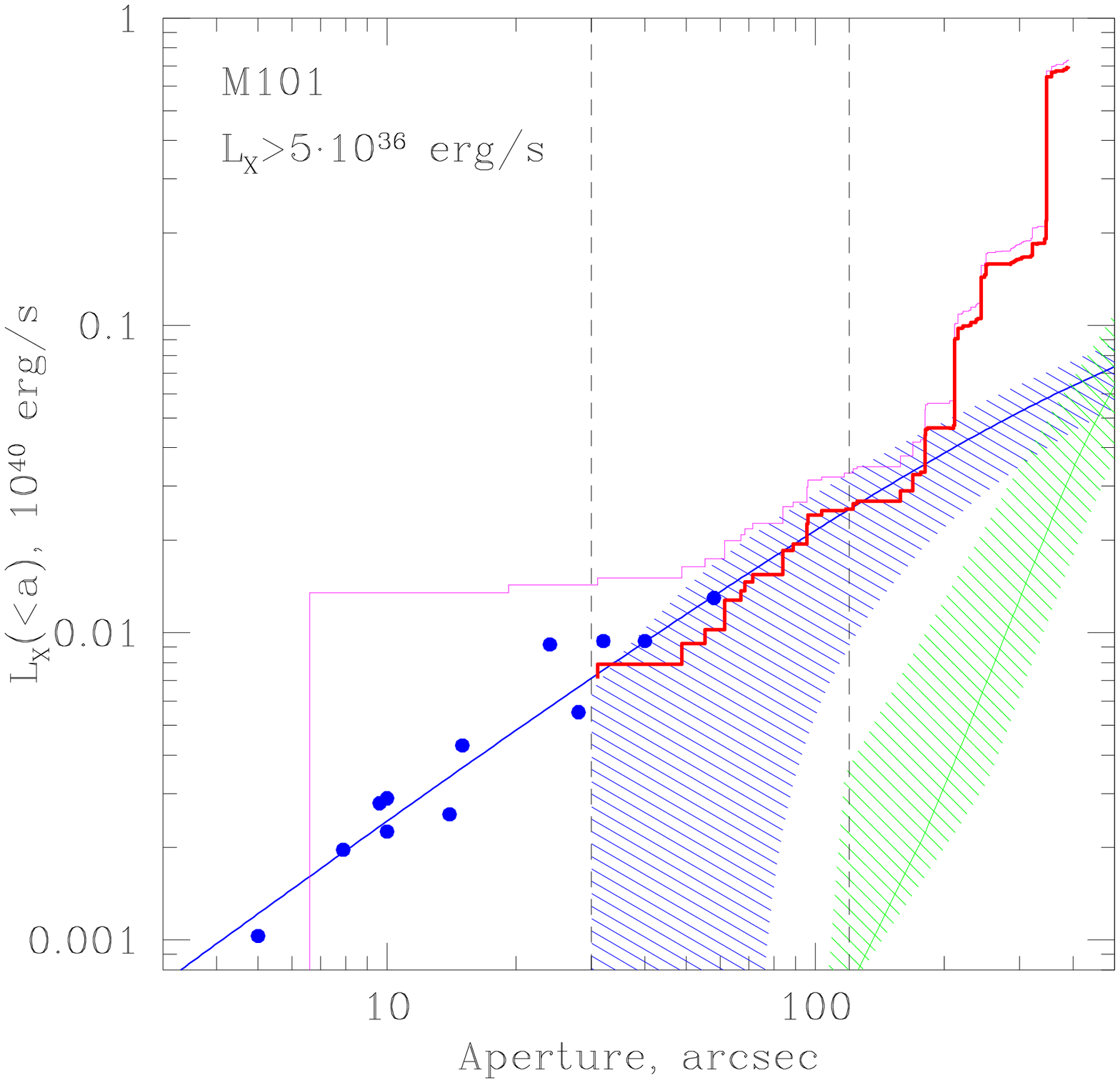}}
}
}
}
\caption{The same as Fig.\ref{fig:gcurves_nx}, but for the total X-ray
luminosity $L_X(<a)$. See caption to Fig.\ref{fig:gcurves_nx} for details.
}
\label{fig:gcurves_lx}
\end{figure*}

\begin{figure*}
\centerline{
\hbox{
\resizebox{0.4\hsize}{!}{\includegraphics{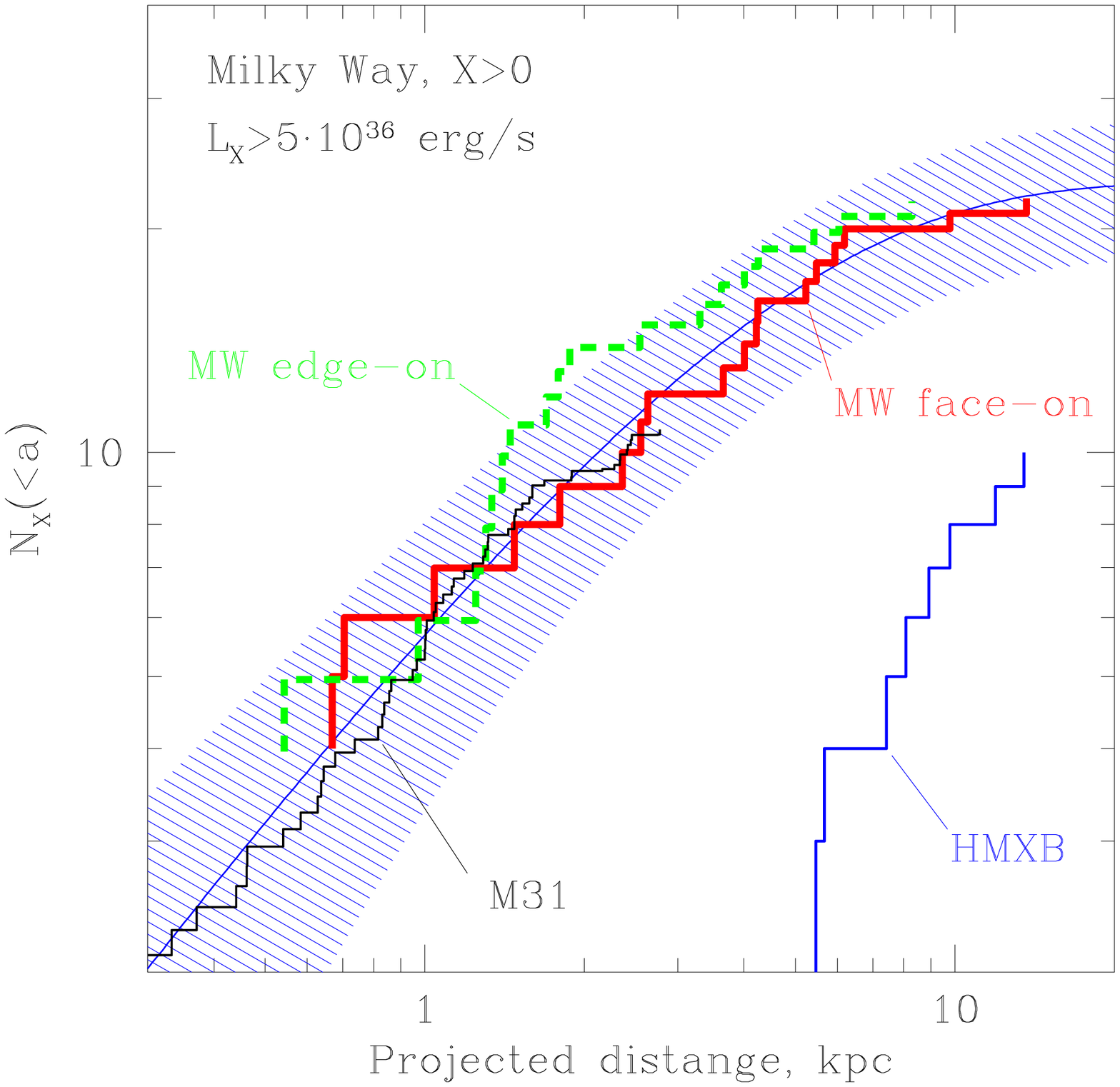}}
\resizebox{0.4\hsize}{!}{\includegraphics{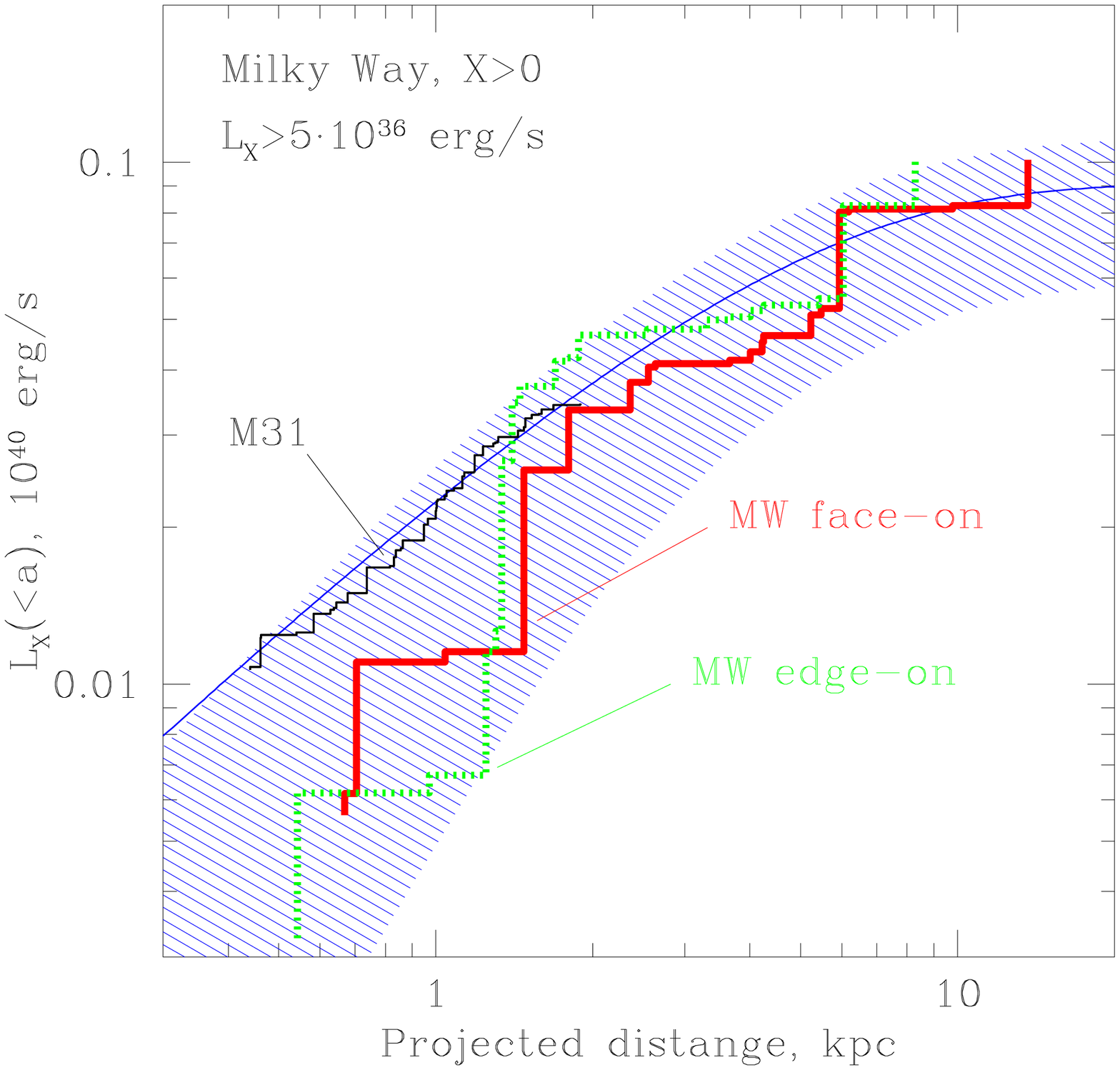}}
}
}
\caption{Growth curves for the number and total luminosity of
LMXBs in the Milky Way compared with X-ray and near-infrared growth
curves of M31. The growth curve of HMXBs in the Milky Way is also
shown for comparison. Due to $\sim 10$ times smaller number of HMXB 
sources in the Galaxy, a lower value of the threshold luminosity  was
chosen for them, $L_X>5\cdot 10^{35}$ erg/s. In plotting the growth
curves, only the sources in the half of the Galaxy, corresponding to
$X>0$, were selected (see section \ref{sec:mw_gcurve}).
The X-ray and NIR growth curves of M31 were normalized
according to X-ray/NIR ratio of the Milky Way. 
}
\label{fig:gcurves_mw}
\end{figure*}

\subsection{Comparison of the growth curves}
\label{sec:nir_x_gcurve}

The X-ray and NIR growth curves are presented in
Figs.\ref{fig:gcurves_nx} and \ref{fig:gcurves_lx}. 
The solid symbols are the multi-aperture NIR photometry data, the
smooth solid curve going through the points is the best fit NIR growth
curve (subsection \ref{sec:nir_gcurve}). Both are scaled by the
X-ray/NIR ratios determined as described below. 
The X-ray growth curves were constructed for the number of sources
$N_X(<a)$, Fig.\ref{fig:gcurves_nx}, and their collective luminosity
$L_X(<a)$, Fig.\ref{fig:gcurves_lx}, inside aperture 
(diameter) $a$. The thin histograms show apparent (observed) X-ray
growth curves plotted using unbinned data, so that each step of the
histogram corresponds to one source. The thin solid line in the
lower-right part of the figures presents the growth curve for the CXB
sources estimated as described in subsection \ref{sec:cxb}. The shaded
areas around NIR and CXB growth curves correspond to the 67\%
statistical uncertainty calculated assuming Poisson distribution
for the number of sources and results of subsection \ref{sec:stat} for
their total luminosity. The field-to-field variations of the CXB
$\log(N)-\log(S)$ were not taken into account in computing the
uncertainties.

The growth curves were analyzed and the X-ray/NIR ratios were
determined in the annulus with the boundaries $a_{X,1}-a_{X,2}$
indicated by the dashed vertical lines in 
Figs.\ref{fig:gcurves_nx}, \ref{fig:gcurves_lx}.
The inner boundary of the annulus $a_{X,1}$ was chosen to avoid the
incompleteness effects 
in the central parts of the galaxies as discussed in subsection
\ref{sec:xray_gcurve}. It was fixed at the value of $30\arcsec$
for the majority of the galaxies. 
For three galaxies larger values of $a_{X,1}$ were adopted. 
A strong diffuse source is observed in the case of NGC4472
\citep{ngc4472} -- we used inner diameter of $50\arcsec$. 
NGC5846 is one of the two most distant and gas-rich \citep{ngc5846}
galaxies in our sample.  Its X-ray growth curve shows an apparent
decline inside $\sim  100\arcsec$ (Fig.\ref{fig:gcurves_nx}). We
attributed this decline to incompleteness effects and, in part, to
confusion effects and assumed $a_{X,1}=100\arcsec$. 
In the case of Cen A, the search for point X-ray sources in the
inner  part is complicated by the powerful jet emission. We therefore
assumed  $a_{X,1}=100\arcsec$.  
The outer diameter $a_{X,2}$  was constrained 
by either contribution of the CXB sources or by the boundaries of the
CCD chip used in the analysis.
It typically varied between $\sim 150-300\arcsec$. For the small bulge
in the Scd galaxy M101 \citep{m101bulge}, the outer boundary
was chosen at $a_{X,2}=120\arcsec$ in order to minimize contribution
of HMXBs \citep[cf. ][]{m101} in the derived X-ray/NIR ratios.

In calculating X-ray/NIR ratios, the number and luminosity of X-ray
sources were corrected for the contribution of CXB sources. The
near-infrared luminosity was computed integrating the best fit NIR
growth curves. The boundaries $a_{X,1}-a_{X,2}$ and derived X-ray and 
near-infrared parameters are listed in Table \ref{tab:xray}. 

As evident from Fig.\ref{fig:gcurves_nx} the growth
curves for the number of X-ray sources agree well with the
near-infrared ones. It is confirmed quantitatively by the results of
the Kolmogorov--Smirnov test -- the minimum values of K-S probabilities
are 14\% (NGC 1553) and 30\% (M84) and are in the $\sim 46-97\%$ range
for other galaxies. 
To facilitate visual comparison of the growth curves, we plot as a thick
histogram the X-ray growth curves shifted vertically to match the NIR
curves at the inner boundary of the annulus $a_{X,1}$:
\begin{eqnarray}
N_X(a)=N_X^{\rm obs}(a)-N_X^{\rm obs}(a_{X,1})+
N_X^{\rm  pred}(a_{X,1})\\
-\left (N_{X,CXB}(a)-N_{X,CXB}(a_{X,1})\right ) \nonumber
\end{eqnarray}
i.e. corrected for the expected number (luminosity) of the sources
inside $a_{X,1}$ and for the contribution of the CXB sources.

With few exceptions, the X-ray luminosity growth curves also agree
well with the distribution of the near-infrared light
(Fig.\ref{fig:gcurves_lx}). 
Significant deviations are observed in the case of Cen A and
NGC1553, and, less prominent, in NGC1291. 
In the Cen A galaxy, the deviations are caused by a bright transient 
source  
with $L_X\sim 10^{39}$ erg/s  located at $r\approx 150\arcsec$ from
the  nucleus, which changed its luminosity by a factor of $\ga 500$ 
between two Chandra  observations separated by  
half a year \citep{cena}. If this source is excluded, the good 
agreement between X-ray and NIR growth curves is restored (dashed
thick histogram in Fig.\ref{fig:gcurves_lx}). 

In the case of NGC1553 and NGC1291, the X-ray growth curves appear
somewhat ``under-luminous'' inside $\sim$ effective radius of the
galaxy. Interestingly, at larger radii the agreement with the NIR
growth curves seems to be restored.  
The deviations are smaller in NGC1291, $\sim 40\%$, and are especially
pronounced  in NGC1553, reaching a factor of $\sim2$. For the latter,
we give in the Table \ref{tab:xray}  X-ray and NIR parameters for two
different values of the outer radius $a_{X,2}$. 
The behavior of the $N_X$ and $L_X$ growth curves in these galaxies
might indicate the presence of variations in the luminosity
distribution 
of the sources with distance from the nucleus.  As will be discussed
in the following sections,  luminosity distributions  of compact
sources in both galaxies appear to deviate somewhat from the  average
XLF (section \ref{sec:individual}).

\begin{figure*}
\centerline{
\vbox{
\hbox{
\resizebox{0.39\hsize}{!}{\includegraphics{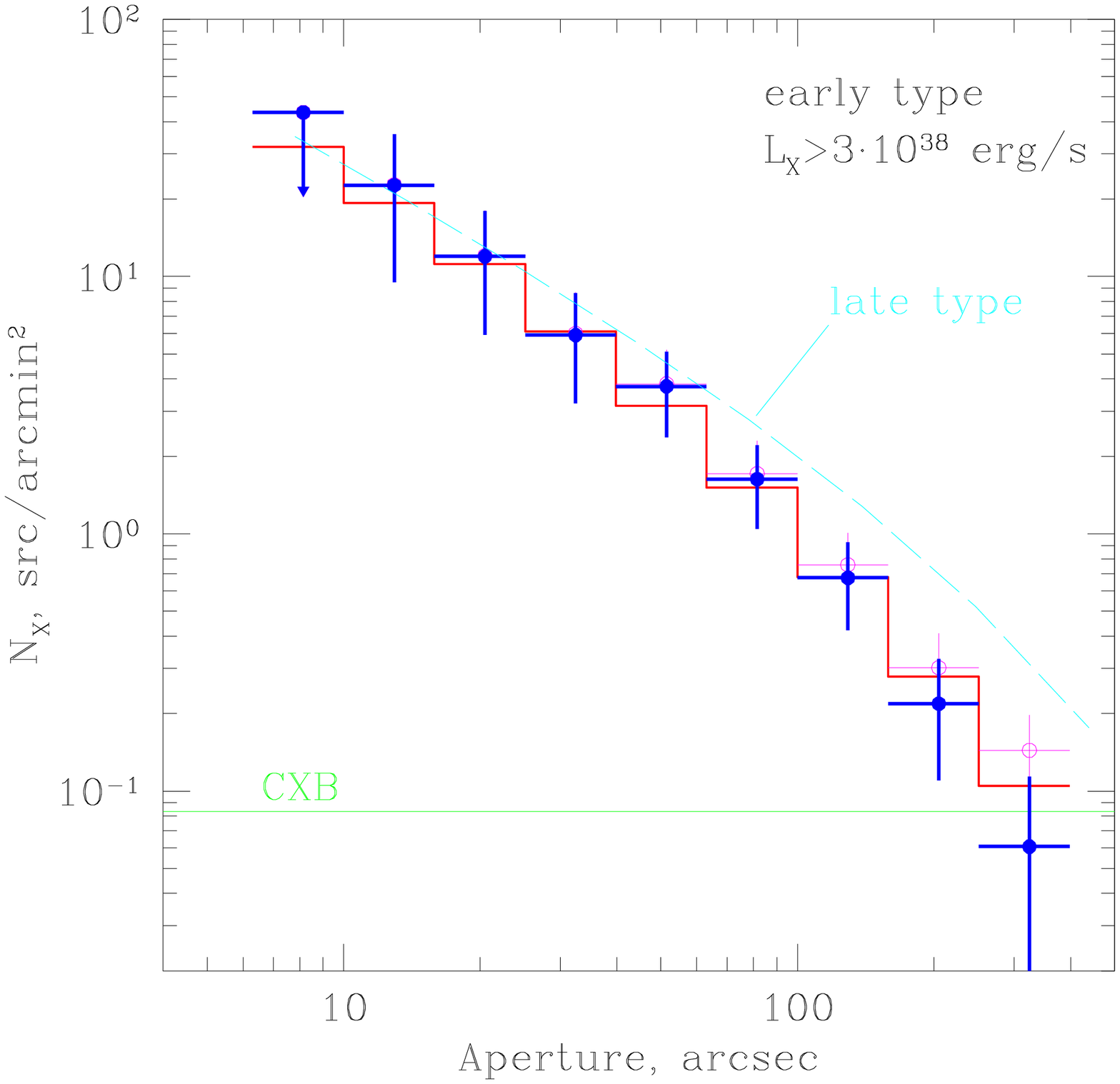}}
\resizebox{0.39\hsize}{!}{\includegraphics{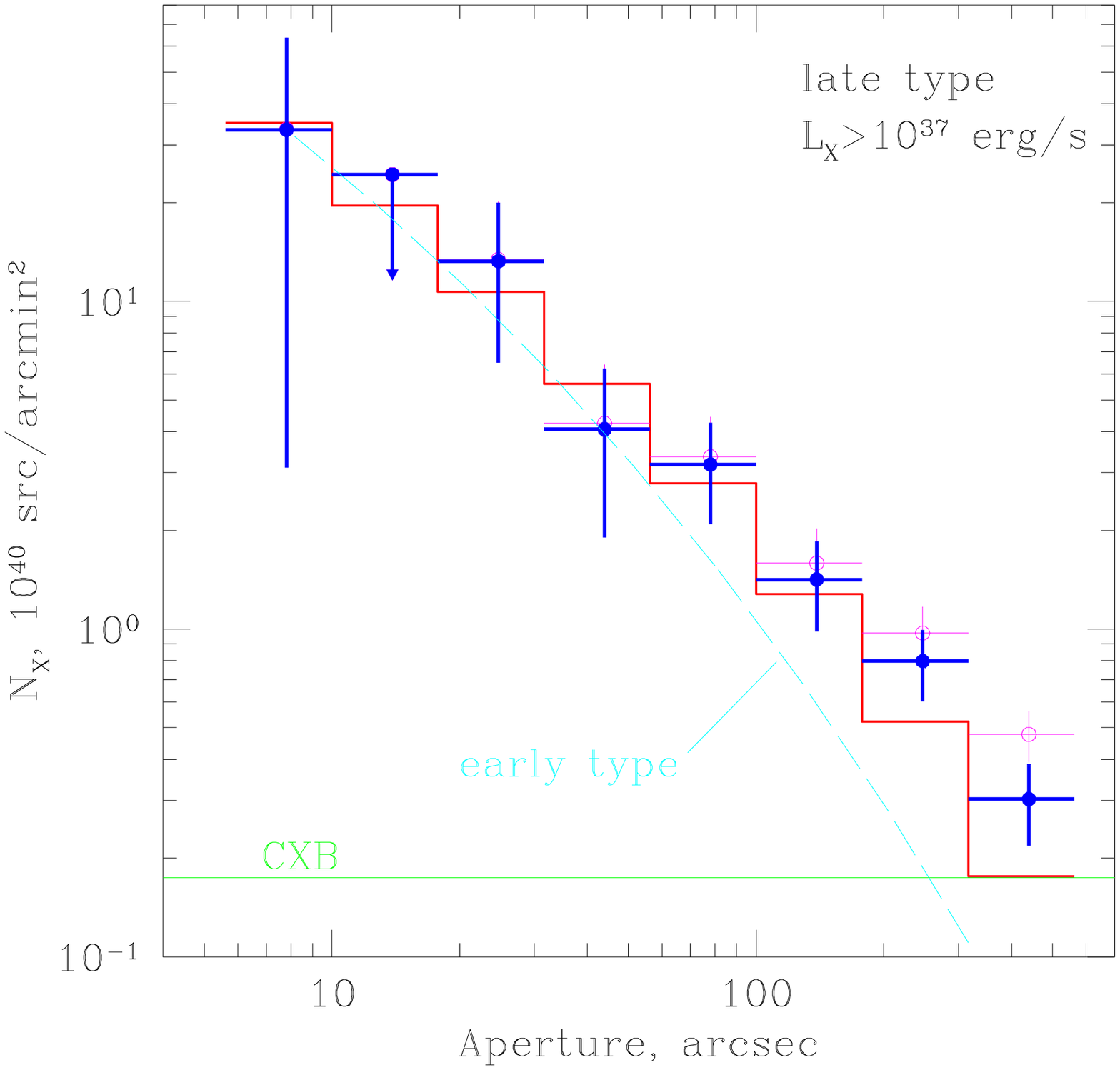}}
}
\hbox{
\resizebox{0.39\hsize}{!}{\includegraphics{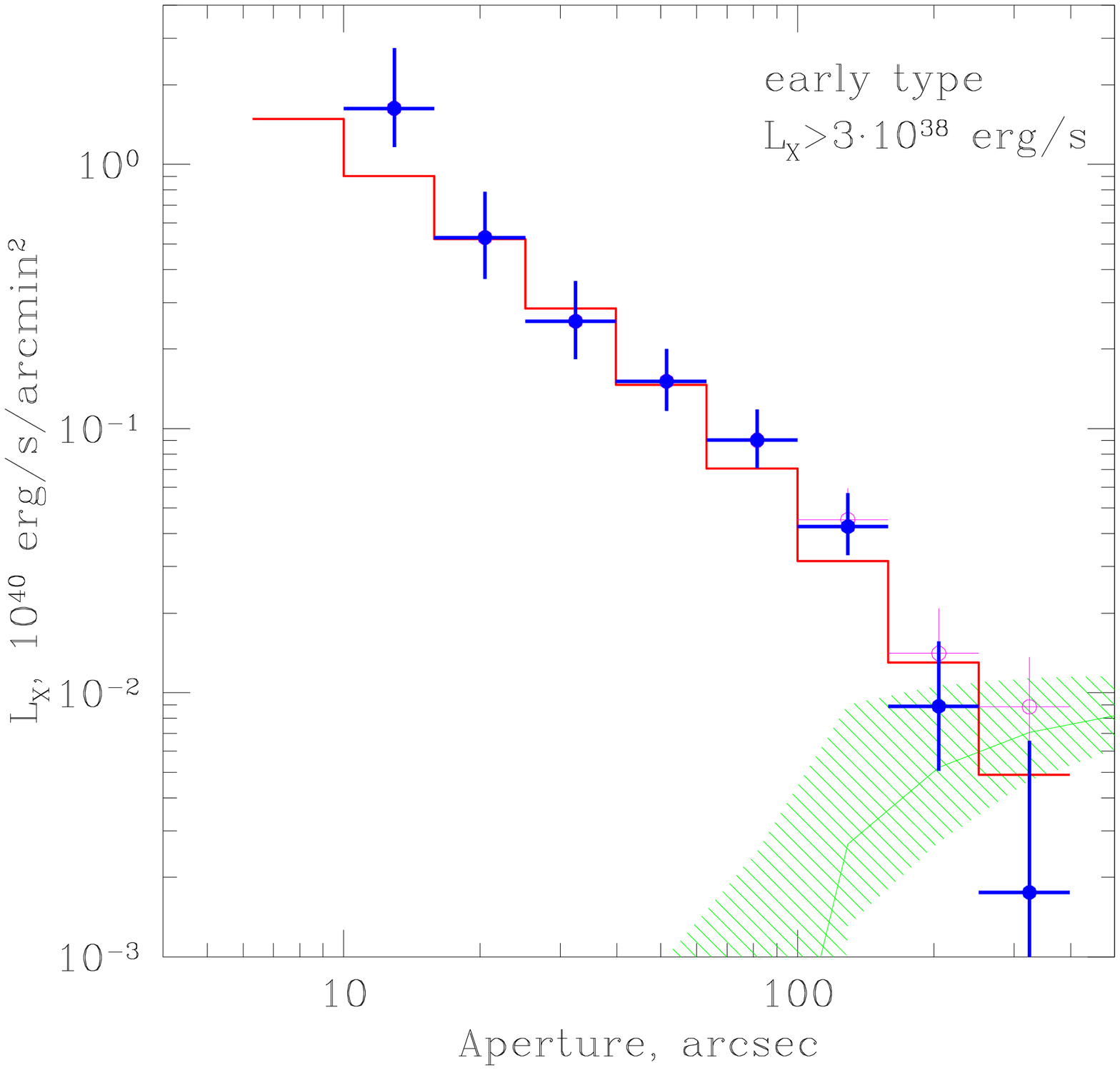}}
\resizebox{0.39\hsize}{!}{\includegraphics{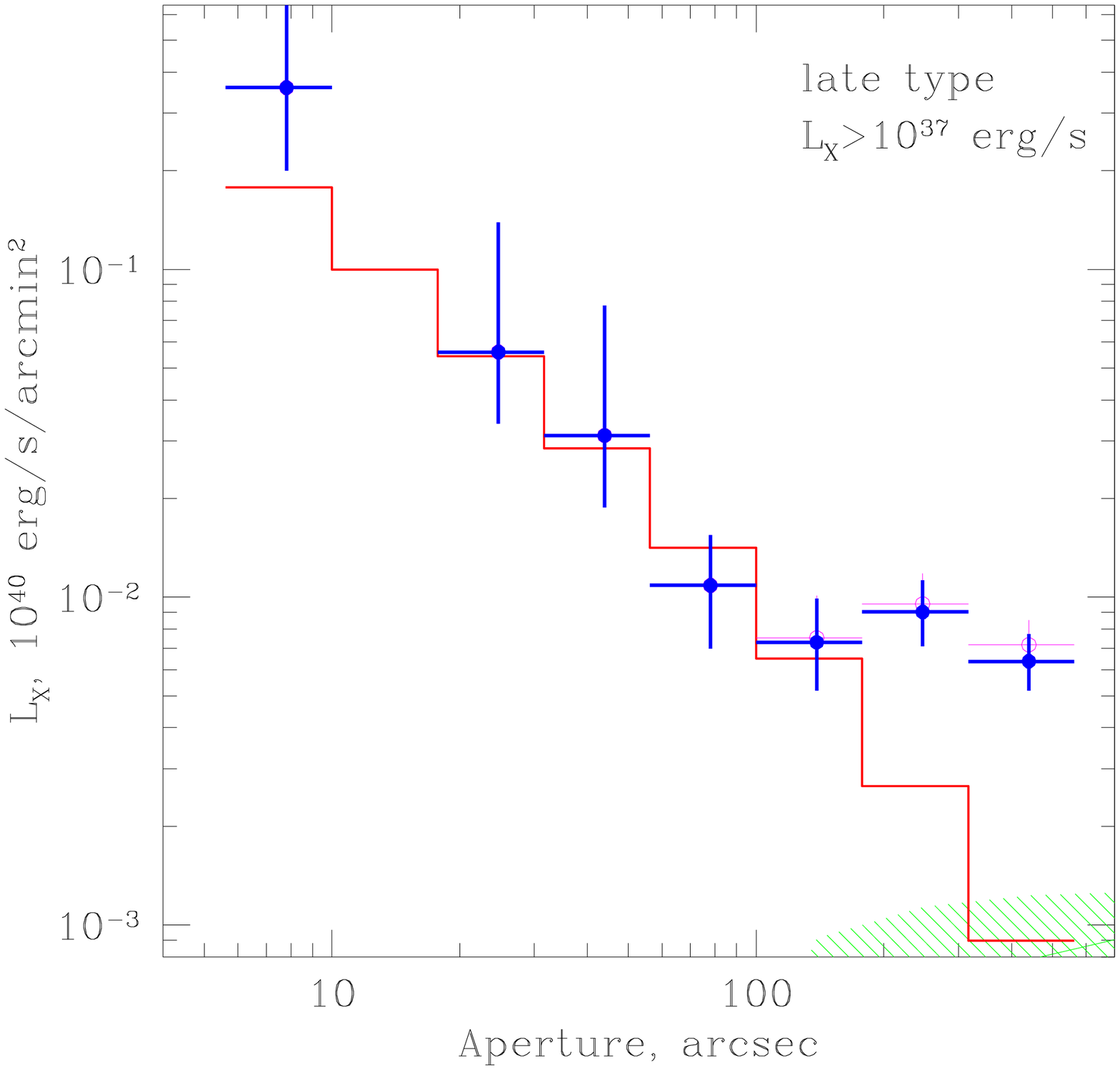}}
}
}
}
\caption{The differential radial profiles for the number of compact
sources $N_X$ and their collective luminosity $L_X$. 
{\em Left:} combined data for 4 early type  galaxies (NGC4472,
NGC4697, M84 and NGC1316). 
{\em Right:} combined profiles of two spiral galaxies (M81
and M101) with evidence of star formation outside the bulge.
The thin open circles with error bars show observed
profiles. The thick solid points with error bars show the profiles
corrected for contribution of CXB sources. The latter is shown by thin
solid line in the lower part of each panel, the shaded area in the
lower graphs shows 67\% statistical uncertainty.  The histograms show
radial profiles of the NIR light. For comparison, the radial profiles
of late and early type galaxies are shown by dashed lines. The upper
limits are 90\% confidence. 
}
\label{fig:profiles}
\end{figure*}

\subsection{Late type galaxies}
\label{sec:gcurves_spir}

The properties of X-ray sources in spiral galaxies appear to correlate
with the properties of the surrounding stellar population
\citep[see][ for a review]{fabbiano2003}. The old stellar systems of
bulges are dominated by low mass X-ray binaries, whereas the disk
population might be significantly affected by ongoing star
formation and, hence, a contribution of HMXB sources should be
expected. These expectations are supported by the spatial distribution
of the compact X-ray sources in the disks of spiral galaxies, which
often show a concentration towards the spiral arms
\citep[e.g.][]{m101,grimm1}.  

As the spatial distribution of HMXBs does not necessarily follow the
stellar mass and, rather, correlates with the regions of star
formation, the X-ray growth curves for spiral galaxies deviate 
significantly from the  distribution of the near-infrared light
outside the bulge (Fig.\ref{fig:gcurves_nx} and
\ref{fig:gcurves_lx}). These deviations are 
mostly apparent for M81 and M101, having disk star formation at the
level of $\sim 1$ M$_{\odot}$/year. The bulge size  
corresponds to the aperture of $\sim 100\arcsec -150\arcsec$ for M101
\citep{m101bulge}) and $\sim 450\arcsec  \times 240\arcsec$ ellipsoid
in the case of M81 \citep{m81bulge}.  As the luminosity
function of HMXBs has a cut-off at $\sim 10$ times higher luminosity
than LMXBs (section \ref{sec:xlf}, \citet{grimm2}),
the deviations are most apparent in the X-ray luminosity growth
curves. The deviations are significantly less pronounced in the case
of M31 which has a larger angular bulge size, 
$a\ga 1000\arcsec$ 
\citep{m31bulge, fabbiano2003}, and smaller star formation rate within
the Chandra field of view.

\subsection{The Milky Way}
\label{sec:mw_gcurve}

Unlike external galaxies, the growth curves for the Milky Way cannot be
directly constructed, because different sources are located at
different distances. During the past decade, significant progress has
been achieved in studying the bright X-ray binaries in the Galaxy. In
particular, distances have been determined by various methods for
a significant fraction of X-ray binaries brighter than $\sim 10^{36}$
erg/s. This allowed \citet{grimm1} to reconstruct  a 3-dimensional
picture of the galactic X-ray sources. Progress in optical
observations also has led to determination of the nature of the
optical companion in the majority of these systems. As has been argued
by  \citet{grimm1}, although residual incompleteness effects are still
in play, the sample of X-ray binaries with $L_X>10^{35.5-36}$ erg/sec
located within $\sim 10-12$ kpc from the Sun should be sufficiently
complete.

In order to study the growth curve of the low mass X-ray binaries in
the Milky Way, we use  X-ray luminosities and the compilation of 
source distances from \citet{grimm1}. The X-ray luminosities were
determined by \citet{grimm1} from the 5 year average of the the
RXTE/ASM light curves of individual sources. We selected the LMXBs
in the half of the Galaxy corresponding to $X>0$, where the origin of
the Cartesian coordinate system coincides with the Galactic Center and
the X-axis is directed towards the Sun \citep{grimm1}. 
As the similar problem of different distances to the
different emitting regions arises in the near-infrared band as well,
we used as a model for the Milky Way the NIR light distribution
in M31 which is sufficiently similar
to our Galaxy in size and morphological type. 
The resulting growth curves for face-on and edge-on views of the Milky
Way are plotted  and compared with the NIR and X-ray growth curves of
M31 in Fig.\ref{fig:gcurves_mw}. Also shown for comparison is the growth
curve for HMXB sources in the Milky Way. Good agreement in the Milky
Way and M31 growth curves is apparent. On the other hand, it is
obvious that the HMXB sources have a significantly different spatial 
distribution.   

As the accuracy of the distance determination is unlikely to exceed
$\sim 10-20\%$ even in the best studied cases, we excluded the central 1
kpc from our calculation of the X-ray/M$_*$ ratios. 
The X-ray and near-infrared parameters listed in Table \ref{tab:xray}
were calculated for a face-on view of the Galaxy using the LMXB
sources located at 
$X>0$ and  $1{\rm ~kpc} <R_{\rm proj}<10{\rm ~kpc}$, where  
$R_{\rm proj}$ is the projected distance from the Galactic Center for 
a face-on view. 
The edge-on projection gives somewhat higher values, by $\sim 10\%$
for the number of sources and by $\sim 30\%$ for their total luminosity.
Note that the values of the near-infrared luminosity, stellar mass,
number of low mass X-ray binaries and their total luminosity,  
given in Table \ref{tab:xray}, refer to the  $\la$ half of the Galaxy.

\begin{figure}
\resizebox{\hsize}{!}{\includegraphics{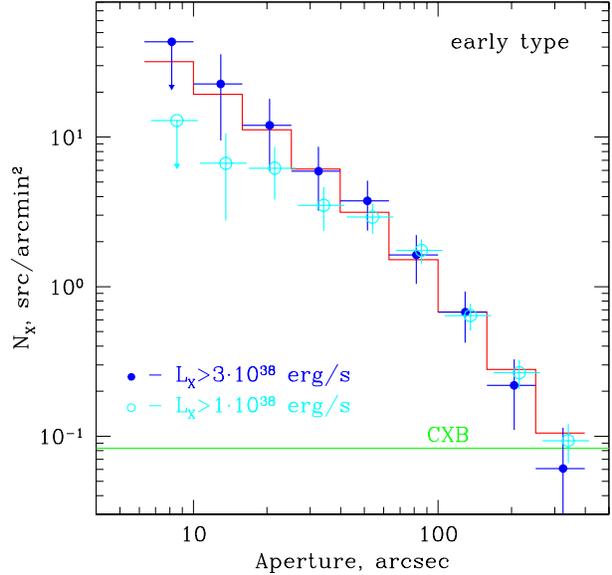}}
\caption{Illustration of incompleteness effects in the central part of
elliptical galaxies. The points with error bars show  differential
radial profile of the number of sources for two different values of
cut-off luminosity $L_{X,min}$. Contribution of CXB sources
subtracted. The X-ray radial profile for
$L_{X,min}=10^{38}$ erg/sec was rescaled according to the
$L_X/L_{NIR}$ ratio and shifted horizontally by a small offset for
clarity. The histogram shows radial distribution  
of the NIR light. The horizontal line shows density of the CXB sources
for cut-off luminosity $L_{X,min}=3\cdot 10^{38}$ erg/sec.
}
\label{fig:incompl}
\end{figure}

\subsection{Differential radial profiles}

The total number of sources above the completeness limit in  most
of the galaxies is insufficient to construct meaningful differential
radial profiles. Therefore, in order to study the differential
distribution 
of X-ray binaries, we combined the data for several galaxies
of similar morphological type. The resulting differential profiles for
the number of sources and their total luminosity are shown in
Fig.\ref{fig:profiles} using combined data for 4 early type galaxies
having sufficiently low completeness limit (NGC4472, NGC4697, M84 and 
NGC1316), and late type galaxies M81 and M101. The near-infrared
profiles were computed accordingly, combining the individual NIR
growth curves.   

As was discussed in section \ref{sec:nir_x_gcurve}, the growth curves,
especially for the early type galaxies, appear to have a deficit of the
sources in the central $\sim20\arcsec-40\arcsec$ (diameter), which was
attributed to a combination of various factors affecting the
completeness of the source samples. In order to check this assumption
we compare the radial profiles for the combined data for early type
galaxies constructed using different low luminosity limits $L_{\rm min}$ 
(Fig.\ref{fig:incompl}). As is evident from Fig.\ref{fig:incompl}, the
radial profile for $L_{\rm min}=10^{38}$ erg/s shows a clear deviation
from the NIR profile inside $\sim 40\arcsec$. Increasing  $L_{\rm
min}$ to $3\cdot 10^{38}$ erg/s leads to a radial profile
consistent with the NIR profile. Although the luminosity dependent
effects can not  be presently excluded,  it seems unlikely that such
effects are responsible for distortion of the apparent radial 
profiles in the central part of galaxies. This conclusion is further
supported by analysis of \citet{m84} and \citet{ngc1316} showing that 
incompleteness effects due to the centrally concentrated diffuse
emission can cause an apparent decrease of the number of detected
sources in the central parts of the galaxies.

As might be expected from the growth curve analysis, there is  good
agreement between X-ray radial profiles and the  distribution of
the near-infrared light. For the spiral galaxies M81 and M101,
deviation of the radial distribution from the NIR profile is evident
outside the bulge, where star formation and HMXBs become important.

\begin{figure*}
\hbox{
\resizebox{0.5\hsize}{!}{\includegraphics{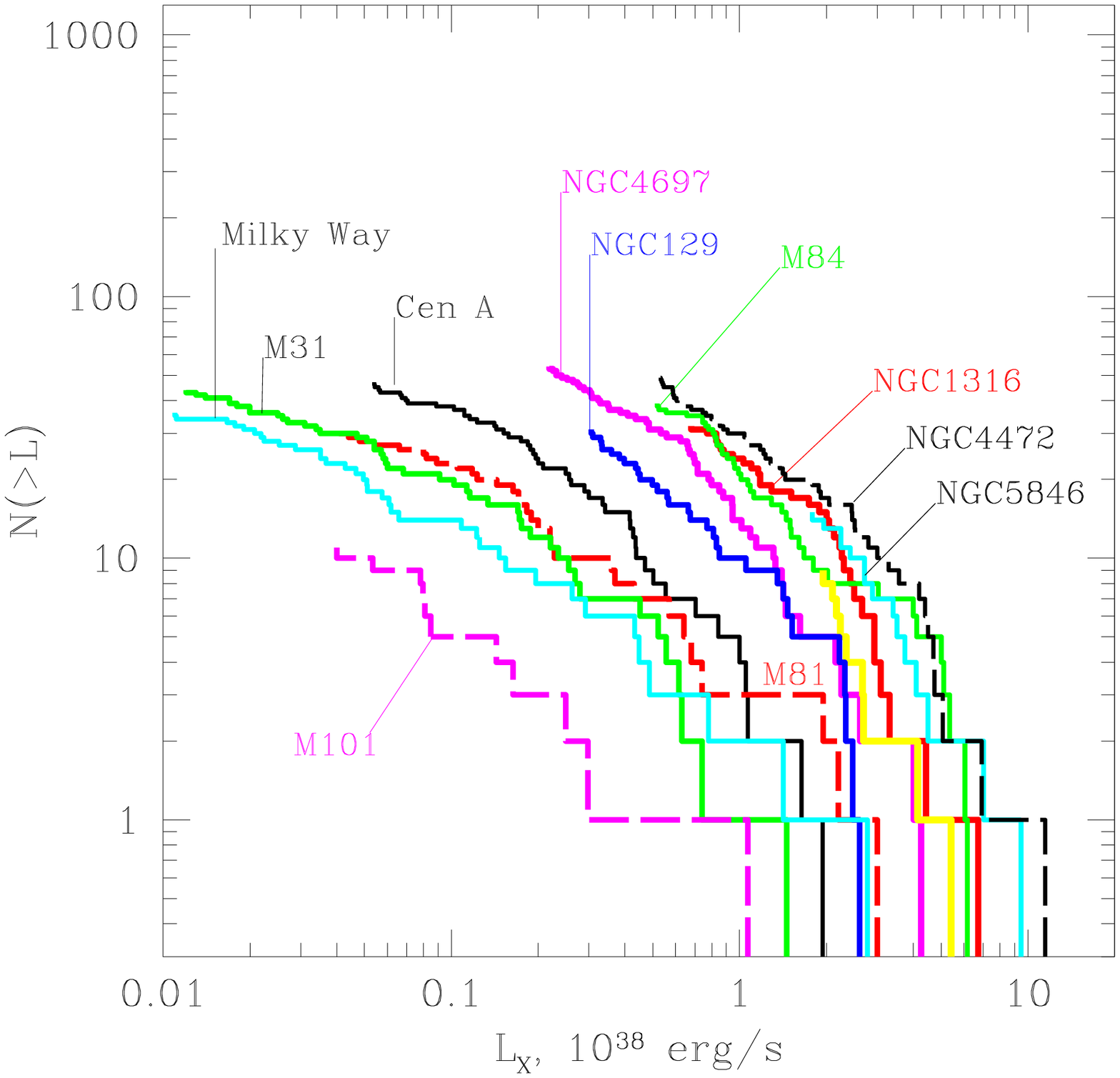}}
\resizebox{0.5\hsize}{!}{\includegraphics{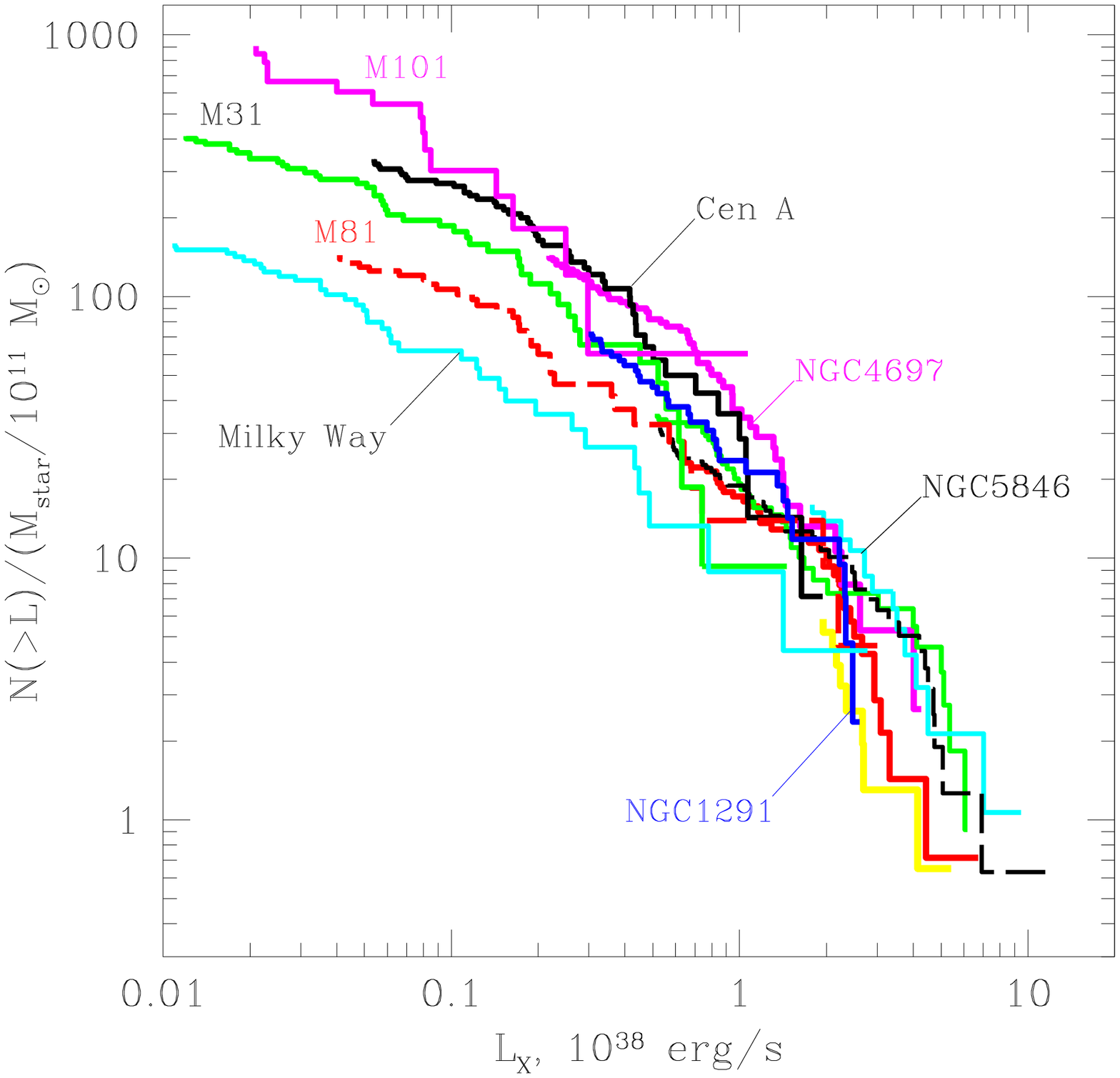}}
}
\caption{Cumulative X-ray luminosity functions for galaxies from our
  sample. {\em Left:} observed; {\em Right:} scaled by stellar mass.
  The luminosity functions were extracted in the annuli  
  defined in the Table \ref{tab:sample2}. 
}
\label{fig:xlf_cum}
\end{figure*}

\begin{figure*}
\vspace{0.5cm}
\hbox{
\resizebox{0.5\hsize}{!}{\includegraphics{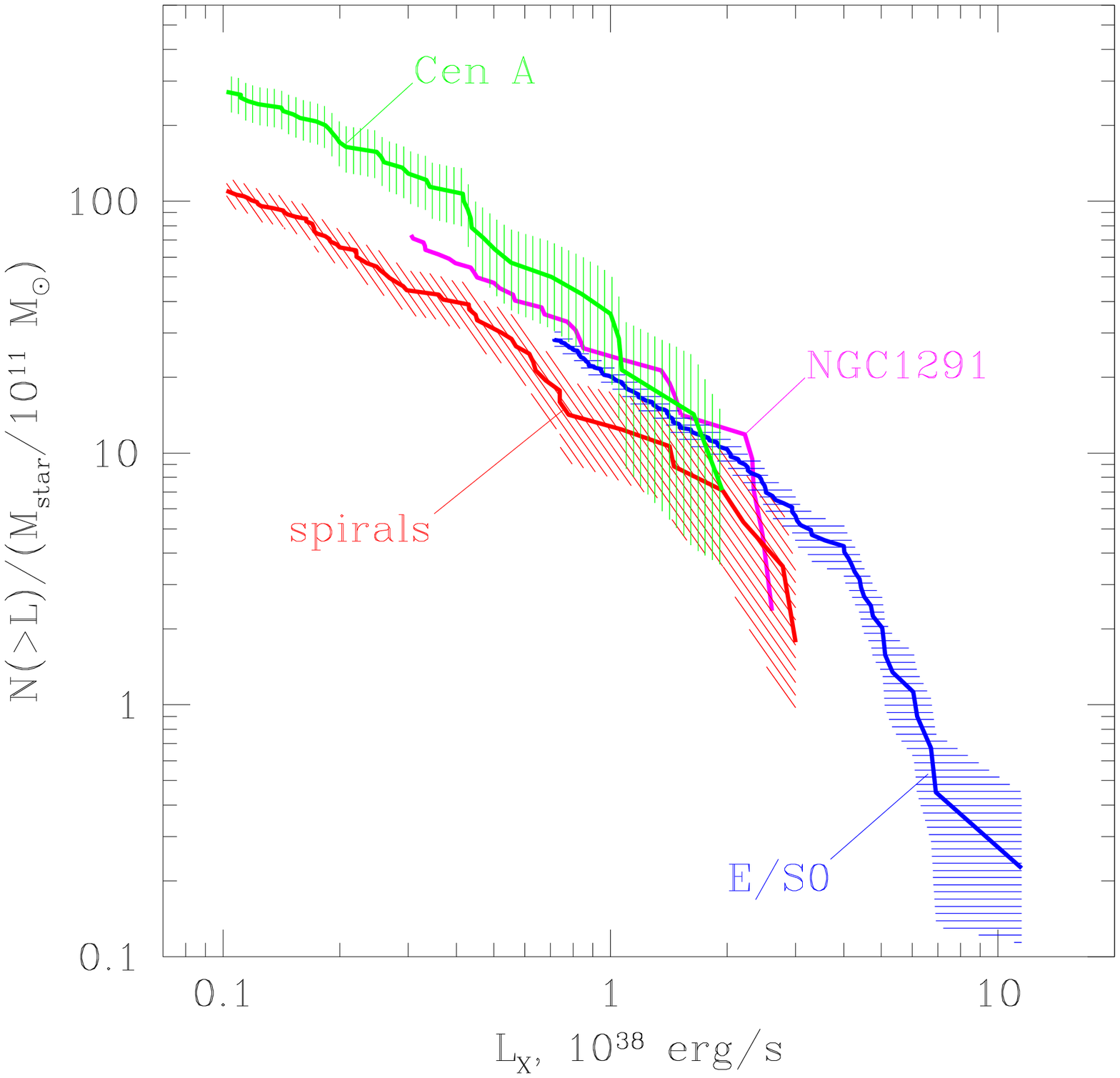}}
\resizebox{0.5\hsize}{!}{\includegraphics{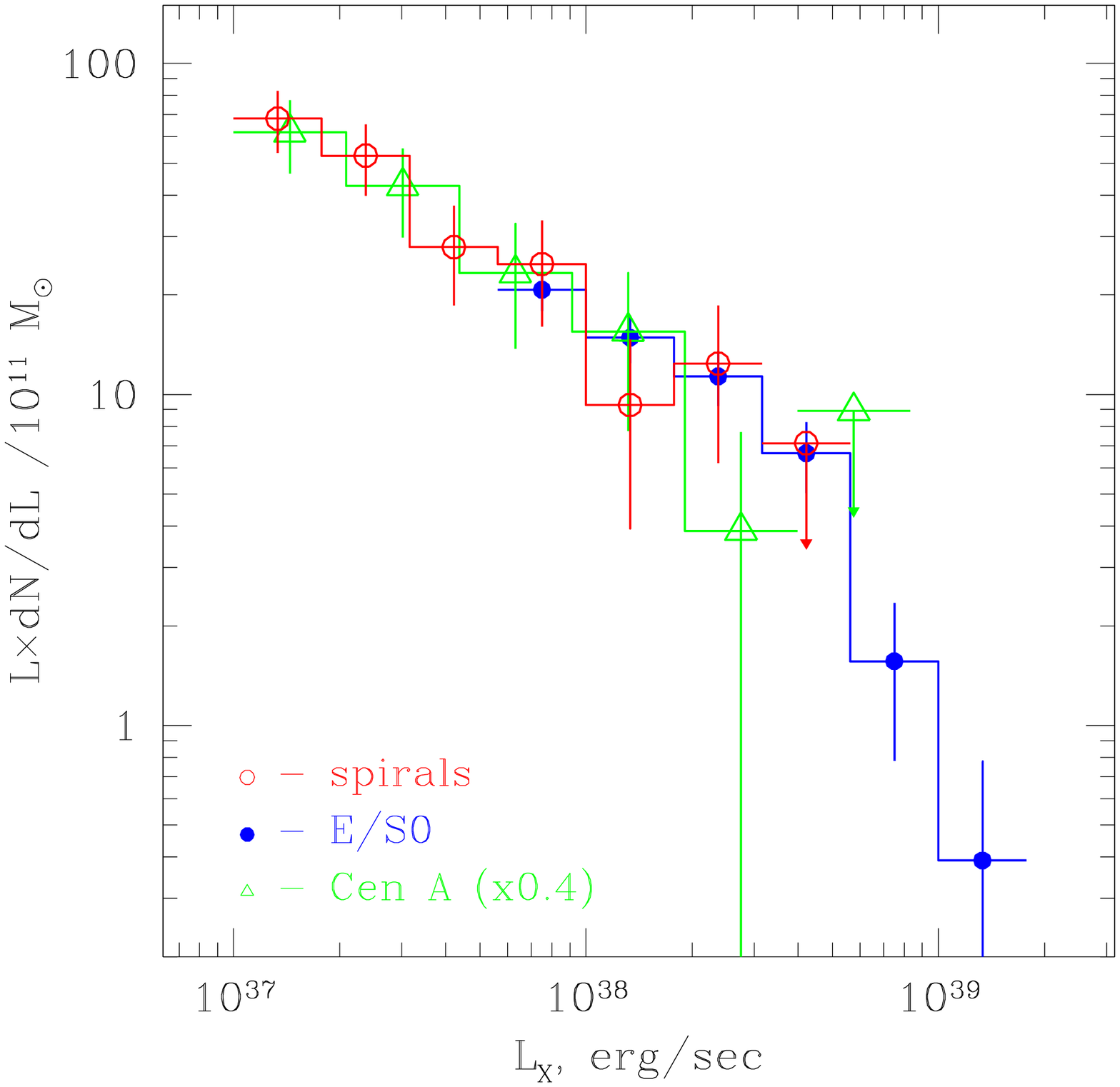}}
}
\caption{{\em Left:} Combined cumulative  X-ray luminosity function
  for  early type galaxies (NGC4472, NGC4697, M84 and NGC1316),
  spiral galaxies (M31, M81, M101 and the Milky Way) and Cen A galaxy,
  normalized to the stellar mass.  The shaded areas 
  indicate 67\% uncertainty in the number of sources assuming
  Poisson distribution. 
{\em Right:} Differential luminosity functions for the same sets  as
in the left panel normalized to the stellar mass. To facilitate
comparison of the shape, rather than normalization, the luminosity
function of Cen A in the right panel is  multiplied by a factor of
0.4. The upper limits in the right panel are 90\% confidence.}
\label{fig:xlf_comb}
\end{figure*}

\subsection{X-ray/NIR ratios}

The X-ray/NIR ratios derived from  the analysis of the growth curves
have the advantage of being relatively free of a number of biases and
incompleteness effects they could suffer otherwise. In particular they
are not subject to incompleteness effects due to diffuse emission and
source confusion in the central parts of the galaxies and are not
affected by the contribution of CXB sources. The disadvantage, however
is that they are obtained for different values of the luminosity
threshold $L_{\rm min}$. This is due to the fact that, for more
distant early type galaxies, the Chandra sensitivity is typically $\sim
10^{38}$ erg/s, whereas, in more nearby late type galaxies, the number
of high luminosity sources is insufficient for meaningful
analysis. For this reason the X-ray/NIR ratios cannot be directly
compared for different galaxies from our sample and have to be
corrected for difference in the threshold luminosity. This will be 
done in the section \ref{sec:x_nir_ratios}.

\section{X-ray luminosity function of LMXBs}
\label{sec:xlf}

In order to study the X-ray luminosity functions, 
we use compact sources located in the annuli defined in section 
\ref{sec:nir_x_gcurve}. The observed luminosity functions are shown in
the left panel in Fig.\ref{fig:xlf_cum}, with the luminosity functions
normalized to the stellar mass in the right panel. 
As evident from Fig.\ref{fig:xlf_cum} 
the normalized XLFs occupy a rather narrow but still finite
width band in the $N(>L) - L$ plane. The difference in the number of
sources is up to a factor of $\ga 3-4$ and is statistically
significant. Note, that the Milky Way galaxy shows the largest
deviation and is significantly below the main group. 
However, the shapes of the XLF for different galaxies are similar to
each other, with the possible exception of NGC1553, which is discussed
in detail in section \ref{sec:individual}. 
This is further illustrated by Fig.\ref{fig:xlf_comb} which
shows the combined XLFs of early, late type galaxies and Cen A in
cumulative and differential forms.
Although the data for early and late type galaxies  overlap in a
rather narrow luminosity range near $L_X\sim 10^{38}$ erg/s, it is
obvious that they have close, although not identical,
normalization. They also appear to have a similar slope below 
$\sim$few$\times 10^{38}$ erg/s. This is further confirmed by the XLF
of Cen A -- the only early type galaxy whose luminosity function
extends down to $\sim 10^{37}$ erg/s. 

Based on this evidence, we tentatively conclude that there is no
significant difference in the shape of the XLFs above $\sim 10^{37}$
erg/s between individual galaxies as well as in the combined data of
early and late type galaxies.

\begin{table*}
\renewcommand{\arraystretch}{1.2}
\caption{Results of the maximum likelihood fits to the observed
luminosity distributions by the template XLF, eq.(\ref{eq:uxlf})} 
\begin{tabular}{|l|c|c|c|c|c|c|c|}
\hline
sample & $L_{\rm min}$ & $\alpha_1$ & $L_{b,1}$ & $\alpha_2$ & $L_{b,2}$ & $\alpha_3$\\
\hline
% 4 ellipticals: 1316+4472+4697+M84
early type & $1\cdot 10^{38}$ & -- & -- & $1.64\pm0.22$ &
$5.1^{+1.4}_{-0.7}$ & $5.0^{+2.3}_{-1.1}$ \\

% 4 ellipticals: 1316+4472+4697+M84, Lxmin=2e38
early type &  $2\cdot 10^{38}$ & -- & -- & $1.80_{-0.53}^{+0.61}$ &
$5.1^{+1.9}_{-0.7}$ & $5.0^{+2.8}_{-1.1}$ \\

% 6 ellipticals: 1316+4472+4697+M84+N1553+N5846
%early type & Table \ref{tab:xray} & -- & -- & $1.68\pm0.22$ &
%$5.0\pm 0.7$ & $4.7^{+1.2}_{-0.9}$ \\

% 6 ellipticals: 1316+4472+4697+M84+N1553+N5846, Lxmin=2e38
%early type & b & -- & -- & $1.94^{+0.48}_{-0.52}$ &
%$5.1^{+1.9}_{-0.7}$ & $4.7^{+1.8}_{-0.9}$ \\

Cen A & $1\cdot 10^{37}$ & $1.0^f$ & $0.2^f$  & $1.96\pm0.23$ &
$5.0^f$ & $5.0^f$ \\

late type & $5\cdot10^{35}-10^{37}$ (a) & $0.98\pm 0.11$ & $0.17^{+0.07}_{-0.03}$ & 
$1.90^{+0.22}_{-0.15}$ &
$5.0^f$ & $5.0^f$ \\

\hline
all & $5\cdot10^{35}-3\cdot 10^{38}$ (b) &  $1.0\pm0.13$ & $0.19^{+0.06}_{-0.04}$ & $1.86\pm 0.12$ & 
$5.0\pm 0.7$ & $4.8\pm 1.1$\\
\hline
\end{tabular}
\flushleft
early type: NGC1316, NGC4472, NGC4697 and M84; 
late type: M81, M31, M101 and Milky Way;\\
all -- all galaxies from Table \ref{tab:sample}, except NGC1553
(section \ref{sec:individual}); \\
$L_{\rm min}$ -- minimum luminosity for ML fit; 
$\alpha_1, \alpha_2, \alpha_3$ -- differential slopes;
$L_{b,1}, L_{b,2}$ -- first and second break in units of $10^{38}$
erg/s (eq.\ref{eq:uxlf});\\
a -- $L_{\rm min}$: $5\cdot 10^{35}$ erg/s (M31 and MW), $5\cdot
10^{36}$ erg/s (M101), $1\cdot 10^{37}$ erg/s (M81);\\
b -- $L_{\rm min}$: the same as in (a) for late type galaxies, see
Table \ref{tab:xray} for other galaxies; \\
f -- parameter was fixed at the quoted value. 

\label{tab:xlf}
\end{table*}

\begin{figure}
\resizebox{\hsize}{!}{\includegraphics{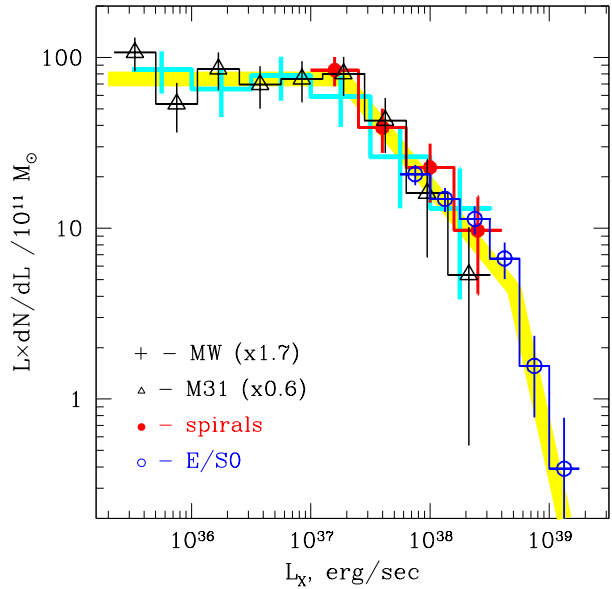}}
\caption{Differential luminosity function in a broad luminosity
range. The combined data for early and late type
galaxies (the same as in Fig.\ref{fig:xlf_comb}) plotted
along with XLFs of the central bulge of M31 and the Milky Way LMXBs.
The XLFs of the Milky Way and M31 are multiplied by various factors  
(indicated in the parenthesis) to match each other in
normalization. Note, that XLFs of the M31 and the Milky Way were 
obtained from Chandra and RXTE/ASM data respectively and are subject
to different systematic effects. Since the XLF is plotted in units
of $LdN/dL$, the flat part below $\log(L_X)\sim 37$ corresponds
to a differential slope of $\sim 1$. The thick grey line shows
the template XLF defined in  subsection \ref{sec:uxlf}.
}
\label{fig:xlf_universal}
\end{figure}

\subsection{Low luminosity end of XLFs}

Study of the X-ray luminosity function at low luminosities is
complicated by the fact that for only two galaxies of late type -- M31 
and the Milky Way, does the data extend sufficiently below $\sim
10^{37}$ erg/s. This range of luminosities is not covered at all
in the observations of early type galaxies. Therefore any 
conclusion regarding the behavior of the XLF at low luminosity relies
on a certain degree of extrapolation of the observational data.

We show in Fig.\ref{fig:xlf_universal} the XLFs of the galaxies from
our sample in the differential form in a broad luminosity range from 
$\sim 5\cdot 10^{35}$ erg/s to $\sim 10^{39}$ erg/s.
As can be seen from the figure, there is  significant flattening
of the luminosity function below $L_X\sim 10^{37}$ erg/s, observed in
both M31 and the Milky Way. We emphasize that the luminosity
functions for these two galaxies were constructed from entirely
different datasets and therefore are subject to different systematic
effects. The M31 luminosity function was obtained by Chandra
\citep{m31} whereas the Milky Way luminosity function is based on the
RXTE ASM data \citep{grimm1}.  For both galaxies the completeness
limit is substantially below $\sim 10^{36}$ erg/s.

\subsection{Average XLF of LMXBs}
\label{sec:uxlf}

In the following we assume that the luminosity functions of M31 and
the Milky Way at the low luminosities are representative of the
behavior 
common for all morphological types. As stressed  in the previous
subsection, this assumption involves a certain degree of
extrapolation. NGC1553, showing somewhat peculiar behavior both in
the $L_X$ growth curves and in the X-ray luminosity distribution, is
excluded from the analysis presented below in this subsection and will
be discussed separately in the section \ref{sec:individual}.

The data plotted in Fig.\ref{fig:xlf_universal} indicate a rather
complex shape of the combined luminosity function of LMXBs. 
To describe it  quantitatively we define a template XLF as a power law
with two breaks:
\begin{eqnarray}
\frac{dN}{dL_{38}}=\left\{ \begin{array}{ll}
\renewcommand{\arraystretch}{3}
K_1 \left(L_{38}/L_{b,1}\right) ^{-\alpha_1}	
			& \mbox{\hspace{0.9cm} $L_{38}<L_{b,1}$}\\
K_2 \left(L_{38}/L_{b,2}\right)^{-\alpha_2}	
			& \mbox{$L_{b,1}<L_{38}<L_{b,2}$}\\
K_3 \left(L_{38}/L_{cut}\right)^{-\alpha_3}	
			& \mbox{$L_{b,2}<L_{38}<L_{cut}$}\\
0			& \mbox{\hspace{0.9cm} $L_{38}>L_{cut}$}\\
\end{array}
\right.
\label{eq:uxlf}
\end{eqnarray}
where $L_{38}=L_X/10^{38}$ erg/s and normalizations $K_{1,2,3}$ are
related by
\begin{eqnarray}
K_2=K_1 \left(L_{b,1}/L_{b,2}\right)^{\alpha_2}\nonumber \\
K_3=K_2 \left(L_{b,2}/L_{cut}\right)^{\alpha_3}\nonumber 
\end{eqnarray}
The value of the high luminosity cut-off was fixed at
$L_{cut}=500$. Due to the steep slope of the luminosity
function above $L_{b,2}$, the results are insensitive to the actual
value of $L_{cut}$. 

\begin{figure}
\hbox{
\resizebox{\hsize}{!}{\includegraphics{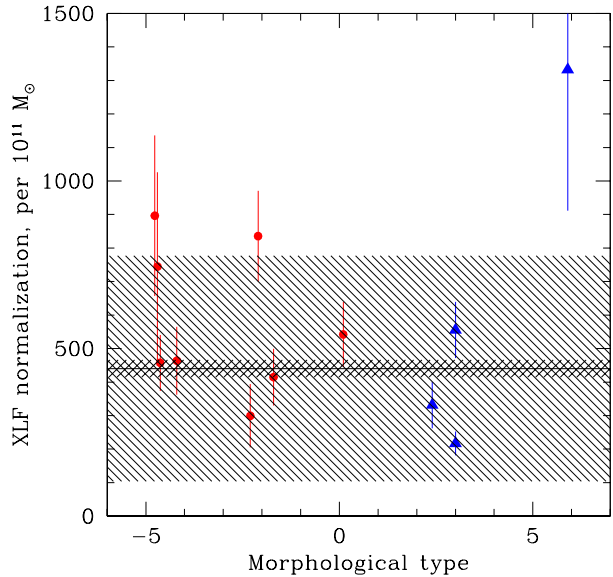}}
}
\caption{Normalization of XLF vs morphological type.  Normalization is
expressed in the same units as in eq.(\ref{eq:xlfnorm}). The solid
line and narrower shaded area show the average and its formal $1\sigma$
error from eq.(\ref{eq:xlfnorm}). The bigger shaded area shows the rms
of the  points with respect to the average. Note that this is
essentially the 
same as the $N_X/M_*$ plot shown in Fig.\ref{fig:x2mass_vs_type}. The
only  
difference is that lower values of $L_{min}$ were adopted  for MW, M31
and M101, whose points consequently have smaller error bars.
}
\label{fig:xlfnorm_vs_type}
\end{figure}

In order to determine the XLF parameters, we do Maximum Likelihood
fits to 
the unbinned data for all galaxies from our sample, leaving the relative
normalizations free. The best fit parameters are listed in the  
bottom line of Table \ref{tab:xlf} and the luminosity function is plotted
as a thick grey line in Fig.\ref{fig:xlf_universal}. 
The normalizations $K_1$ for individual galaxies are shown as a
function of the morphological type in Fig.\ref{fig:xlfnorm_vs_type}. The
best fit value of the average normalization is: 
\begin{eqnarray}
K_1=440.4\pm25.9 {\rm ~ per~ 10^{11}~ M_\odot} 
\label{eq:xlfnorm}
\end{eqnarray}
With this normalization, the cumulative number of sources computed
integrating eq.(\ref{eq:uxlf})
\begin{eqnarray}
N_X(>L)=\int_{L}^{L_{cut}} \frac{dN}{dL_{38}} dL_{38}
\end{eqnarray}
gives the number of sources per $10^{11}$ M$_\odot$. In cumulative form
this normalization corresponds to the following values of the total
number of sources with luminosity exceeding $10^{37}$ erg/s and their
collective luminosity: 
\begin{eqnarray}
N_X(>10^{37} {\rm ~ erg/s})= 142.9\pm 8.4 
{\rm ~sources ~ per~ 10^{11}~ M_\odot}~~~\\
L_X(>10^{37} {\rm ~ erg/s})= (8.0\pm 0.5)
{\rm ~10^{39}~ erg/s ~ per~ 10^{11}~ M_\odot}  \nonumber
\label{eq:x_nir_from_uxlf}
\end{eqnarray}

To check for a possible dependence of the shape of the luminosity function
on the morphological type, we also performed M-L fits to early and late
type galaxies and Cen A separately. 
The best fit parameters are given in Table \ref{tab:xlf}.
As the late type galaxies data are insensitive to the XLF behavior
above $\ga {\rm few}\cdot 10^{38}$ erg/s, we fix the second break and
slope at the values derived from all data combined
together. Similarly, for the Cen A galaxy, we fix the first slope and
break. As seen from Table 
\ref{tab:xlf}, the early type galaxies show a somewhat flatter slope
$\alpha_2$ than late type galaxies or Cen A. The statistical
significance of this 
difference is $\sim 1\sigma$. Increasing $L_{\rm min}$ by a
factor of two, from $10^{38}$ erg/s to $2\cdot 10^{38}$ erg/s, reduces
further this difference, indicating that it might be a result of
residual incompleteness effects in the combined XLF of early type
galaxies \citep[c.f.][]{ngc1316}.

\begin{figure*}
\hbox{
\resizebox{0.5\hsize}{!}{\includegraphics{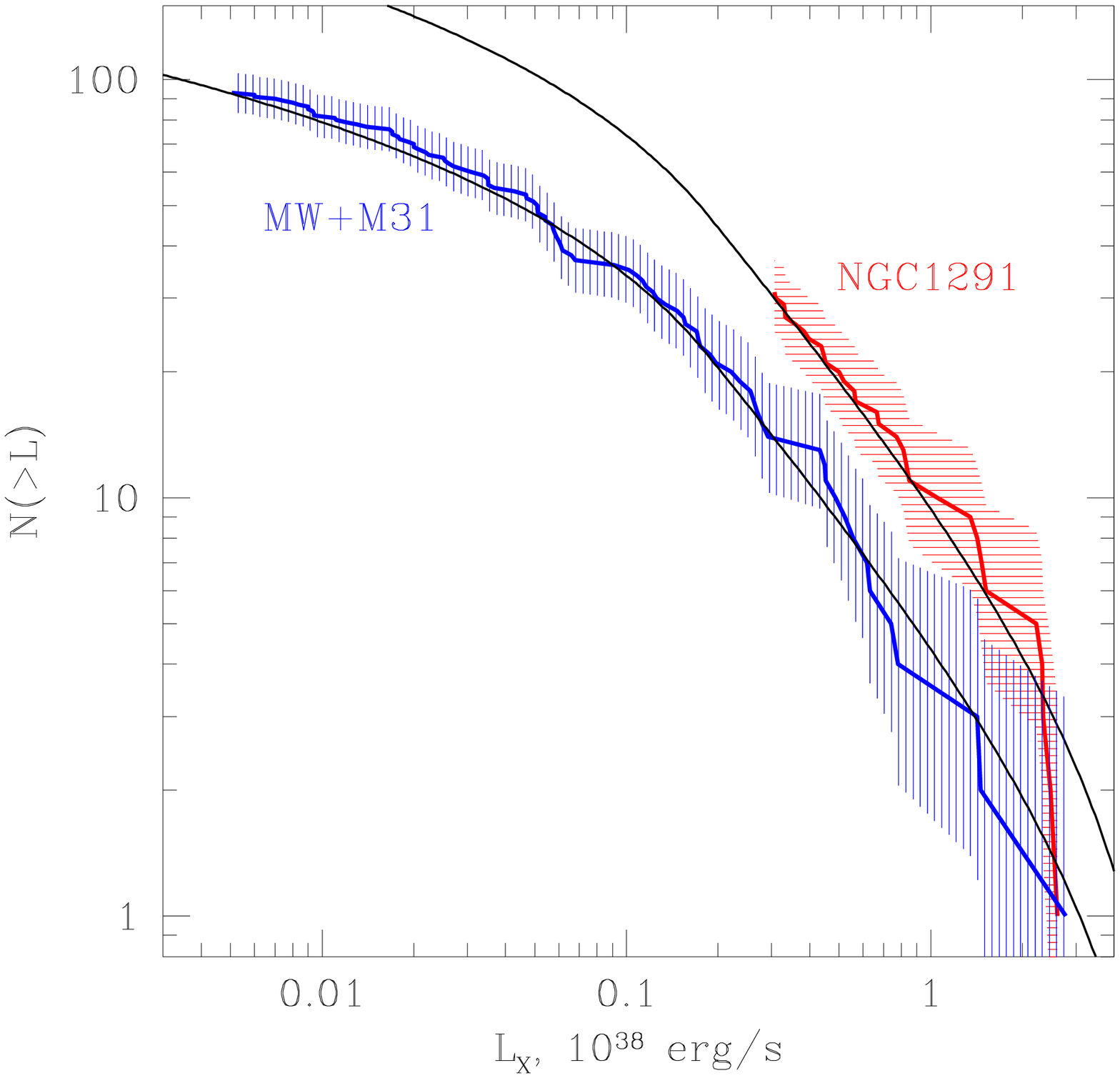}}
\resizebox{0.5\hsize}{!}{\includegraphics{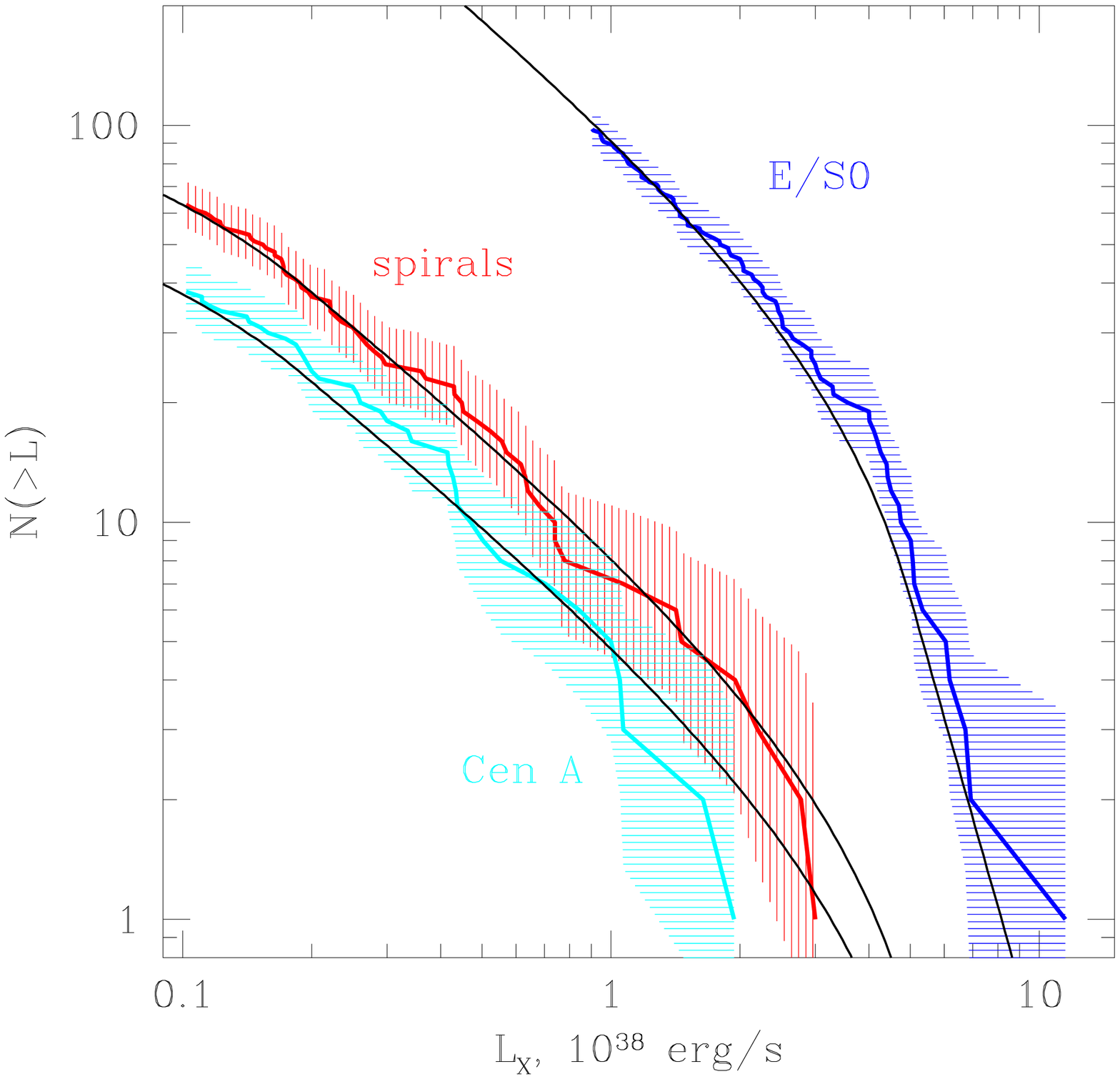}}
}
\caption{Comparison of the template XLF with the data. The shaded
areas around the observed XLFs show the uncertainty of the Poisson
distribution calculated using the prescription of
\citet{gehrels86}. The 
solid lines show the template XLF with the parameters defined in 
subsection \ref{sec:uxlf}. To facilitate comparison of the shape, the
normalization of the template XLF was adjusted to match the observed
number of sources at the low luminosity end of each curve,
irrespective to the stellar mass. 
}
\label{fig:xlf_comb_uxlf}
\end{figure*}

\begin{table}
\renewcommand{\arraystretch}{1.2}
\caption{The Kolmogorov-Smirnov test results for the template XLF}
\begin{tabular}{|l|c|c|c|c|}
\hline
sample & $L_{\rm min}$ & $N_{\rm src}$ $^a$ & K-S prob.$^b$ \\
\hline
% 4 ellipticals: 1316+4472+4697+M84
early type & $1\cdot 10^{38}$ & 89 & 0.65\\
NGC 5846 & $3\cdot 10^{38}$ & 7 & 0.89\\
NGC1553  & $2\cdot 10^{38}$ & 22 & 0.16\\
Cen A & $1\cdot 10^{37}$ & 38 & 0.92\\
NGC1291 & $3\cdot 10^{37}$ & 31 & 0.94\\
late type & $1\cdot 10^{37}$ & 63 & 0.99\\
M31 + Milky Way & $5\cdot 10^{35}$ & 95 & 0.74\\
\hline
\end{tabular}
\flushleft
The template XLF is defined by eq.(\ref{eq:uxlf}) with the parameters
from the bottom line of Table \ref{tab:xlf}.
early type: NGC1316, NGC4472, NGC4697 and M84; 
late type: M81, M31, M101 and Milky Way;\\
a -- number of sources;
b -- probability that observed XLF deviates from the model due to
statistical fluctuations.
\label{tab:ks_test}
\end{table}

The results of the Kolmogorov-Smirnov test of the observed luminosity
functions against the average XLF  are presented in Table
\ref{tab:ks_test}. The good agreement between the observed and the
average luminosity functions is further demonstrated in 
Fig.\ref{fig:xlf_comb_uxlf}, which can be regarded as an approximate 
graphical representation of the Kolmogorov-Smirnov test. 

\begin{figure}
\resizebox{\hsize}{!}{\includegraphics{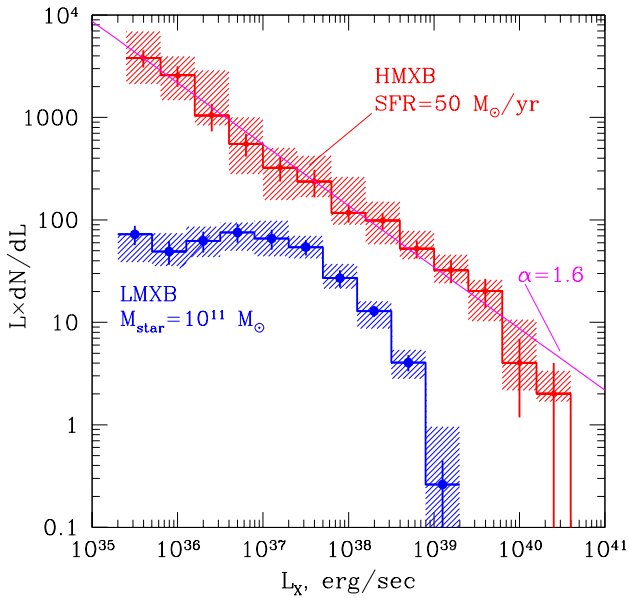}}
\caption{Average differential luminosity functions of high and low
mass X-ray binaries. The HMXB XLF was adopted from \citet{grimm2} and
scaled to the SFR=50 M$_{\odot}$/year. The average LMXB XLF was
constructed as described in section \ref{sec:uxlf}. The shaded
areas around the luminosity functions 
illustrate the amplitude of systematic errors (90\% confidence level)
due to uncertainties in the source distance (assuming 20\% relative 
uncertainty), mass-to-light ratios (30\%) and star formation rates
(30\%).   
}
\label{fig:xrb_xlf}
\end{figure}

Finally, we construct the average LMXB luminosity function combining
the data for all galaxies from our sample. We bin all the sources into 
logarithmically spaced bins and normalize the result by the sum of the
stellar masses of all galaxies contributing to the given or previous
luminosity bins. The advantage of this approach is that it allows us
to 
combine the data with different completeness limits. The disadvantage,
however, is that, due to significantly different luminosity ranges of
the individual luminosity functions,  uncertainties in the galaxy
distance, stellar mass  and intrinsic variations of
the shape and normalization of the individual luminosity distributions
may lead to the appearance of artificial features in the combined
luminosity function.  With that in mind, we plot the combined
luminosity function in Fig.\ref{fig:xrb_xlf}. 

In order to assess the amplitude of possible systematic 
uncertainties we performed Monte-Carlo simulations. In each run  the
distance of each galaxy and its mass-to-light ratio were
replaced by random numbers drawn from a Gaussian distribution. The
mean of the distribution was equal to the default value of the
parameter and its $1\sigma$ width was set to 20\% and 30\% of the mean
for the distance and the mass-to-light ratio respectively. If a
negative number was drawn, it was rejected and the run repeated. For
the Milky Way, the distance to each source was varied independently,
assuming 20\% 
relative uncertainty. For each bin of the luminosity function, the
$1\sigma$ systematic error was estimated as the rms of the values
obtained in individual runs. The resulting 90\% uncertainty regions
are shown as the shaded areas in Fig.\ref{fig:xrb_xlf}.

\section{X-ray -- stellar mass relation}
\label{sec:x_nir_ratios}

The average XLF of low mass X-ray binaries can be used to
transform the X-ray luminosity and number of sources, computed with
different completeness limits, to the same luminosity range. We chose  
the range $L_X>10^{37}$ erg/s in order to avoid  the luminosity 
region where XLF shape is based on the late type galaxy data
only. Such a choice helps avoid too large an extrapolation for early 
type galaxies, typically having higher completeness limits, and, on
the other hand, allows a sufficient number of sources for late type
galaxies, having smaller mass.  
The correction was done in the following way. Using the   
number of sources detected above the completeness limit for the given
galaxy, we first determine the normalization of XLF. Then we use this
normalization to calculate the expected number of sources and their
total luminosity between the completeness limit and $10^{37}$
erg/s. These quantities are listed in Table \ref{tab:xray} and are
added to the observed number of sources and to the sum of their
luminosities.    

For the derived shape of the average luminosity function, 
the luminosity correction does not exceed a factor $\la 2$ for 
most of the galaxies, except for NGC5846, in which case it equals 
$\approx 3.5$. The correction factor is significantly larger for the
total number of 
sources, generally up to a factor of $\approx 15-20$ ($\approx 40$ for
NGC5846), making the estimates of the total number of sources
significantly less robust. Note, that for the derived shape of the
LMXB XLF, the luminosity of the sources  above $10^{37}$
erg/s accurately represents the total X-ray luminosity
of the compact sources -- the correction factor to the total
luminosity is $\approx 1.1$ (assuming that the luminosity function
does not steepen at lower luminosities, below $10^{35}-10^{36}$
erg/s).

\begin{figure*}
\hbox{
\resizebox{0.5\hsize}{!}{\includegraphics{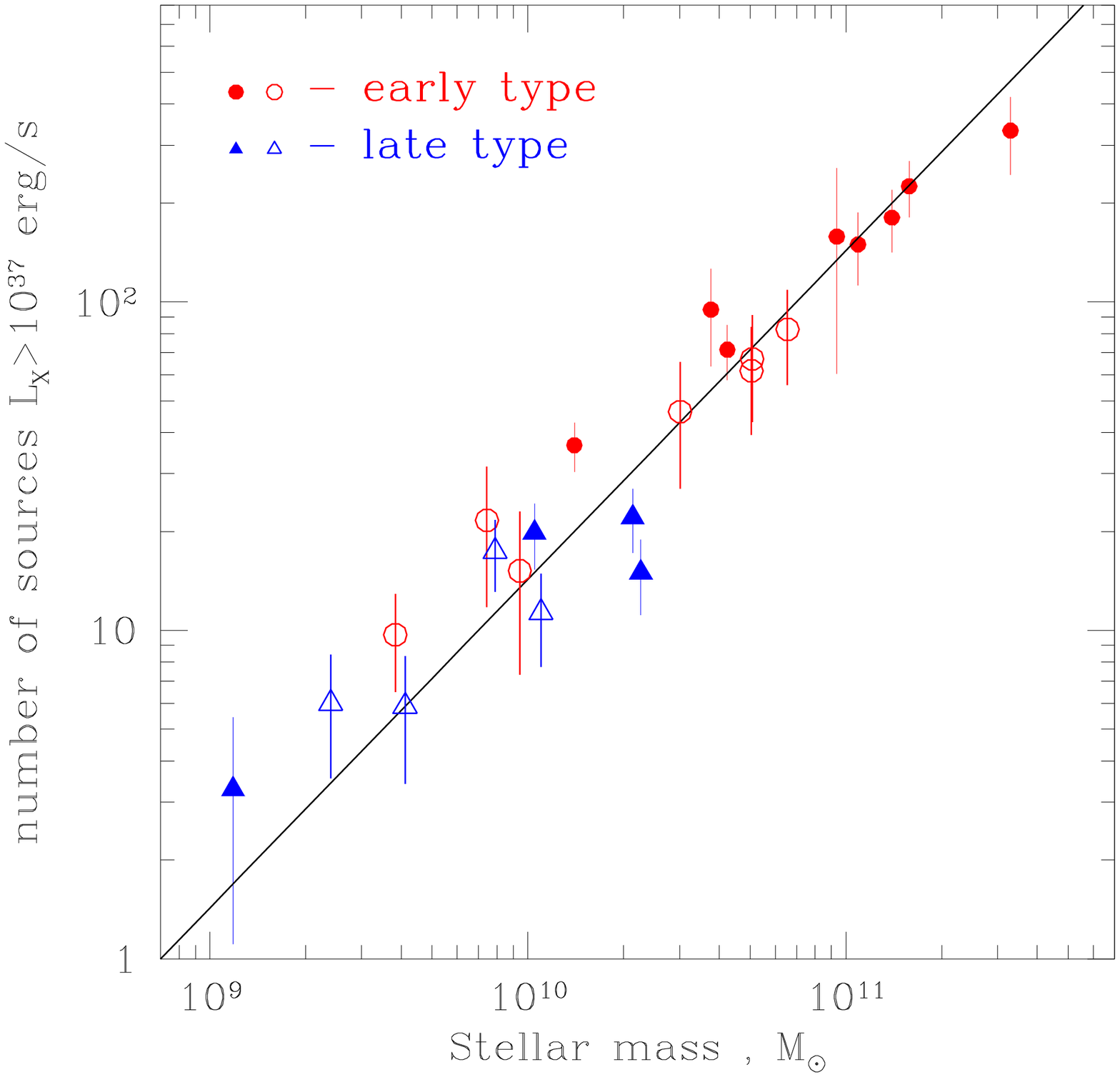}}
\resizebox{0.5\hsize}{!}{\includegraphics{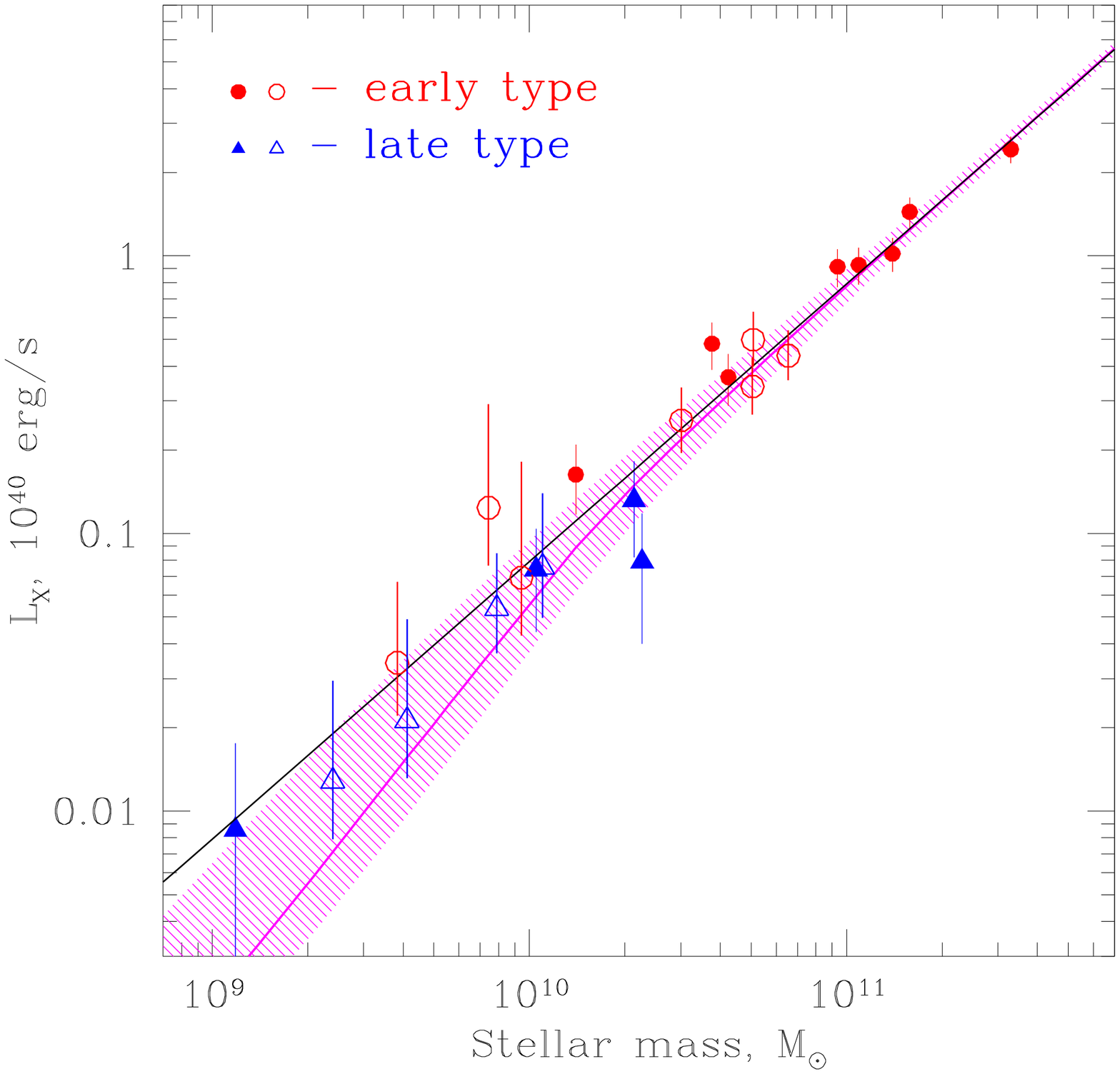}}
}
\caption{Number of sources with luminosity $L_X>10^{37}$ erg/s and
their collective X-ray luminosity vs stellar mass. The data for
galaxies from our sample (Table \ref{tab:xray}) are shown by
solid circles and triangles for early and late type galaxies
respectively. To further facilitate comparison of early and late type
galaxies, several galaxies were divided in to annuli of smaller
size (i.e. containing smaller mass). These are shown by open symbols.
Solid lines are linear relations given by eq.(\ref{eq:x_nir_from_uxlf}).
On the right panel, the thick solid curve and the shaded area around
it represent the relation between the stellar mass and the most
probable value of the total luminosity and its 67\% intrinsic
uncertainty, obtained from the average XLF derived in subsection 
\ref{sec:uxlf}.  
}
\label{fig:x_mass}
\end{figure*}

\begin{figure*}
\hbox{
\resizebox{0.5\hsize}{!}{\includegraphics{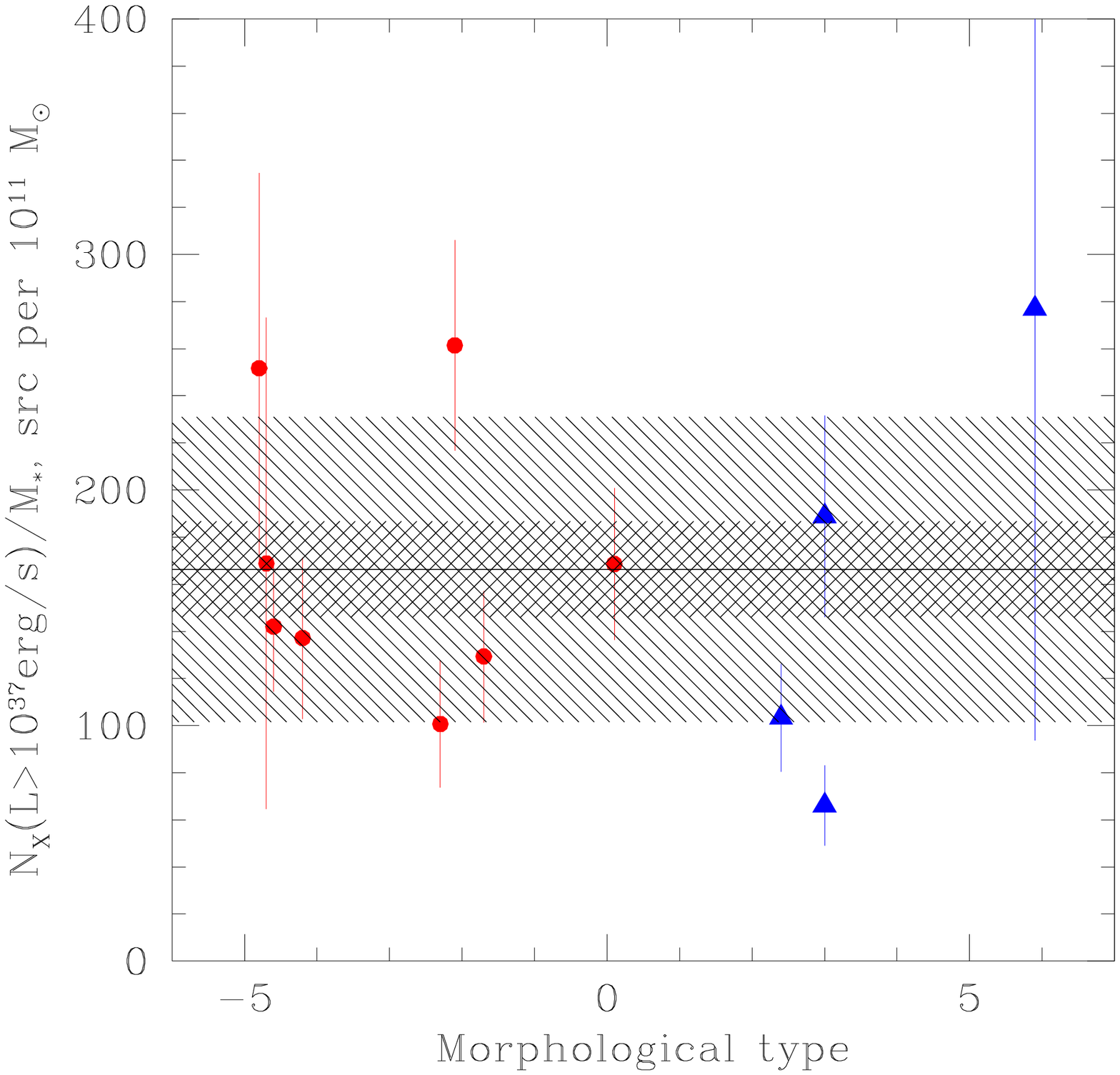}}
\resizebox{0.5\hsize}{!}{\includegraphics{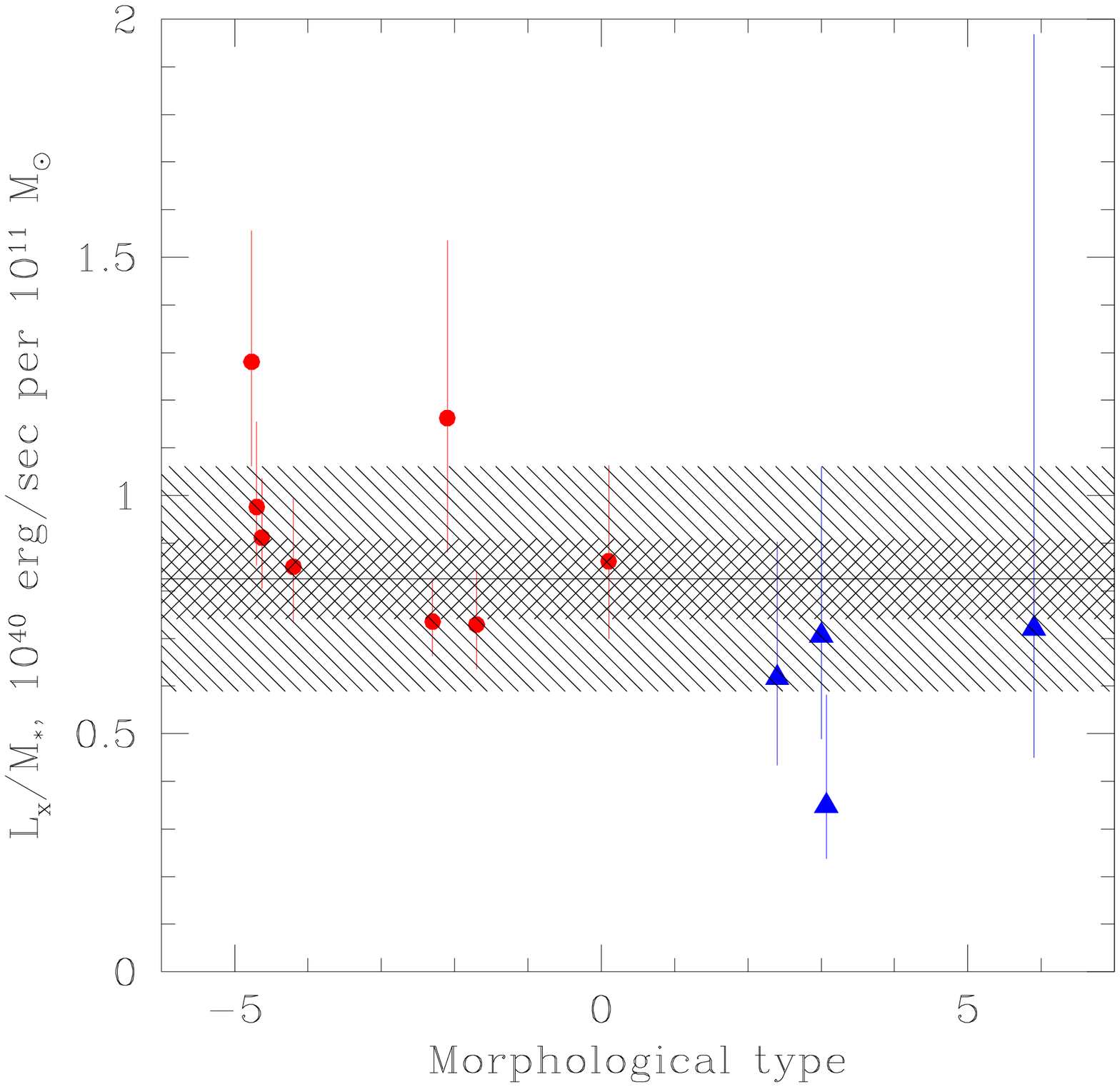}}
}
\caption{$L_X/M_*$ and $N_X/M_*$ ratios vs morphological
type. The solid lines and narrower shaded areas show average and its
formal $1\sigma$ error from Table \ref{tab:x2mass}. The error was
computed using statistical errors of the individual data points. The
bigger shaded areas show rms of the points with respect to the
average. $L_X/M_*$ are not corrected for the effects of statistics.
}
\label{fig:x2mass_vs_type}
\end{figure*}

\begin{table*}
\renewcommand{\arraystretch}{1.2}
\caption{Average X/M$_*$ and X/$L_{\rm NIR}$ ratios}
\begin{tabular}{|l|c|c|c|c|c|}
\hline
sample & $N_X/L_{\rm NIR}$ & $L_X/L_{\rm NIR}$ (a) & $N_X/M_*$ &
$L_X/M_*$ (a) & $L_X/M_*$ (b)\\ 
\hline
early type& $135.0\pm 15.9$ & $0.75\pm0.05$ & $170.0\pm19.5$ &
$0.94\pm0.07$ & $0.98\pm0.07$\\
late type & $81.2\pm 19.6$ & $0.33\pm0.10$ & $158.8\pm47.6$ &
$0.60\pm0.21$ & $1.04\pm0.22$\\
\hline
all & $117.0\pm 12.4$ & $0.61\pm0.05$ & $166.3\pm20.5$ 
& $0.83\pm0.08$ & $1.00\pm0.08$\\ 
\hline
\end{tabular}
\\
%\flushleft
a -- observed; b -- corrected for the effects of statistics;\\
$N_X/M_*(L_{\rm NIR})$ -- in units of src per $10^{11}$ M$_\odot$ 
(L$_\odot$);
$L_X/M_*(L_{\rm NIR})$ -- in units of $10^{40}$ erg/s per $10^{11}$
M$_\odot$(L$_\odot$);\\
$N_X$ and $L_X$ -- number and total luminosity of X-ray sources
brighter than $10^{37}$ erg/s  
\label{tab:x2mass}
\end{table*}

The number of sources with luminosity above $10^{37}$ erg/s and their
total luminosity is plotted as a function of the stellar mass in 
Fig.\ref{fig:x_mass}. Early and late type galaxies are plotted by
different symbols. To further illustrate that there is no 
significant difference between early and late galaxies we divide the
early type galaxies into smaller annuli, whose mass is comparable to
the typical masses  of bulges in spiral galaxies. These are plotted in 
Fig.\ref{fig:x_mass} as open symbols.

We fit the $N_X$--mass and $L_X$--mass relations with a power law 
$X=A M_*^{\alpha}$ using all galaxies from our sample (Table
\ref{tab:xray}, solid symbols in Fig.\ref{fig:x_mass}). We use
$\chi^2$ minimization, which is strictly applicable in the limit of
a large number of sources per galaxy. This is correct for most of 
the galaxies from our sample (Table \ref{tab:xray}), therefore we do
not expect any significant bias in the best-fit slope. 
The best fit values of the slope are consistent with unity: 
$\alpha_{LX}=1.01\pm 0.06$ and $\alpha_{NX}=1.00\pm 0.07$ for the
luminosity and the number of sources respectively. We
therefore fix the slopes at $\alpha=1$ and consider average $N_X/M_*$ and
$L_X/M_*$ ratios. To estimate these we use the unweighted average
$\left<X/M\right>=\sum (X_i/M_i)/n$. 
The results are given in Table \ref{tab:x2mass}
separately for early and late type galaxies and for all galaxies. Note
that the average ratios given in Table \ref{tab:x2mass} are not identical
to those inferred from eq.(\ref{eq:x_nir_from_uxlf}), obtained
integrating the average XLF, especially for $N_X/M_*$
ratio. This discrepancy is due to difference between the expectation
values of  $\left<X/M\right>=\sum (X_i/M_i)/n$ and  
$\left<X\right>/\left<M\right>=\sum X_i/ \sum M_i$ in the
presence of systematic errors. The systematic errors are present in
both $X_i$, due to transformation to the same $L_{min}$ based on the
average XLF, and in $M_i$, as they are derived from near-infrared
luminosity using an average color-based correction to mass-to-light
ratio. The fact that we are dealing with rather small samples (e.g. 4
galaxies of late morphological type) also plays a role. 
For the same reason, yet different values would be obtained if one used 
weighted estimates for  $\left<X/M\right>$, e.g. $\chi^2$ minimization
technique.  These factors should be taken into consideration when
comparing the  $X/M_*$ ratios.

We show in Fig.\ref{fig:x2mass_vs_type} the $N_X/M_*$ and $L_X/M_*$
ratios for individual galaxies as a function of the morphological
type. As can be seen, there is statistically significant dispersion,
of the order of $\sim 25\%$ and $\sim 40\%$ from the mean value for
$L_X/M_*$ and $N_X/M_*$ 
respectively. A larger dispersion for $N_X/M_*$ can probably
be explained by the larger correction factors for the number of
sources 
and, respectively, the stronger dependence of the correction factors
on the 
details of the individual XLFs.

\subsection{Dependence of the X/{\boldmath ${ M_*}$} ratio on the
morphological type ?} 

Fig.\ref{fig:x2mass_vs_type} and  Table \ref{tab:x2mass} indicate a
possible dependence of the $L_X/M_*$ ratio on the morphological type.  
This can be in part caused by the statistical effects
(section \ref{sec:stat}). Indeed, the bulges of spiral galaxies
have smaller mass than early type galaxies, consequently, smaller 
numbers of luminous LMXB sources and, therefore, are more subject to
the effects of statistics.
That these effects can contribute to the 
observed trend is demonstrated by the last column in Table
\ref{tab:x2mass} where average $L_X/M_*$ ratios, corrected for the
statistical effects, are listed. An additional argument in favor of
this explanation is that a similar trend is not observed for the
$N_X/M_*$, 
ratio (Fig.\ref{fig:x2mass_vs_type}, Table \ref{tab:x2mass}). 
On the other hand, the $N_X/M_*$ ratios are more sensitive to the
details of the individual luminosity functions and therefore are less  
robust.

Another factor might be insufficiently accurate calibration of the
near-infrared mass-to-light ratios. As discussed in section
\ref{sec:nir_mass}, the color-based correction was obtained under
the assumption of the universal initial mass function and might fail,
if the morphological type dependent variations in the IMF are present
\citep{m2l}. Such variations cannot be excluded a priori, given the
broad range of morphological types in our sample.  
Yet another source of uncertainty is that in computing the $M_*/L_K$
ratios integrated colors of the galaxies were used. As the galaxies do
show color gradients \citep{col_grad, leda}, whose amplitude is type
dependent and is largest for late types, $2\la T \la 5$, this
introduces additional  uncertainty in the relative values of the
$M_*/L_K$ ratio derived for  early and late type galaxies.

To conclude, due to a number of uncertain factors involved, 
the possibility of a weak dependence of the X-ray/mass ratios  on
the morphological type can be neither rejected nor confirmed with
sufficient confidence.  However, if present, it does not exceed a
factor of $\sim 1.5-2$.  

\begin{figure}
\resizebox{\hsize}{!}{\includegraphics{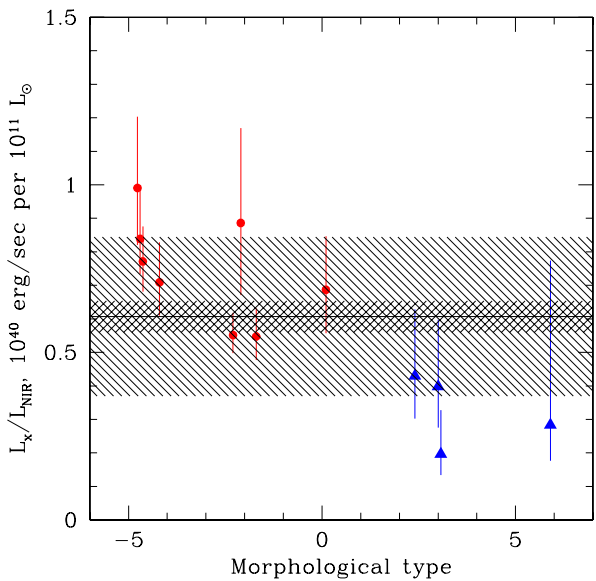}}
\caption{$L_X/L_{\rm NIR}$ ratios vs morphological type. The solid
line and narrower shaded area show the average and its 
formal $1\sigma$ error. The error was computed propagating the errors
of the individual data points. The bigger shaded area shows the rms of
the points with respect to the average. The $L_X/L_{NIR}$ ratios are
not corrected for the effects of statistics.}
\label{fig:lx2lnir_vs_type}
\end{figure}

Interestingly,  the X-ray/$L_{\rm NIR}$ ratios do show clear
dependence on the morphological type
(Fig.\ref{fig:lx2lnir_vs_type}). This is 
obviously caused by dependence of the near-infrared mass-to-light
ratio on the morphological type of the galaxy. The correlation of
the X-ray to optical light ratios with the morphological type should be
significantly more pronounces at shorter wavelength, e.g. in the
B-band, due to significantly stronger type dependent variations  of
the mass-to-light ratio in the optical band \citep{m2l}.

\section{Discussion}
\label{sec:discussion}

\subsection{Is the average LMXB XLF universal?}

Obviously, the shape of the luminosity function is defined by a number
of factor which can vary from galaxy to galaxy. 
Of these effects, most important should be the effects of binary
evolution.  Despite of a number of population synthesis models
developed, 
there is no a clear prediction about the evolution of the luminosity
distribution of X-ray binaries with time. 
Based on simple arguments, one might expect that these
effects are mostly pronounced at the high luminosity end of the
XLF. Indeed, a luminosity of $10^{38}$ erg/s requires a mass accretion
rate of $\sim 10^{-8}$ M$_{\odot}$/yr, which can be sustained by a low
mass star for less than $\la 10^8$ yrs \citep{pods2002}.  This value
is significantly shorter than the life time of a galaxy. It is yet
shorter for the most luminous LMXB systems with $L_X\sim10^{39}$ 
erg/s. In order to have the presently observed shape of
the luminosity distribution, with a moderate fraction of sources with
$L_X\ga 10^{37-38}$ erg/s, in the stellar systems of the age of 
$\sim 10^{10}$ yrs, a continuous replenishment of the high luminosity
sources is required.\footnote{ 
Another possibility, considered by \citet{piro02} is that   the
brightest LMXBs are transient sources with sufficiently low duty
cycle,  $\sim 10^{-2} - 10^{-3}$.}   
Such replenishment can be maintained for example due to binary systems
with initially less massive companion stars, reaching the X-ray active 
phase at later times.  
Clearly, evolution of the luminosity function with time must be
present and it should be more pronounced at the high luminosity end of
XLF. Consequently, no universal luminosity function in the strict
sense should be expected, from which the source populations observed in
different galaxies are drawn.

However, the results of the present study suggest that there are no
significant variations of the shape of the luminosity distribution of
low mass X-ray binaries in majority of nearby galaxies -- they are
broadly consistent with having the same shape. This fact might be
somewhat  surprising in the view of the arguments presented above. An
obvious explanation  is that the nearby  galaxies constituting our
sample have similar age of the stellar population, with sufficiently
long time  passed since the last starburst events.
On the other hand, the  departures from the average XLF possibly
observed in NGC1553 and NGC1291 (section \ref{sec:individual}), if
real, might be a manifestation of the XLF dependence on the age and
star formation history of the host galaxy.

In interpreting the absence of pronounced systematic variations of
the shape of the luminosity distribution, one should take into
consideration the sensitivity limitations of the present analysis.   
Indeed, in order to minimize contamination by the background sources
and interference of the CCD boundaries, we analyzed the central parts
of the galaxies, within $\sim 1.5-2$  effective radii for the early type 
galaxies, and inner bulges for spirals. This resulted in a rather
limited number of sources in our sample, $\sim 350$  in
total, and, consequently, to a limited sensitivity to possible XLF
variations. The amplitude of the statistical uncertainties is
illustrated by the shaded area in Fig.\ref{fig:xlf_comb_uxlf}.  
With increased sample and detection sensitivity, more complex behavior 
and more subtle effects can be discovered. 
We should note, however, that the analyzed annuli encompass 
from $\sim 30\%$ to $\sim 70\%$ of the total near-infrared luminosity
(and stellar mass) of the galaxies. Typically, more than a half of the
remaining stellar mass is contained in the central parts 
of the galaxies, inside the inner boundary of the analyzed
annuli. On the other hand, for many galaxies the outer boundary was
sufficiently close to the maximum value allowed by CXB contribution,
especially for the luminosity growth curves.
Therefore a significant increase of the number of X-ray
binaries suitable for analysis would require an adequate improvement
in the sensitivity and an accurate treatment of the incompleteness 
effects in the centers of the galaxies.

Due to typically larger distances to the elliptical galaxies, the low
luminosity end of the XLF was studied using the data of only two late
type galaxies, M31 and the Milky Way. Generalization of these 
results on the galaxies of all morphological types involves certain
degree of extrapolation of the data. On the other hand, the old
stellar systems in the ellipticals and spirals are similar in many
respects and such an extrapolation might be justified.

\subsection{X-ray luminosity function of low and high mass X-ray
binaries} 

As was shown by \citet{grimm2}, in a very broad  range of
luminosities, $\log(L_X)\sim 35.5-40.5$, the luminosity function of
HMXBs is consistent, to the first approximation, with a single slope
power law  with the differential slope of  $\alpha_{\rm HMXB}\approx 1.6$. 
The average luminosity function of LMXBs, on the contrary, has a
complex shape. It is a power law with the differential slope of
$\approx 1$ at low luminosities, steepens significantly at
$\log(L_X)\sim 37.5$ and has a cut-off at $\log(L_X)\sim 39.0-39.5$
(Fig.\ref{fig:xrb_xlf}).  

Different shape of the luminosity functions of
the high and low mass binaries reflects, obviously, the difference in
the accretion regimes. The majority of high mass systems are wind
accretors. It has been shown by \citet{postnov}, that the expected
XLF slope is $\sim 1.5$, i.e. close to the observed value. Note, that
Postnov's derivation was based on the assumption, that the mass
distribution of the donor stars in HMXBs can be described by the
Salpeter IMF. The low mass systems, on the other hand, are
close binaries fed via Roche lobe overflow. The slope of the average
LMXB XLF at $\log(L_X)\ga 37$ is qualitatively similar to that obtained
by \citet{pfahl2003} from X-ray binaries population synthesis,
$\alpha\sim 1.5-2$. At lower
luminosities, however, the population synthesis model predicts
significantly more sources than actually observed.
As discussed by \citet{pfahl2003}, the number of low luminosity
sources is substantially reduced, if the irradiation of the donor
star by the X-ray emission from the primary is taken into account.
Indeed, their illustrative $\dot{M}$ distribution with simplified
account of the irradiation effect is significantly flatter below 
$\log(L_X)\sim 37$ and is qualitatively similar to the observed
average XLF of LMXBs.

\subsection{Diagnostics of on-going star formation}

The different  shape of XLFs of HMXBs and LMXBs can be used
to diagnose  on-going star formation in relatively nearby galaxies
($\sim 20-30$ Mpc for the Chandra angular resolution). Two
possibilities can be exploited. 

The cut-off in the HMXB luminosity function occurs at $\sim 10$ times
higher luminosity than in LMXBs (Fig.\ref{fig:xrb_xlf}). 
In our sample of $\sim 350$ sources in the old stellar systems with
total stellar mass of $\sim 10^{12}$ M$_{\sun}$ two brightest sources,
in NGC1553, have luminosity $\sim 3.3\cdot 10^{39}$ and $\sim 2\cdot
10^{39}$ erg/s (see sections \ref{sec:nir_x_gcurve} and
\ref{sec:ngc1553}). These two sources excluded, the luminosity of the
other sources does not exceed $\sim 10^{39}$ erg/s
(Fig.\ref{fig:xlf_cum}). The brightest high mass X-ray binaries, on
the other hand, have luminosities up to $\sim (2-3)\cdot 10^{40}$
erg/s. Detection of even one source with luminosity of the order of
$\sim 10^{40}$ erg/s or greater implies that the star  
formation process is taking place. From the number of such luminous
sources the star formation rate can be estimated, as it is directly 
proportional to the SFR \citep{grimm2}:
\begin{eqnarray}
\nonumber
N(L>10^{40}{\rm ~ erg/s})
\approx 1.2\times \frac{\rm SFR}{10{\rm ~M_{\odot}/yr}}~~~~
\\
\nonumber
N(L>5\cdot 10^{39}{\rm ~ erg/s})
\approx 2.9\times \frac{\rm SFR}{10{\rm ~M_{\odot}/yr}}
\end{eqnarray}
where SFR refers to the formation rate of stars more massive
than $\sim 5$M$_{\odot}$. This, obviously, can be used for sufficiently
high star formation rates, exceeding $\sim 10$ M$_{\sun}$/yr.  
%There is 2.9 sources per $10$ M$_{\odot}$/yr
%for $L>5\cdot 10^{39}$ erg/s. 

Non-detection of luminous sources, on the other hand, immediately
constrains the star formation rate, e.g. absence of sources with
$L_X>5\cdot 10^{39}$ or $L_X>10^{40}$ erg/s results in upper limit on the 
star formation rate of SFR$\la 8$ and SFR$\la 20$ M$_{\odot}$/yr
respectively (90\% confidence level).

At lower star formation rates one can use the fact, that the XLF
slopes at the low luminosity end are substantially different, $\approx
1.0$ and $\approx 1.6$ for LMXBs and HMXBs respectively
(Fig.\ref{fig:xrb_xlf}). Therefore detection of a population of
compact sources with sufficiently steep $\log(N)-\log(S)$ distribution
at $L_X\la 10^{37}$ erg/s might indicate on-going star
formation. Conversely, $dN/dS\propto L^{-1}$ 
distribution in the low
luminosity limit would indicate that the X-ray emission is dominated by
an old stellar population.

\subsection{Relative contributions of high and low mass X-ray binaries}
\label{sec:l2m_m2sfr}

Relative contributions of low and high mass X-ray binaries to the
observed population of compact X-ray sources is defined by the ratio
of the stellar mass to the star formation rate. Due to different
shapes of the luminosity functions (Fig.\ref{fig:xrb_xlf}) it also
depends on the considered luminosities range. 

Using the calibrations of $L_X$--SFR and $L_X-M_*$ relations
obtained by \citet{grimm2} and in the present paper, the ratio of the
total luminosities can be estimated:
\begin{eqnarray}
\frac{L_{\rm LMXB}}{L_{\rm HMXB}}\sim 0.13 \ 
\frac{M_*/{\rm SFR}}{10^{10}{\rm ~yrs}} 
\end{eqnarray}
In the low SFR limit, SFR$\la 4-5$ M$_{\odot}$/yr, the luminosity
of high mass X-ray binaries will be subject to effects of statistics
(section \ref{sec:stat}) and the ratio of the most probable values is
given by:  
\begin{eqnarray}
\frac{L_{\rm LMXB}}{L_{\rm HMXB}}\sim 0.34 \ 
\frac{M_*/{\rm SFR}}{10^{10}{\rm ~yrs}} \ 
{\rm SFR}^{-2/3},{\rm ~~ SFR}\la 4.2 {\rm M_{\odot}/yr}
\end{eqnarray}
The two above dependences intersect at SFR$\approx 4.2$ M$_{\odot}$/yr,

The ratio of the number of sources depends non-monotonically  on the
threshold luminosity $L_{\rm min}$. For the sources brighter than
$10^{38}$ erg/s it equals to:
\begin{eqnarray}
\frac{N_{\rm LMXB}}{N_{\rm HMXB}}\sim 0.35 \
\frac{M_*/{\rm SFR}}{10^{10}{\rm ~yrs}} 
\end{eqnarray}
The coefficient in the
above formula is $0.66, ~0.37, ~0.14$ for 
$L_{\rm min}=10^{37},~10^{36},~10^{35}$ erg/s respectively. Note, that
the last number is based on extrapolation of the  luminosity 
functions beyond the luminosity range  covered by data.

\begin{figure}
\resizebox{\hsize}{!}{\includegraphics{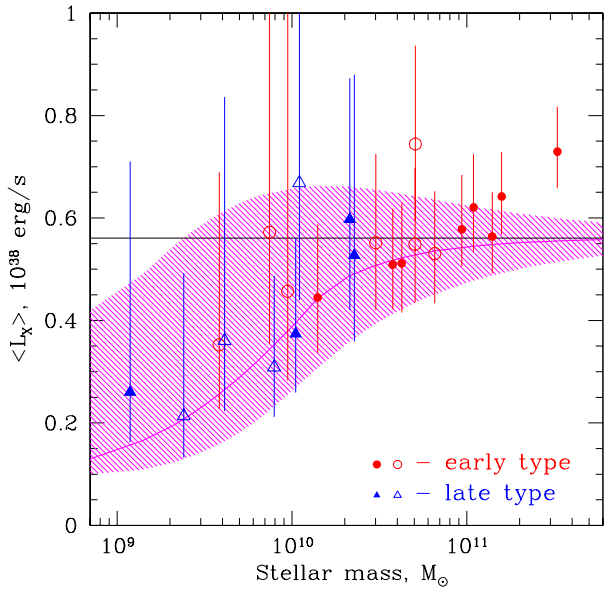}}
\caption{The average X-ray luminosity vs. stellar mass.
The average luminosity, $<L_X>=L_{\rm tot}/N_X$, was computed for the
sources with $L_X>10^{37}$ erg/s. The filled symbols are the data
from Table \ref{tab:xray}, open symbols -- smaller non-overlapping
annuli in the same galaxies. Early and late type galaxies are shown by
different symbols, as indicated in the legend. 
The thick solid line and the shaded area around it show the most
probable value (section \ref{sec:stat}) of the average luminosity and
its 67\% uncertainty predicted from the average luminosity
function. The horizontal line shows the value expected in the 
limit of large number of sources. The effects of statistics lead to
smaller value of the average source luminosity for less massive (parts
of) galaxies.  
}
\label{fig:lx2nx_vs_mass}
\end{figure}

\begin{figure*}
\centerline{\hbox{
\resizebox{0.5\hsize}{!}{\includegraphics{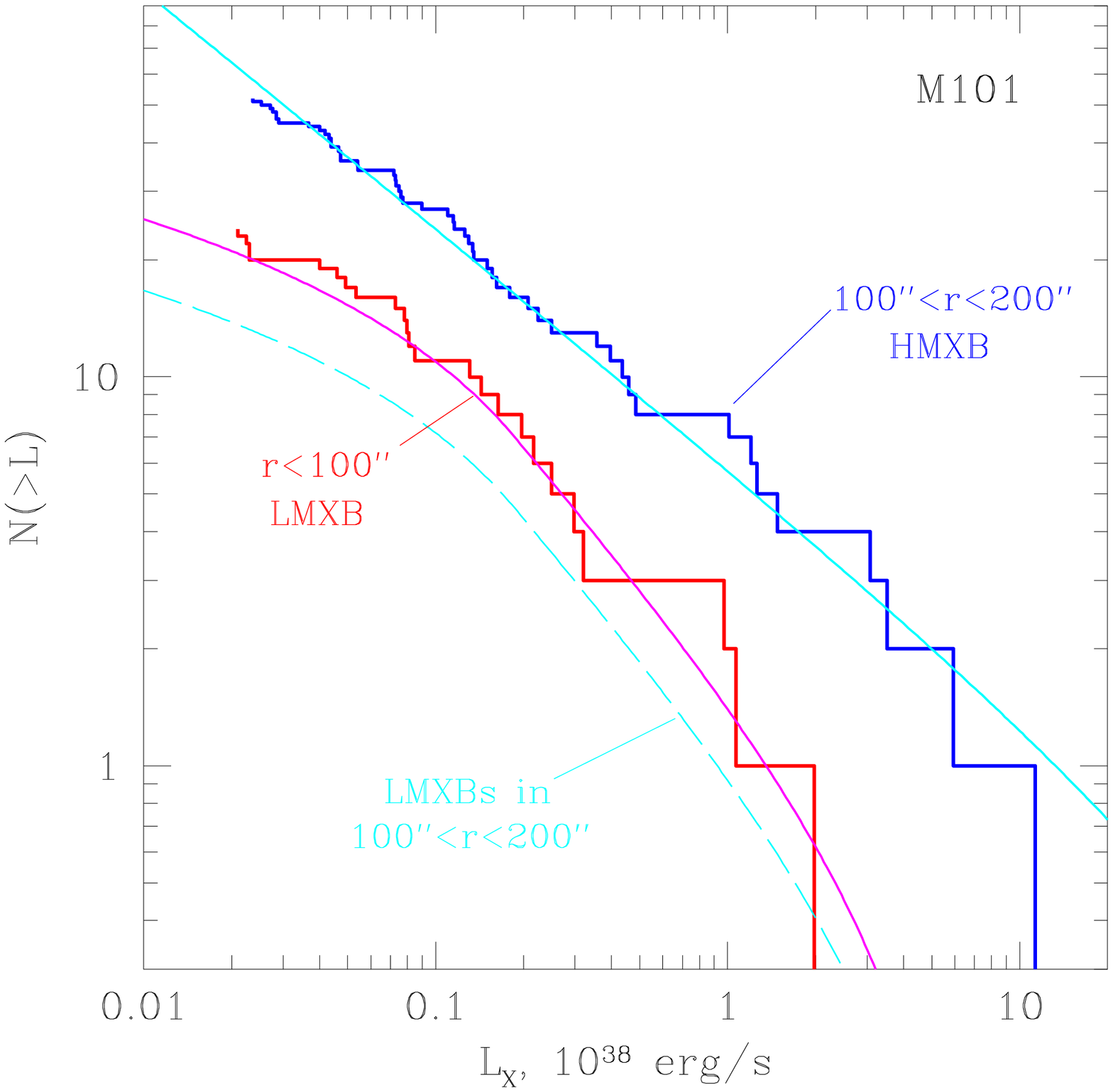}}
\resizebox{0.5\hsize}{!}{\includegraphics{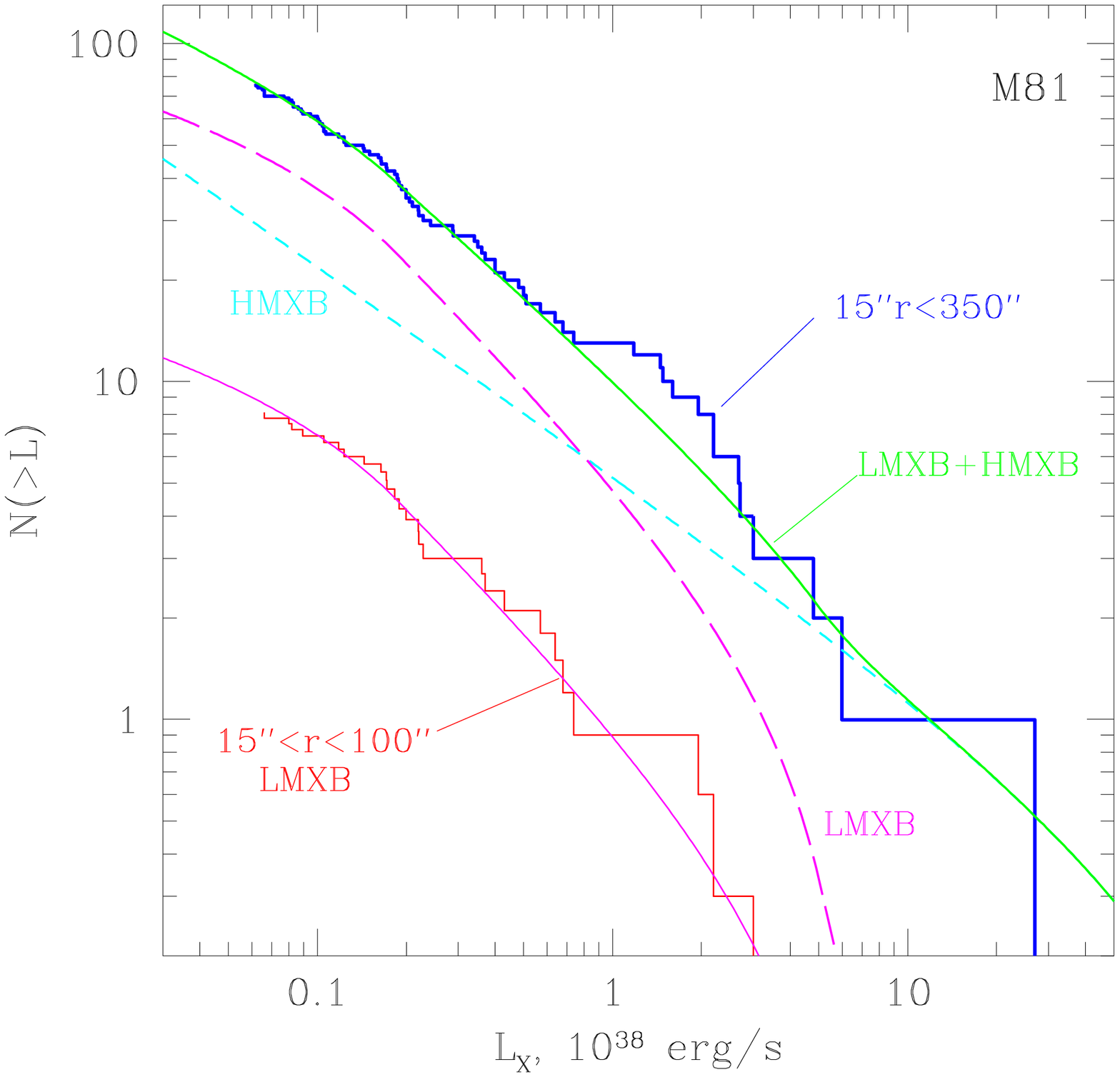}}
}}
\caption{{\em Left:}  XLF of bulge ($r<100\arcsec$) and disk ($100\arcsec <r<
200\arcsec$) of M101. The upper solid line shows predicted XLF of
HMXBs corresponding to star formation rate of $1.1$ M$_{\odot}$/year,
the lower solid line -- predicted XLF of LMXBs with the normalization 
equal to the best fit value for M101
(e.g. Fig.\ref{fig:xlfnorm_vs_type}). The dashed line shows predicted
contribution of the LMXBs to the disk population of X-ray
binaries. It was computed using corresponding near-infrared
luminosity and the same X-ray/NIR ratio as determined from the bulge
population. 
{\em Right:} XLF of the inner bulge ($15\arcsec <r<100\arcsec$) and
bulge+disk ($15\arcsec<r<350\arcsec$) populations in M81. The lower
solid line is predicted XLF of LMXBs in the inner bulge, the
short-dashed, long dashed and upper solid lines show  contribution of
HMXBs (SFR=1 M$_{\odot}$/yr) and LMXBs in the XLF of bulge+disk and
their sum.  In calculating the LMXB XLF for the bulge+disk the X/mass
ratio of the bulge was multiplied by a factor of 0.75.  
The observed and predicted XLFs of the bulge are scaled down by a
factor of 0.3 for clarity.   
}
\label{fig:xlf_m101}
\end{figure*}

\subsection{Average luminosity of LMXBs}

Given the shape of the LMXB XLF (Fig.\ref{fig:xrb_xlf}), the
effects of statistics are not very pronounced in the considered range
of the stellar mass, $\log(M_*)\sim 9-11.5$, and  do not affect
significantly the $L_X-M_*$ relation (thick solid line and shaded
area around it in Fig.\ref{fig:x_mass}). They are more noticeable when
considering average luminosity of X-ray sources in a
galaxy and lead to an artificial (unphysical) dependence of the
average luminosity on the stellar mass
(Fig.\ref{fig:lx2nx_vs_mass}). As illustrated by 
Fig.\ref{fig:lx2nx_vs_mass},  the observed dependence agrees well with
the theoretical prediction based on the shape of the average
XLF.

\begin{figure*}
\hbox{
\resizebox{0.5\hsize}{!}{\includegraphics{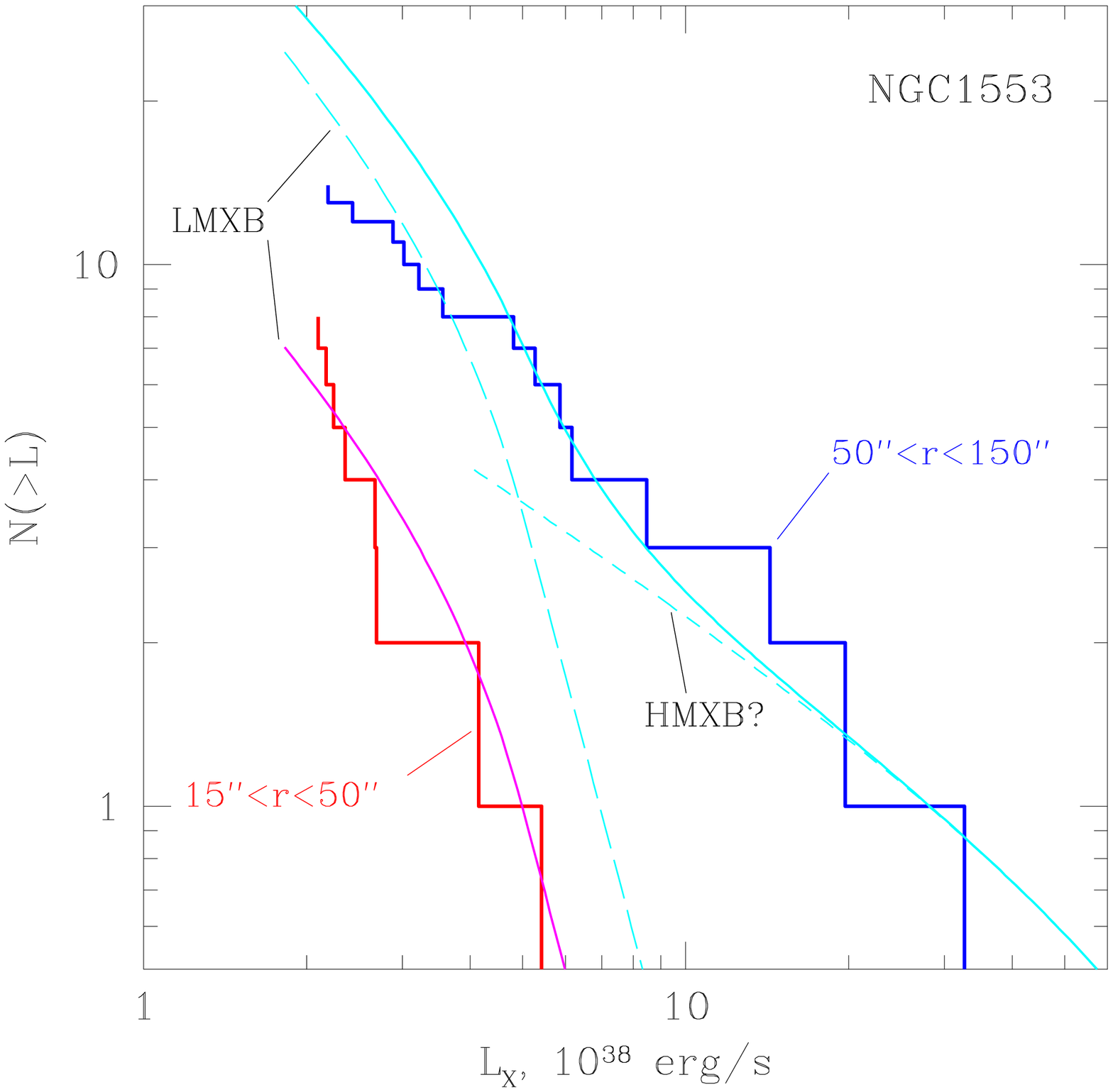}}
\resizebox{0.5\hsize}{!}{\includegraphics{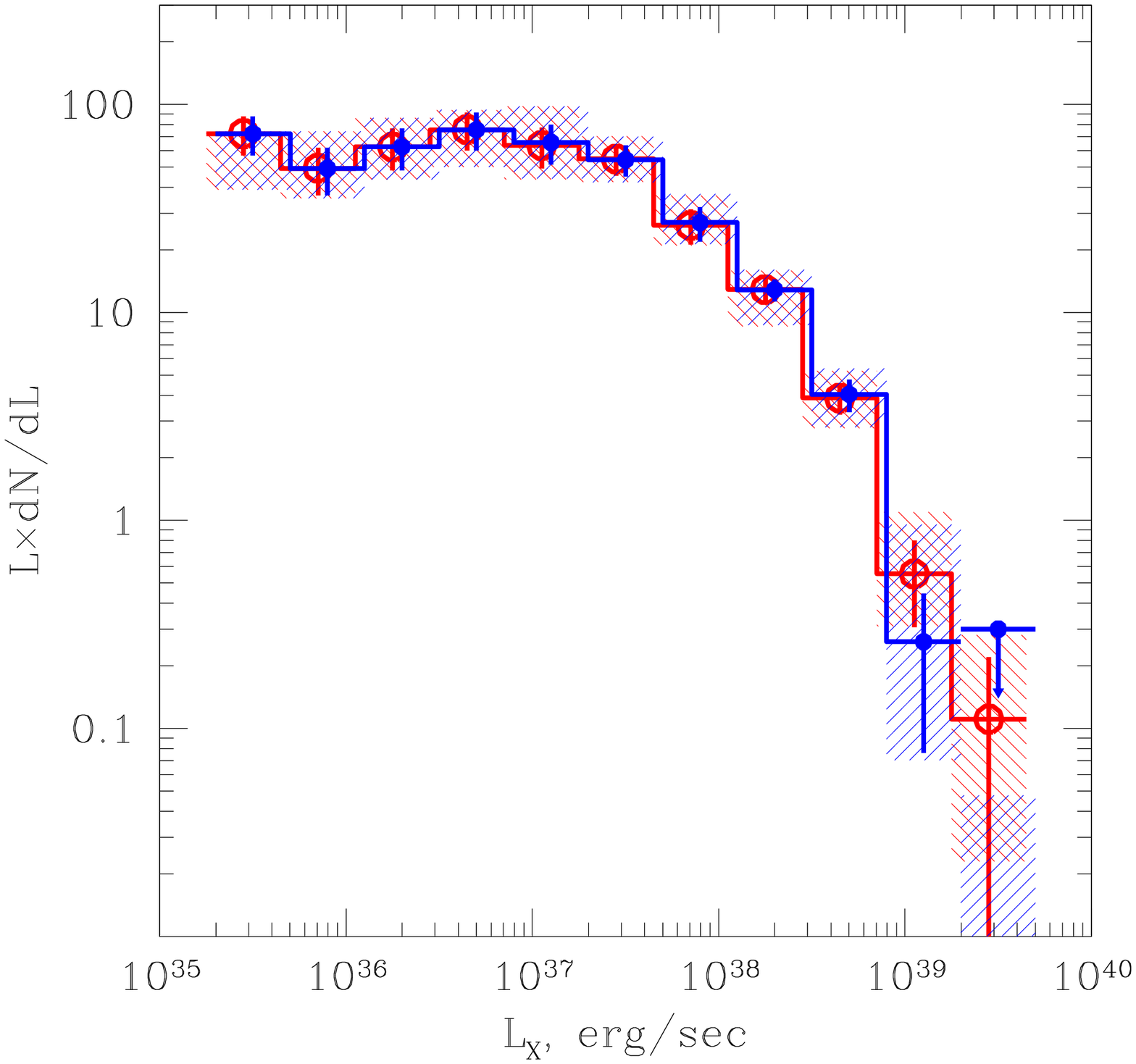}}
}
\caption{NGC1553: {\em Left:} XLFs in two annuli, $15\arcsec
<r<100\arcsec$  and $50\arcsec <r< 150\arcsec$. The histograms show
the observed distributions. The lower solid curve shows expected
distribution of LMXBs with the normalization equals to the best fit
value for the inner region. The dashed curve shows expected LMXB
population for the outer region, its normalization was computed using
respective near-infrared luminosity and the same X-ray/NIR ratio as
determined for the inner region. The short-dashed curve shows
luminosity distribution of the HMXBs expected for SFR=2
M$_{\odot}$/yr, the solid line -- the sum of LMXB and HMXB
populations.  
{\em Right:} Effect of NGC1553  on the average LMXB XLF.
Solid and grey open symbols show the average XLF computed
without and with the NGC1553 sources. The upper
limit is at 90\% confidence. The shaded areas have the same meaning as
in Fig.\ref{fig:xrb_xlf}. The two histograms are shifted horizontally
with respect to each other by a small offset for clarity.
}
\label{fig:xlf_ngc1553}
\end{figure*}

\subsection{Bulge and disk populations in spiral galaxies}
\label{sec:bulge_disk}

The deviation of X-ray growth curves from the distribution of the
near-infrared light observed in the disks of spiral galaxies has been
attributed in section \ref{sec:gcurves_spir} to the contribution of
high mass X-ray binaries. In Fig.\ref{fig:xlf_m101} we further support
this suggestion by comparing the X-ray luminosity functions of the
bulge and disk population in M101. The observed
difference in the number of sources and the shapes of their luminosity 
distributions agree both qualitatively and quantitatively with that
expected from the ``universal'' luminosity functions of high and
low mass X-ray binaries. Also shown in Fig.\ref{fig:xlf_m101} is the
expected contribution of the low mass X-ray binaries in the
disk of M101, predicted from its near-infrared luminosity and using
the X-ray/NIR ratio derived for the bulge. 
The smallness of the expected number of LMXBs in the disk is in a good 
agreement with the fact, that outside the bulge of M101, the sources
exhibit strong concentration towards the spiral arms, with the source
density in the inter-arm regions being close to the density of
background sources \citep{m101}.

The disk of M101 has a rather small ratio of the stellar mass to the star
formation rate, $M_*/{\rm SFR}\sim 2.4\cdot 10^{9}$
yrs. Correspondingly (section \ref{sec:l2m_m2sfr}), the LMXB
contribution to the disk population of the compact X-ray sources is
small and the luminosity function of the disk sources is close to that
of HMXBs. In a general case, the population  of the compact
X-ray sources can be a superposition of the high and low mass X-ray
binaries with comparable relative contributions, resulting to a
complex shape of the luminosity distribution, coinciding neither 
with LMXB nor with HMXB XLF. This might explain a variety of the shapes
of the luminosity functions observed by Chandra in different regions
of nearby late type galaxies \citep[e.g.][]{m31b}, as illustrated  
by right panel of  Fig.\ref{fig:xlf_m101}, showing the luminosity
distribution in a large region of M81, including both bulge and disk
population. Due to rather large  value of the 
$M_*/{\rm SFR}\sim 4.6\cdot 10^{10}$ yrs, comparable fractions of
low and high mass X-ray binaries are expected. The total luminosity function
of all sources can be presented as a sum of LMXB and HMXB 
contributions (Fig.\ref{fig:xlf_m101}), in agreement with different
luminosity distributions of the sources near the spiral arms and in
the inter-arm regions observed by Chandra \citep{m81}.

\subsection{Individual galaxies}
\label{sec:individual}

The $L_X$ growth curves of two galaxies, NGC1553 and NGC1291, show
significant deviations from the distribution of the NIR
light (Fig.\ref{fig:gcurves_lx}).  
The luminosity distributions for these two galaxies also exhibit
largest deviations from the average XLF, although consistent with the
latter within statistical uncertainties, according to the
Kolmogorov-Smirnov test (Table \ref{tab:ks_test}). 

These two galaxies might indeed present an example of the population
age dependent variations in the shape of the luminosity distribution.
In both cases the deviations from the average LMXB XLF occur at the
high luminosity end as might be expected for such
variations.  Below we assess the statistical significance of the
observed deviations and discuss alternative interpretations.

\subsubsection{NGC1553}
\label{sec:ngc1553}

Comparison of $N_X$ and $L_X$ growth curves for this S0 galaxy
indicates possible variations of the luminosity distribution with
the distance from the nucleus. This is confirmed by
Fig.\ref{fig:xlf_ngc1553}, presenting luminosity 
functions for the inner and outer regions.
In the inner ring, the luminosity distribution of the sources is
consistent with that of other early type galaxies (solid smooth line
marked ``LMXB''). In the outer ring, on the contrary, it deviates
significantly from the average (long-dashed line), both in  the slope
of the distribution and in the luminosity of the brightest sources.  
Although the luminosity function of all sources from 
$15\arcsec < r < 150\arcsec$ is consistent with the average LMXB XLF
(Table \ref{tab:ks_test}), the outer ring taken separately gives the
value of the Kolmogorov-Smirnov probability of $\approx 2.1\cdot
10^{-2}$. 
In particular, 3 sources with $L_X\ga 10^{39}$ erg/s are observed at
the radial distance $50\arcsec\la r\la 150\arcsec$ from the nucleus. 
Although their luminosities are not extreme -- still by a factor of
$\sim 10$ below that of the brightest ULXs in star forming
galaxies, they nevertheless stand out in our sample. 
The nature of these sources is unclear.
They do not coincide with any of the known globular clusters
in the galaxy \citep{ngc1553}.  At the time of writing,
there were no repeated observations of NGC1553, therefore their
transient nature can not be established. 

The shape of the XLF is reminiscent of that observed in the disks of
the spiral galaxies  (cf. Fig\ref{fig:xlf_m101}). Hypothetically it
could be  understood as a superposition of the LMXB and HMXB
populations (solid line in Fig.\ref{fig:xlf_ngc1553}). 
However, such an explanation would require star formation on
the level  of $\sim 1-2$ M$_{\odot}$/yr. 
Although on-going star formation in NGC1553 can not be entirely
excluded -- it forms an interacting pair with NGC1549
\citep{ngc1553_gc}, the required value of SFR
exceeds by an order of magnitude the  star formation rates detected in
some of the S0 galaxies \citep[e.g.][]{s0_sfr}. Moreover, the value of
the far-infrared flux observed by IRAS, 
$F_{60\mu}=0.55$ and $F_{100\mu}=1.14$ Jy (NED,
http://nedwww.ipac.caltech.edu), constrains the  present star
formation rate in NGC1553 by $\la 0.1$ M$_{\odot}$/yr.

Another possibility is that the field of NGC1553 is contaminated by a
fluctuation  in the density of the CXB sources. 
An indirect evidence in favor of this interpretation is provided by
the detection of another very bright source in the periphery of
NGC1553 located at $\approx  3\farcm5$ from the nucleus (the source
no.38 in the Table 1 of \citet{ngc1553}). Its luminosity is $L_X\sim
9.4\cdot 10^{39}$ erg/s assuming that it is a member of galactic
population of the compact sources. An optical counterpart has been
found in the DSS  plate \citep{ngc1553}. The blue magnitude, listed in
the USNO-2 
catalog, $m_{\rm B}=20.4$, would correspond to the luminosity
$L_{\rm B}\sim 6\cdot 10^6$ L$_{\odot}$ 
if the source was located at the distance of NGC1553. 
Obviously, this source is a background AGN (a foreground object in
the Galaxy is excluded by the $F_X/L_{\rm opt}\sim 0.5$). 
As the expected number of such bright background objects is 
$N_{\rm CXB}\sim 0.08$, 
its presence might indicate an enhancement in the number density of
CXB sources in the filed of NGC1553.  
Detection of optical counterparts of other apparently luminous sources
would prove this hypothesis, however optical observations inside $\sim
1-2$ effective radii of the galaxy are complicated by its optical
emission. 
Note, that in order to account for the number of bright sources 
observed in the outer ring (Fig.\ref{fig:xlf_ngc1553}), a significant
over-density of the CXB sources, by a factor of $\sim 2-3$, on
arcmin angular scale is required, which is probably too large to be
explained in terms of the average angular correlation function of CXB
sources \citep{vikhl95}.

In conclusion, we note, that although NGC1553 clearly stands out in our
sample from the point of view of the $L_X$ growth curve and the shape
of the luminosity function, its growth curve for the number of sources
and X/M$_*$ ratios are within the dispersion of the values observed in
other nearby galaxies (Table \ref{tab:xray},
Figs.\ref{fig:gcurves_nx}, \ref{fig:x_mass}, \ref{fig:x2mass_vs_type}). 
We give in Table \ref{tab:xray} the X-ray and near-infrared parameters
for both the inner ring and the entire $15\arcsec < r < 150\arcsec$ 
region.  NGC1553 data were not included in constructing the
average luminosity function (Fig.\ref{fig:xrb_xlf}) and determination   
of its best fit parameters (Table \ref{tab:xlf}). Its influence on the
shape of the average luminosity function is illustrated in the right
panel in Fig.\ref{fig:xlf_ngc1553}, showing that associated changes
are well within the statistical and systematic uncertainties.

\subsubsection{NGC1291}

The deviations of the $L_X$ growth curve from the NIR profile are
significantly less pronounced and are within about $\la 2\sigma$. 
The examination of the luminosity  
distribution in different annuli did not reveal any peculiarities,
similar to those observed in NGC1553. The overall luminosity
distribution (Fig.\ref{fig:xlf_comb_uxlf}) agrees very well with the
average LMXB XLF at low luminosities, $L_X\la 2.5\cdot 10^{38}$ erg/s,
but appears to have a too abrupt cut-off above this value. To
characterize it quantitatively, we note that the
average XLF, normalized to the total number of sources at the the
adopted completeness limit, $3\cdot 10^{37}$ erg/s, predicts $\approx
2.7$ sources above the luminosity of the brightest observed source,
$2.6\cdot 10^{38}$ erg/s. For a Poisson distribution with
expectation value of 2.7, the probability to draw zero is
6.7\%, i.e. the case of NGC1291 presents $\la 2\sigma$
deviation. This value of probability is not small enough to claim
presence of statistically significant deviations from the average
XLF.

\subsection{The Milky Way}

The luminosity function of LMXB sources in the Milky Way and
X-ray/M$_*$ ratios are by a factor of $\sim 1.5-2$ lower than the main
group of galaxies  on all plots (Fig. \ref{fig:xlf_cum},
\ref{fig:x_mass}, \ref{fig:x2mass_vs_type}). 
As the X-ray/M$_*$ ratios for the Milky Way were derived in a different
way than for other galaxies, they are subject to different systematic
uncertainties. We discuss below various factors, affecting its
X-ray/NIR ratios. 

The stellar mass of the Milky Way was estimated using the K-band
growth curve and mass-to-light ratio of M31, which is sufficiently
similar to our Galaxy in the morphological type. In the growth curve
analysis we used the low mass X-ray binaries located at $X>0$, where
the origin of the Cartesian coordinate system is located at the
Galactic Center and the X-axis is directed towards the Sun
\citep{grimm1}. Thus, to the accuracy defined by the completeness of
the LMXB catalog, our analysis should include $\sim 
1/2$ of the Galaxy. Correspondingly, the NIR growth curve of M31 was
scaled down by a factor of 2, resulting in the K-band luminosity of
the half of the Galaxy  $L_{K}\approx 4.5\cdot 10^{10}$ L$_{\odot}$
(calculated from the best fit total K-band magnitude of M31,
$m_{tot}=0.5$, Table \ref{tab:sample}). 
This value is consistent with accuracy better than $\la 10\%$ with
that expected from the the total K-band luminosity of the Milky Way,   
$L_{\rm MW,K}\approx 9.6\cdot 10^{10}$ L$_{\odot}$,  
obtained from 3D modeling of the DIRBE data \citep{malhotra96}. 
The K-band mass-to-light ratio obtained from the optical color of M31
is $M_*/L_K\approx 0.68$. This value is close, but not identical to
the value obtained by \citet{m2l_mw} for the solar neighborhood, 
$M_*/L_K\approx 0.78$. It should be mentioned, though, that the latter
value was derived for the disk population of the Galaxy and might not
be appropriate for the older stellar population of the bulge.
Note, that higher mass-to-light ratio would increase the stellar mass
estimate, decrease of the X-ray/mass ratio, and therefore 
would result in a larger discrepancy between the Milky Way and other
galaxies.

An important factor is the completeness of the LMXBs catalog.
\citet{grimm1} used the sources with known distances and reliable
optical identification. Comparing with the stellar mass distribution
in the Galaxy, they concluded, that their sample is reasonably
complete above $\sim 10^{36}$ erg/s and within $\sim 10-12$ kpc from
the Sun. In computing the final luminosity function and total X-ray
luminosity of the Galaxy they introduced luminosity dependent volume
correction using the mass model of the Galaxy.  On the other hand we
computed the X-ray/NIR ratios using the sources located at $X>0$ 
in the range of projected galactocentric distances 
$1{\rm ~kpc}< R_{\rm proj}< 10{\rm ~kpc}$, which, strictly speaking, is
not identical to the definition of the completeness region in 
\citet{grimm1}. It is not clear, whether this procedure could result
in an underestimate of the Milky Way luminosity by a factor 
of $\sim 1.5-2$. On the other hand, the value of the X-ray-to-mass
ratio, derived  in this paper, $L_X/M_*\approx 3.5\cdot 10^{28}$
erg/s/$M_*$ is close but somewhat lower than the one obtained by
\citet{grimm1},  using different method, $L_X/M_*\approx 5\cdot
10^{28}$ erg/s/$M_*$. 

Finally, \citet{grimm1} used the source luminosities averaged over 5
year period of the ASM/RXTE observations.
Depending on the properties of the collective X-ray light
curve of the Galaxy this might lead to a difference between their
average value and the value to be  most probably observed in a
snapshot.

\subsection{Total energy output of LMXBs and HMXBs}

The calibration of the $L_X-$SFR and $L_X-M_*$ relations obtained by
\citet{grimm2} and in the present paper allows us to estimate the
total energy output of X-ray binaries throughout the life time of a 
galaxy. 

The total energy output of HMXBs in the Chandra passband can be
estimated as: 
\begin{eqnarray}
E_{\rm HMXB}=\int \frac{L_X}{\rm SFR} \ {\rm SFR}(t) \ dt = 
%\frac{L_X}{\rm SFR} \int {\rm SFR}(t) \ dt =
\frac{L_X}{\rm SFR}\ \alpha \eta M_*
\label{eq:e_hmxb}
\end{eqnarray}
where ${\rm SFR}(t)$ is the time dependent star formation rate of
stars more massive than 5 M$_{\odot}$ and describes the star formation
history of the galaxy,  $M_*$ -- its present stellar mass,  
$\alpha$ -- fraction of the total mass of stars formed during the
life time of the galaxy, which presently resides in stars (of all
masses), $\eta$ is the IMF-dependent mass fraction of stars more
massive than  5 M$_{\odot}$ and accounts for the fact, that  $L_X$--SFR
was calibrated for formation rate of stars $M>5$ M$_{\odot}$: 
\begin{eqnarray}
\eta=\frac{\int^{M_{\rm max}}_{5 M_{\odot}} \xi(M)\ M\ dM}
{\int^{M_{\rm max}}_{M_{\rm min}} \xi(M) \ M \ dM}
\end{eqnarray}
For the ``extended'' Miller-Scalo IMF \citep{imf}, 
$\xi(M)=M^{-1.4}$ for $0.1M_{\odot}\le M\le 1M_{\odot}$, and
$\xi(M)=M^{-2.5}$ for $1M_{\odot}\le M\le 100M_{\odot}$,
$\eta\approx 0.23$.

Studying the intermediate  redshift (up to $z\approx 1.3$) starburst
galaxies observed by CHANDRA in the Hubble Deep Field North
\citet{grimm2} showed that the calibration of the $L_X/$SFR ratio
based on the local galaxies is valid for these distant galaxies as
well. For this reason the $L_X/$SFR ratio can be taken
outside the integration in eq.(\ref{eq:e_hmxb})
With $L_X/{\rm SFR}\approx 6.7\cdot 10^{39}$ erg/s per M$_{\odot}$/yr
and assuming $\alpha\sim 0.3-0.5$, the total energy output of HMXBs is 
\begin{eqnarray}
E_{\rm HMXB}\approx 2.4 \cdot 10^{57}\ \left( \frac{\alpha}{0.5}
\right ) \ 
\left( \frac{M_*}{10^{11}\ M_{\odot}}\right) {\rm~erg}
\label{eq:e_hmxb2}
\end{eqnarray}

The energy output of LMXBs is:
\begin{eqnarray}
E_{\rm LMXB}=\int \frac{L_X}{M_*}(t) \ {M_*}(t) \ dt \sim
 \frac{L_X}{M_*} \ {M_*} \ t_{\rm gal}
\label{eq:e_lmxb}
\end{eqnarray}
Unlike for high mass X-ray binaries, eq.(\ref{eq:e_hmxb}), the
$L_X$--mass relations was 
measured for nearby galaxies only. Therefore the last
equality in the above equation relies on the extrapolation of this
relation to high redshifts. As the evolution effects might play an
important role in the case of LMXBs, this estimate is significantly
less robust than that for HMXBs. However, the details of the evolution
of the luminosity distribution of LMXBs are yet unexplored, and
even a crude estimate might be of certain interest. Using
$L_X/M_*\approx 8\cdot 10^{39}$ erg/s per $10^{11}$ M$_{\odot}$ and
assuming $t_{\rm gal}\sim 10^{10}$ yrs:
\begin{eqnarray}
E_{\rm LMXB}\sim 2.5\cdot 10^{57}
\left( \frac{M_*}{10^{11}\ M_{\odot}}\right) 
\left( \frac{t_{\rm gal}}{10^{10}\rm yrs} \right)
{\rm ~erg} 
\label{eq:e_lmxb2}
\end{eqnarray}
which is close to the value obtained for HMXBs. As low mass X-ray
binaries were brighter in the past, this number is a lower limit on
their energy output in the Chandra passband.

\subsection{Clusters of galaxies}

For the typical value of the stellar mass in the rich clusters
of galaxies, $M_*\sim 10^{13-14}$  M$_{\odot}$, the expected X-ray
luminosity of low mass X-ray binaries is $L_{\rm LMXB}\sim 10^{42-43}$
erg/s. Taken at face value, this is a negligible fraction, at the
level of $\la$ few per cent, of the luminosity of the X-ray emitting
gas, $L_{\rm gas}\sim 10^{43-45}$ erg/s.  

However, due to different
energy spectra of X-ray gas and of X-ray binaries, contribution of the
latter might become significant in certain energy ranges. 
It is well known that the hard tails of Comptonized radiation 
might be present in the spectra of
X-ray binaries. Depending on the spectral state, the spectrum of the
Comptonized component extends to $\sim 100-200$ keV (low state) or
to the $\sim $ MeV energies in the high state of black hole binaries.
Importantly, in the low state, the hard spectral component is observed
in the spectra of both black hole and neutron star binaries
\citep[e.g.][]{gc_sigma, sax1808}.
The hard X-ray spectral component  might amplify the relative  
contribution of X-ray binaries at high X-ray energies, $\sim 20-150$
keV and higher. 
An accurate calculation of this contribution would require knowledge
of the X-ray binaries luminosity distribution in hard X-rays, far
beyond the Chandra bandpass. 
For a crude, an order of magnitude estimate we note, that in the low
spectral state, corresponding to the luminosities below $\sim
(2-3)\cdot 10^{37}$ erg/s,  the luminosity above  $\sim 30$ keV is at
the very least equal to that emitted in the Chandra bandpass.  Given
the shape of the LMXB luminosity function, the sources with
$L_X\la (2-3)\cdot 10^{37}$ erg/s contribute $\sim 25\%$ to the
combined X-ray luminosity of LMXBs. Thus, only due to X-ray binaries
in the low  spectral state the luminosity above $\sim 20$ keV should
be of the order of $\sim 10^{42-43}\times 0.25\sim {\rm few}\times
10^{41-42}$ erg/s. This is, obviously, a lower limit to the  
actual value of the total hard X-ray luminosity due to LMXBs. 
On the other hand, for typical temperatures of the X-ray gas in the
clusters of galaxies, $T_{\rm gas}\sim 5$ keV, of the order of $\la
10^{-2}$ of the total luminosity is emitted above $\sim 20$ keV,
i.e. $\la 10^{41-43}$ erg/s which is comparable to the lower limit on
the luminosity of LMXBs. At higher energies the situation becomes
even more favorable for LMXBs. As the emissivity of X-ray gas falls
off with radius quicker, than the number density of the galaxies, the
relative contribution of LMXBs in hard X-ray energy domain  will be
further  enhanced in the outer parts of the clusters. We should note,
however, that the hard X-ray emission may be dominated by the low
luminosity AGNs, depending on their actual frequency and the
luminosity distribution.

\section{Summary}
\label{sec:summary}

We analyzed population properties of compact X-ray sources in 
nearby external galaxies of various morphological types and of the LMXB
sources in the Milky Way. We focused our analysis on the old stellar
systems -- early type galaxies and the bulges of spiral
galaxies. This ensures that the compact X-ray sources in the
external galaxies are dominated by low mass X-ray
binaries. In the case of the Milky Way we explicitly selected LMXB  
sources based on the results of \citet{grimm1}. Our findings can be
summarized as follows.
\begin{enumerate}
\item
For all galaxies the azimuthally averaged spatial distribution of the
number of LMXBs follows closely the distribution of the near-infrared
light (Fig.\ref{fig:gcurves_nx}).
Our analysis covered the central parts of the galaxies, out to
$\sim 1.5-2$ effective radii in early type galaxies and
$\sim$inside the inner bulge in spiral galaxies.
With the exception of NGC1553 and, to lesser extend, NGC1291 
(section \ref{sec:individual}), the same is true for the
distribution of their combined luminosity in the Chandra bandpass,
$\sim 0.5-8$ keV  (Fig.\ref{fig:gcurves_lx}). 

In the disks of spiral galaxies, on the contrary, the distribution of
the number and combined luminosity of the compact sources deviates
significantly from the NIR profile. The deviation is caused 
by the contribution of HMXBs, as evidenced  by the
example of the Milky Way (Fig.\ref{fig:gcurves_mw}) and by the
comparison of the luminosity functions of the disk and bulge in M101
and M81 (Fig.\ref{fig:xlf_m101}). 

\item
Considering galaxies as whole, 
the total number of LMXBs and their collective luminosity are directly  
proportional to the stellar mass of the host galaxy (Fig.\ref{fig:x_mass}), 
the latter calculated from the K-band luminosity using
color-based correction of the mass-to-light ratio (eq.(\ref{eq:m2l})).   

There is statistically significant dispersion between 
values of X/M$_*$ ratios, with the fractional rms of $\sim 25\%$ and 
$\sim 40\%$ for  $L_X/M_*$ and $N_X/M_*$  
respectively (Fig.\ref{fig:x2mass_vs_type}).
The accuracy of the color-based correction of the K-band mass-to-light
ratio in the present analysis is insufficient to establish reality of
the observed dispersion. Neither it is sufficient to confirm
or rule out the possibility of a morphological type-dependent trend in
the X/M$_*$ ratios.  If present, its effect does not exceed a 
factor of $\sim 1.5-2$  (Fig.\ref{fig:x2mass_vs_type}).  

On average, the number of sources with $L_X>10^{37}$ erg/s and their
combined luminosity are related to the stellar mass as follows:
\begin{eqnarray}
N_X(>10^{37} {\rm ~ erg/s})= 142.9\pm 8.4 
{\rm ~sources ~ per~ 10^{11}~ M_\odot}~~~~~\\
L_X(>10^{37} {\rm ~ erg/s})= (8.0\pm 0.5)\cdot
{\rm ~10^{39}~ erg/s ~ per~ 10^{11}~ M_\odot}  \nonumber
\label{eq:x_mass_from_uxlf_concl}
\end{eqnarray}
The precise values of the coefficients in these formulae depends
somewhat on the weighting method (section \ref{sec:x_nir_ratios}).

\item The X-ray/$L_{\rm NIR}$ ratios show clear dependence on the
morphological type (Fig.\ref{fig:lx2lnir_vs_type}). 
The average values are:
\begin{eqnarray}
N_X(>10^{37} {\rm ~ erg/s})= (81-135)
{\rm ~sources ~ per~ 10^{11}~ L_\odot}~~~~~~\\
L_X(>10^{37} {\rm ~ erg/s})= (3.3-7.5)\cdot
{\rm ~10^{39}~ erg/s ~ per~ 10^{11}~ L_\odot}  \nonumber
\label{eq:x_nir_from_uxlf_concl}
\end{eqnarray}
where the first and the second number in the parenthesis correspond to
average values for late and early type galaxies respectively. The
average values for all galaxies are $117\pm12.4$ sources and 
$(6.1\pm 0.5)\cdot 10^{39}$ erg/s for $N_X$ and $L_X$
respectively. These numbers obviously depend on the content of our
sample. 

Even stronger dependence on the morphological type should be
observed at shorter wavelength, e.g. in the B-band, often used to
characterize relation between X-ray and optical properties of the
galaxies. 

\item
The shape of the luminosity function of the LMXBs is similar in
different galaxies (Fig.\ref{fig:xlf_cum}, \ref{fig:xlf_comb}, 
\ref{fig:xlf_universal}). The data on individual galaxies, with a
possible exception of NGC1553 (section \ref{sec:individual}), are
broadly consistent with the average luminosity function of LMXBs,
derived using all galaxies from our sample
(Fig.\ref{fig:xlf_comb_uxlf}, Table \ref{tab:xlf}, 
\ref{tab:ks_test}).   
The average LMXBs  XLF has a complex shape and differs significantly
from that of HMXBs (Fig.\ref{fig:xrb_xlf}, Table \ref{tab:xlf}).  
It is consistent with a power law with differential slope of 
$\approx 1$  at low luminosities, gradually steepens above
$\log(L_X)\sim 37.0-37.5$ and has a rather abrupt 
cut-off at $\log(L_X)\sim 39.0-39.5$. 
In the $\log(L_X)\sim 37.5-38.7$ luminosity range it is approximately
represented by a power law with differential slope of $\sim 1.7-2.0$. 
The normalization of the
luminosity function found in individual galaxies is proportional to
the stellar mass  (Fig.\ref{fig:xlfnorm_vs_type}). 

\item
The luminosity of the  brightest sources in our sample of old stellar
systems (with total stellar mass of $\sim 10^{12}$ M$_{\sun}$) does
not exceed the value  of $L_X\sim (2-3)\cdot 10^{39}$ erg/s. The maximum
luminosity of HMXB sources is by a factor of $\sim 10$ larger. 
This can be used to diagnose on-going star formation.

\item 
Relative contributions of low and high mass X-ray binaries to the
population of compact sources in a galaxy is defined by the ratio of
its stellar mass to the present value of star formation rate,
according to the formulae given in section \ref{sec:l2m_m2sfr}.

\item
Calibration of the $L_X-M_*$ relation derived in the present paper allows
one to use the X-ray luminosity of a galaxy due to low mass X-ray
binaries as an independent stellar mass indicator. Applicability of
this calibration to distant galaxies at larger redshifts is yet to be
established and  might require proper account for the X-ray binary
evolution effects.

\item
The total energy outputs of LMXBs and HMXBs in the Chandra passband
during the life time of a galaxy are of the same order, 
$\sim {\rm few}\times 10^{57}\times (M_*/10^{11}M_{\odot})$ erg. In
the case of LMXBs, this number does not take in to account the binary
evolution effects and, most likely, is a lower limit.
\end{enumerate}

{}


\begin{thebibliography}{}



\bibitem[\protect\citeauthoryear{Bell \& de Jong}{2001}]{m2l}
Bell E. \& de Jong R., 2001, ApJ, 550, 212

\bibitem[\protect\citeauthoryear{Blanton, Sarazin \& Irwin}{Blanton et
al.}{2001}]{ngc1553} Blanton E., Sarazin C. \& Irwin J., 2001, ApJ,
552, 106 

\bibitem[\protect\citeauthoryear{Bridges \& Hanes}{1990}]{ngc1553_gc}
Bridges T. \& Hanes D., 1990, AJ, 99, 1100

\bibitem[\protect\citeauthoryear{Brinchmann \& Ellis}{2000}]{jarle2001}
Brinchmann J. \& Ellis R., 2000, ApJ, 536, L77

\bibitem[\protect\citeauthoryear{David, Jones \& Forman}{David et al.}
{1992}]{david92} David L., Jones C. \& Forman W., 1992, ApJ, 388, 82

\bibitem[\protect\citeauthoryear{de Vaucouleurs \& Corwin}{de
Vaucouleurs \& Corwin} {1977}]{col_grad} de Vaucouleurs G. \& Corwin
H.G., 1977, ApJS, 33, 219


\bibitem[\protect\citeauthoryear{de Vaucouleurs et al.}{de Vaucouleurs et al.}
{1991}]{rc3} de Vaucouleurs G. et al., 1991, Third Refererence Catalog
of Bright Galaxies. Springer-Verlag (RC3)

\bibitem[\protect\citeauthoryear{Fabbiano \& White}{2003}]
{fabbiano2003} Fabbiano G. \& White N., 2003, in: ``Compact Stellar
X-ray Sources", Cambridge University Press (eds., W. Lewin \& M. van
der Klis), astro-ph/0307077


\bibitem[\protect\citeauthoryear{Finoguenov \& Jones}{Finoguenov \&
Jones} {2002}]{m84} Finoguenov A. \& Jones C. 2002, ApJ, 574, 754

\bibitem[\protect\citeauthoryear{Fioc \& Rocca-Volmerange}{Fioc \&
Rocca-Volmerange} {1999}]{fioc99} Fioc M. \& Rocca-Volmerange B.,
1999, A\&A, 351, 869

\bibitem[\protect\citeauthoryear{Fitzpatrick} {1999}]{extinction}
Fitzpatrick L., 1999, PASP, 111, 63


\bibitem[\protect\citeauthoryear{Gehrels}{Gehrels} {1986}]{gehrels86} 
Gehrels N. 1986, ApJ, 303, 336


\bibitem[\protect\citeauthoryear{Gezari, Pitts \& Schmitz}{Gezari et
al.} {1999}]{gezari99} Gezari D.Y., Pitts P.S. \& Schmitz M., 1999,
``Catalog of Infrared Observations, Edition 5''

\bibitem[\protect\citeauthoryear{Ghosh \& White}{Ghosh \& White}
{2001}]{ghosh01} Ghosh P. \& White N., 2001, ApJ, 559, L97

\bibitem[\protect\citeauthoryear{Gilfanov et al.}{1998}]{sax1808}
Gilfanov M. et al.,  1998, A\&A, 338, L83

\bibitem[\protect\citeauthoryear{Gilfanov, Grimm \& Sunyaev}{Gilfanov
et al.}{2003}]{stat} Gilfanov M., Grimm H.-J. \& Sunyaev R., 2003, in
preparation 

\bibitem[\protect\citeauthoryear{Griffiths \& Padovani}{Griffiths \&
Padovani} {1990}]{griffiths90} Griffiths R. \& Padovani P., 1990, ApJ, 360,
483 

\bibitem[\protect\citeauthoryear{Grimm, Gilfanov \& Sunyaev}{Grimm et
al.}{2002}]{grimm1} Grimm H.-J., Gilfanov M.\& Sunyaev R., 2002, A\&A,
391, 923 

\bibitem[\protect\citeauthoryear{Grimm, Gilfanov \& Sunyaev}{Grimm et
al.}{2003}]{grimm2} Grimm H.-J., Gilfanov M.\& Sunyaev R., 2003, MNRAS,
339, 793

\bibitem[\protect\citeauthoryear{Irwin, Sarazin \& Bregman}{Irwin et
al.}{2002}]{ngc1291} Irwin J.A., Sarazin C.L. \& Bregman J.N., 2003,
ApJ, 570, 152   

\bibitem[\protect\citeauthoryear{Jensen et al.}{2003}]{dist_5846} 
Jensen J., et al., 2003, ApJ, 583, 712 


\bibitem[\protect\citeauthoryear{Kauffmann et al.}{2003}]{sdss} 
Kauffmann G. et al., 2003, MNRAS, 341, 33

\bibitem[\protect\citeauthoryear{Kennicutt}{1983}]{imf} 
Kennicutt R., 1983, ApJ, 272, 54

\bibitem[\protect\citeauthoryear{Kim \& Fabbiano}{Kim \&
Fabbiano}{2003}]{ngc1316} Kim D.-W. \& Fabbiano G., 2003, ApJ, 586, 826 

\bibitem[\protect\citeauthoryear{Kong et al.}{Kong et al.}{2002}]{m31}
Kong A. et al., 2002, ApJ, 577, 738  

\bibitem[\protect\citeauthoryear{Kong et al.}{Kong et al.}{2003}]{m31b}
Kong A. et al., 2003, ApJ, 585, 298  

\bibitem[\protect\citeauthoryear{Kraft et al.}{Kraft et
al.}{2001}]{cena} Kraft R.P. et al., 2001, ApJ, 560, 675  

\bibitem[\protect\citeauthoryear{Maccarone, Kundu \& Zepf}{Maccarone et al.}
{2003}]{ngc4472} Maccarone T.J., Kundu A. \& Zepf S.E., 1996, ApJ,
586, 814 

\bibitem[\protect\citeauthoryear{Malhotra et al.}{Malhotra et al.}
{1996}]{malhotra96} Malhotra S. et al., 1996, ApJ, 473, 687

\bibitem[\protect\citeauthoryear{Manners et al.}
{2003}]{elais} Manners J.C. et al., 2003, MNRAS in press,
astro-ph/0207622 

\bibitem[\protect\citeauthoryear{Mendez et al.} {2001}]{dist4697}
Mendez R., 2001, ApJ, 563, 135 

\bibitem[\protect\citeauthoryear{Morton, Andereck \& Bernard}{Morton
et al.} {2001}]{m31bulge} Morton D.C., Andereck C.D. \& Bernard D.A.,
1977, ApJ, 212, 13  

\bibitem[\protect\citeauthoryear{Okamura, Kanazawa \& Kodaira}{Okamura
et al.} {1976}]{m101bulge} Okamura S., Kanazawa T., Kodaira K., 1976,
PASJ, 28, 329

\bibitem[\protect\citeauthoryear{Pahre}{Pahre}
{1999}]{pahre99} Pahre M.A., 1999, ApJSS, 124, 127

\bibitem[\protect\citeauthoryear{Pence et al.}{Pence et al.}
{2001}]{m101} Pence et al., 2001, ApJ, 561, 189

\bibitem[\protect\citeauthoryear{Pfahl, Rappaport \&
Podsiadlowski}{Pfahl et al.} {2002}]{pfahl2003} Pfahl E.D., Rappaport
S. \& Podsiadlowski Ph., 2003, ApJ, 597, 1036


\bibitem[\protect\citeauthoryear{Piro \& Bildsten}
{2002}]{piro02} Piro A. \& Bildsten L.,  2002, ApJ, 571, L103

\bibitem[\protect\citeauthoryear{Podsiadlowski, Rappaport \&
Pfahl}{Podsiadlowski et al.} {2002}]{pods2002} Podsiadlowski Ph.,
Rappaport S. \& Pfahl E.D., 2002, ApJ, 565, 1107


\bibitem[\protect\citeauthoryear{Pogge \& Eskridge}
{1987}]{s0_sfr} Pogge R. \& Eskridge P.,  1987, AJ, 92, 291

\bibitem[\protect\citeauthoryear{Postnov} {2003}]{postnov} Postnov K.,
2003, Ast.Lett., 29, 372 

\bibitem[\protect\citeauthoryear{Prugniel \& Heraudeau}
{1998}]{leda} Prugniel Ph. \& Heraudeau  Ph., 198, A\&AS, 128, 299 (HyperLeda)

\bibitem[\protect\citeauthoryear{Reichen et al.}{1994}]{m81bulge}
Reichen M. et al., 1994, A\&AS, 106, 523

\bibitem[\protect\citeauthoryear{Rosati et al.}{2002}]{rosati02}
Rosati P. et al., 2002, ApJ, 566, 667

\bibitem[\protect\citeauthoryear{Sarazin, Irwin \& Bregman}{Sarazin et
al.} {2001}]{ngc4697} Sarazin C.L., Irwin J.A.\& Bregman J.N,  2001,
ApJ, 556, 533 

\bibitem[\protect\citeauthoryear{Sunyaev et al.}{1991}]{gc_sigma}
Sunyaev R. et al., A\&A, 247, L29

\bibitem[\protect\citeauthoryear{Swartz et al.}{Swartz et al.}
{2003}]{m81} Swartz D. et al., 2003, ApJSS, 144, 213


\bibitem[\protect\citeauthoryear{Trinchieri \& Goudfrooij}{Trinchieri
\& Goudfrooij} {2002}]{ngc5846} Trinchieri G. \& Goudfrooij P., 2002,
A\&A, 386, 472 

\bibitem[\protect\citeauthoryear{Thronson \&Greenhouse}{1988}]
{m2l_mw} Thronson H. \& Greenhouse M., 1988, ApJ, 327, 671

\bibitem[\protect\citeauthoryear{Verbunt \& van den Heuvel}
{1995}]{xrb_evol_rev} Verbunt F. \& van den Heuvel E.P.J, 1995, in:
X-ray Binaries, Eds: W.Lewin, J.van Paradijs \& E.van den Heuvel,
Canbridge Univ.Press, p.457


\bibitem[\protect\citeauthoryear{Vikhlinin \& Forman}
{1995}]{vikhl95} Vikhlinin A. \& Forman W., 1995, ApJ, 455, L109



\end{thebibliography}
\end{document}